\documentclass[prd,nopacs,preprintnumbers,twocolumn,amsmath,nofootinbib,amssymb]{revtex4}
\usepackage{graphicx,color,dcolumn,booktabs,bm}
\usepackage{epstopdf}
\usepackage{longtable,lscape}
\usepackage{txfonts}
\usepackage{overpic}
\usepackage{float}
\usepackage{makecell}
\usepackage{amssymb}
\usepackage{epstopdf}
\usepackage{indentfirst}
\usepackage{feynmf}   
\usepackage{slashed}  
\usepackage{cases}
\usepackage{ulem}
\usepackage{color}
\usepackage{float}
\usepackage{array}
\usepackage{booktabs}
\usepackage{multirow}
\usepackage{tikz}
\usepackage{epstopdf}
\usepackage{epsfig,dsfont,amssymb,amsmath,amsfonts,amsbsy,mathrsfs}
\usepackage{multirow}
\usepackage{graphicx}  
\usepackage{float} 
\usepackage{subfigure} 

\usepackage[colorlinks, citecolor=blue,anchorcolor=red,menucolor=red, linkcolor=red,filecolor=red,runcolor=red,urlcolor=blue,frenchlinks=red]{hyperref}
\makeatletter
\@addtoreset{equation}{section}
\makeatother

\newcommand{\nc}{\newcommand}
\nc{\tj}[1]{\textcolor{red}{Tianjie: #1}}
\allowdisplaybreaks
\begin{document}
\title{Spectroscopic survey of higher-lying states of $B_c$ meson family}
\author{Xue-Jian Li$^{1,2,3,5}$}\email{xjli21@lzu.edu.cn}
\author{Yu-Shuai Li$^{1,2,3,5}$}\email{liysh20@lzu.edu.cn}
\author{Fu-Lai Wang$^{1,2,3,5}$}\email{wangfl2016@lzu.edu.cn}
\author{Xiang Liu$^{1,2,3,4,5}$}\email{xiangliu@lzu.edu.cn}

\affiliation{$^1$School of Physical Science and Technology, Lanzhou University, Lanzhou 730000, China\\
$^2$Lanzhou Center for Theoretical Physics, Key Laboratory of Theoretical Physics of Gansu Province, Lanzhou University, Lanzhou 730000, China\\
$^3$Key Laboratory of Quantum Theory and Applications of MoE, Lanzhou University,
Lanzhou 730000, China\\
$^4$MoE Frontiers Science Center for Rare Isotopes, Lanzhou University, Lanzhou 730000, China\\
$^5$Research Center for Hadron and CSR Physics, Lanzhou University and Institute of Modern Physics of CAS, Lanzhou 730000, China}

\begin{abstract}
In this work, we investigate the complete spectroscopy of the $B_c$ mesons, with a special focus on the consideration of the unquenched effects. To account for such effects, we employ the modified Godfrey-Isgur model and introduce a screening potential. The resulting mass spectrum of the concerned higher $B_c$ states is then presented, showing significant deviations after considering the unquenched effects. 
This emphasizes the importance of considering the unquenched effects when studying of the higher $B_c$ mesons.
Furthermore, we determine the corresponding spatial wave functions of these $B_c$ mesons, which have practical applications in subsequent studies of their decays. These decays include two-body Okuba-Zweig-Iizuka allowed strong decays, dipion transitions between $B_c$ mesons, radiative decays, and some typical weak decays.
With the ongoing high-luminosity upgrade of the Large Hadron Collider, we expect the discovery of additional $B_c$ states in the near future. The knowledge gained from the mass spectrum and the different decay modes will undoubtedly provide valuable insights for future experimental explorations of these higher $B_c$ mesons.
\end{abstract}
\maketitle

\section{Introduction}\label{sec1}

The study of hadron spectroscopy offers a unique avenue to deepen our understanding of the non-perturbative nature of the strong interaction. Over the past few decades, significant progress has been made, both experimentally and theoretically, leading to extensive investigations of the exotic hadronic states, such as the charmonium-like states $XYZ$ and the hidden-charm pentaquark states $P_c/P_{cs}$ \cite{Godfrey:2008nc, Li:2009zu, Liu:2013waa,Yuan:2015kya, Chen:2013wva, HillerBlin:2016odx,Chen:2016qju, Guo:2015umn, LHCb:2019kea, Brambilla:2019esw,Guo:2017jvc,Liu:2019zoy,Meng:2022ozq,Chen:2022asf}. Moreover, notable observations of light flavor hadrons by the BESIII Collaboration \cite{Chu:2016sjc}, as well as heavy flavor hadrons by the LHCb \cite{LHCb:2022sfr,LHCb:2022lzp,Wang:2022zgi, Xu:2022kkh} and Belle collaborations \cite{Onuki:2022ugx,Belle-II:2022fsw,Belle:2022dyc}, indicate that the construction of the conventional hadron family is an ongoing endeavor. The abundance of these phenomena in hadron spectroscopy underscores the field's position at the forefront of precision particle physics.

In contrast to the well-established charmonium and bottomonium families, the $B_c$ meson family remains relatively unexplored, with only a few low-lying $B_c$ states reported in experiments. Pivotal theoretical contributions were made by Refs. \cite{Chang:1992bb, Chang:1992jb}, which proposed the experimental detection of the $B_c$ mesons through the hadron colliders. In 1998, the CDF Collaboration reported the observation of a $B_c$ meson with the mass of $M=(6.40\pm0.39\pm0.13)$ GeV, identified via the $B^\pm\to J/\psi \ell^\pm \nu$ decay \cite{CDF:1998ihx}. This measured mass is consistent with expectations for the ground state of the $B_c$ meson \cite{Lusignoli:1991bn, Braaten:1993jn, Chang:1992jb, Chang:1996jt, Masetti:1995uk}. However, the full establishment of the $B_c$ meson family remains a formidable task, requiring extensive efforts to identify and explore its properties.

The $B_c$ meson family has been the focus of extensive investigations, both experimentally and theoretically, since its initial observation by the CDF Collaboration in 1998 via the $B^\pm\to J/\psi \ell^\pm \nu$ decay \cite{CDF:1998ihx}. Prior to that, searches for the $B_c$ mesons were conducted by the LEP \cite{DELPHI:1996vyn, OPAL:1998gdf, ALEPH:1997oob} and CDF \cite{CDF:1996efe} collaborations. Subsequent experimental efforts by various collaborations have confirmed the existence of the $B_c$ mesons through different decay channels, such as the $B_c \to J/\psi \pi$, $B_c^+ \to J/\psi \pi^+\pi^-\pi^+$, $B^+_c \to B^0_s \pi^+$, and $B_c \to J/\psi K^+ K^- \pi^+$ processes \cite{CDF:1998ihx, CDF:2005yjh, CDF:2007umr, D0:2008bqs, LHCb:2012ag, LHCb:2013vrl, LHCb:2013kwl, LHCb:2013hwj, LHCb:2013xlg, LHCb:2013rud}.

In 2014, the ATLAS Collaboration reported the observation of a structure consistent with the predicted $B_c(2S)$ state, with a mass of $(6842\pm9)$ MeV \cite{ATLAS:2014lga}. Additionally, the CMS and LHCb collaborations observed the excited $B_c(2^1S_0)$ and $B_c(2^3S_1)$ states in the $B^+_c\pi^+\pi^-$ invariant mass spectrum, with masses determined as $(6872.1\pm2.2)$ MeV and $(6841.2\pm1.5)$ MeV, respectively \cite{CMS:2019uhm, LHCb:2019bem}. However, the current Particle Data Group (PDG) includes only two $B_c$ mesons, namely the $B_c(1S)$ and $B_c(2S)$ states \cite{ParticleDataGroup:2022pth}. The limited experimental knowledge of the complete $B_c$ family, particularly the higher excited $B_c$ states, highlights the necessity for further studies in this area.

The study of the $B_c$ meson family plays a crucial role in advancing our understanding of the strong interaction and the non-perturbative regime of Quantum Chromodynamics (QCD). With the forthcoming high-luminosity upgrade of the Large Hadron Collider (LHC), there are promising opportunities to investigate the higher excited states of the $B_c$ meson, and it is anticipated that additional experimental data will become available in the near future.

Some theoretical studies have examined the $B_c$ spectrum using the quenched quark models, as discussed in references \cite{Godfrey:1985xj, Eichten:1994gt, Gershtein:1994dxw, Fulcher:1998ka, Ebert:2002pp, Godfrey:2004ya, Eichten:2019gig}. Among these models, the Godfrey-Isgur (GI) model \cite{Godfrey:1985xj}, proposed by Stephen Godfrey and Nathan Isgur in 1985, has demonstrated notable success in describing the spectra of low-lying hadrons. Additionally, other quenched quark models have exhibited remarkable achievements in predicting the spectra of low-lying mesons. However, it is now widely acknowledged that the unquenched effects play a significant role, as they can resolve the low-mass puzzles of several hadrons, such as the $D_{s0}(2317)$ \cite{vanBeveren:2003kd, Dai:2003yg, Hwang:2004cd, Simonov:2004ar}, $D_{s1}(2460)$ \cite{Dai:2003yg, Ortega:2016mms}, $X(3872)$ \cite{Danilkin:2010cc, Li:2009ad, Kalashnikova:2005ui}, and $\Lambda_c(2940)$ \cite{Luo:2019qkm}. Therefore, when exploring the spectroscopy of higher excited states of the $B_c$ meson, it is imperative to consider the unquenched effects as well.

In this study, we employ the modified Godfrey-Isgur (MGI) model \cite{Song:2015nia, Song:2015fha, Wang:2018rjg, Wang:2019mhs, Wang:2020prx} to account for the unquenched effects. To reflect such effects, we utilize a screening potential introduced in Refs. \cite{Chao:1992et, Ding:1993uy}. {Furthermore, this method has been applied to the study of bottomonia \cite{Wang:2018rjg} and charmonia family \cite{Wang:2019mhs,Wang:2020prx}, and has yielded the predictions, providing valuable guidance for experimental investigations.  It should be noted that the couple channel effect and the screening potential have a similar effect on the hadron masses, i.e. the higher excited state masses are reduced \cite{Li:2009ad,Duan:2020tsx,Duan:2021alw}.} In the following section, we provide a concise overview of the MGI model and present the mass spectrum of the $B_c$ family, along with a comparison to the quenched quark model, in order to elucidate the distinctions between the two models. Furthermore, we present the numerical spatial wave functions obtained from the MGI model, which serve as crucial inputs for investigating various properties. Specifically, we calculate the magnetic moments of the higher excited $B_c$ states to reveal their internal structures. This analysis can be valuable in differentiating between conventional $B_c$ states and exotic $B_c$-like molecular tetraquark states with identical quantum numbers and similar masses.

To provide a comprehensive theoretical analysis, we also calculate the two-body Okubo-Zweig-Iizuka (OZI)-allowed strong decays, dipion decays, radiative decays, and typical weak decays by employing the numerical spatial wave functions of the relevant mesons. In the concrete calculations, the quark pair creation (QPC) model \cite{Micu:1968mk, LeYaouanc:1972vsx, Ackleh:1996yt, Blundell:1996as} is utilized to describe the behavior of the two-body OZI-allowed strong decays, while the electric dipole (E1) and magnetic dipole (M1) radiative transitions are analyzed by considering the radial decays of the higher excited $B_c$ states. For the investigations of dipion transitions, we adopt the quantum chromodynamics multipole expansion (QCDME) method. Additionally, we employ the covariant light-front quark model (CLFQM) to calculate a series of weak transition form factors and their corresponding weak decays. The spatial wave functions play a crucial role in these decay processes, and by utilizing the MGI model, we obtain reliable numerical results. As experimental data continue to accumulate, these decay processes can be further explored and potentially measured with higher precision.

This paper is organized as follows.
In Section \ref{sec2}, we present our analysis of the mass spectrum of $B_c$ mesons using the MGI model with the screening effects, and provide the associated numerical spatial wave functions.
Section \ref{sec3} is dedicated to the predictions of two-body OZI-allowed strong decays of the considered $B_c$ mesons by employing the QPC model.
In Section \ref{sec4}, we investigate the dipion transitions between the $B_c$ states.
Section \ref{sec5} focuses on the predictions of the decay widths for the E1 and M1 radiative transitions and the magnetic moments of these higher $B_c$ mesons.
Furthermore, in Section \ref{sec6}, we explore some concerned weak transition form factors and the corresponding weak decays for the $B_c$ meson.
Finally, our findings are summarized in Section \ref{sec7}.

\section{Mass spectrum and the corresponding spatial wave functions}\label{sec2}

In this study, we employ the MGI model to calculate the mass spectrum of the higher excited $B_c$ states. To accurately account for the screening effects, we incorporate a screening potential into our calculations. Furthermore, we derive the corresponding numerical spatial wave functions, which play a pivotal role in the investigations of the decay properties of the $B_c$ states.

\subsection{The mass spectrum}

To obtain the mass spectrum of the $B_c$ mesons, we apply the MGI model by introducing the screening potential \cite{Song:2015nia,Song:2015fha,Pang:2017dlw,Wang:2018rjg,Wang:2019mhs,Wang:2020prx,Wang:2021abg}.
The involved Hamiltonian is
\begin{align}\label{2.1}
\tilde{H}=(p^2+m_1^2)^{1/2}+(p^2+m_2^2)^{1/2}+V_{\rm eff}(\boldsymbol{p},\boldsymbol{r}),
\end{align}
where $m_{1}$ and $m_{2}$ denote the masses of the $b$ and $c$ quarks, respectively. $V_{\rm eff}(\boldsymbol{p},\boldsymbol{r})={H}^{\rm conf}+{H}^{\rm hyp}+{H}^{\rm so}$ is the effective potential representing the $q'\bar q$ interaction.

In the non-relativistic limit, $V_{\rm eff}(\boldsymbol{p},\boldsymbol{r})$ is transformed into the familiar nonrelativistic potential $V_{\rm eff}(r)$, which can be written as
\begin{align}\label{2.2}
V_{\rm eff}(r)=H^{\rm conf}+H^{\rm hyp}+H^{\rm so},
\end{align}
where
\begin{align}\label{2.3}
H^{\rm conf}=c+\frac{b(1-e^{-\mu r})}{\mu}-\frac{4\alpha_s(r)}{3r}
\end{align}
is the spin-independent potential, which includes the screening potential and the Coulomb-like potential. Here, $\alpha_s(r)$ is a running coupling constant, and $\mu$ is a parameter which stands for the strength of the screening effects. The colour hyperfine interaction is given by
\begin{align}\label{2.4}
H^{\rm hyp}=\frac{4\alpha_s(r)}{3m_1m_2}
		\left[\frac{8\pi}{3}\boldsymbol{S}_1{\!}\cdot{\!}\boldsymbol{S}_2\delta^3(\boldsymbol{r})
		{\!}+{\!}\frac{1}{r^3}\left(\frac{3\boldsymbol{S}_1{\!}\cdot{\!}\boldsymbol{r}\boldsymbol{S}_2{\!}\cdot{\!}\boldsymbol{r}}{r^2}{\!}-{\!}\boldsymbol{S}_1{\!}\cdot{\!}\boldsymbol{S}_2\right)\right],
\end{align}
where $\boldsymbol{S}_{1(2)}$ is the spin of quark or antiquark. And $H^{\rm so}=H^{\rm so(cm)}+H^{\rm so(tp)}$ is the spin-orbit interaction, where
\begin{align}\label{2.5}
H^{\rm so(cm)}=\frac{4 \alpha_{s}(r)}{3} \frac{1}{r^{3}}\left(\frac{ \boldsymbol{S}_{1}}{m^2_{1}}+\frac{\boldsymbol{S}_{2}}{m^2_{2}}+\frac{
\boldsymbol{S}_{1}+\boldsymbol{S}_{2}}{m_{1}m_{2}}\right) \cdot\boldsymbol{L}
\end{align}
is the color-magnetic term resulting from the one-gluon exchange, while 
\begin{align}\label{2.6}
H^{\rm so(tp)}=-\frac{1}{2 r} \frac{\partial H^{\rm conf}}{\partial r}\left(\frac{ \boldsymbol{S}_{1}}{m_{1}^{2}}+\frac{ \boldsymbol{S}_{2}}{m_{2}^{2}}\right) \cdot\boldsymbol{L}
\end{align}
denotes the Thomas precession term with the screening effects, where $\boldsymbol{L}$ is the
orbital angular momentum between quark and antiquark. In addition, we need to smear the screened potential $S(r)=\frac{b(1-e^{-\mu r})}{\mu}+c$ and the Coulomb-like potential $G(r)=-\frac{4\alpha_s(r)}{3r}$ by
\begin{align}\label{2.7}
\tilde{S}(r)/\tilde{G}(r)=\int d^3\boldsymbol{r}'\rho(\boldsymbol{r}-\boldsymbol{r}')S(r')/G(r'),
\end{align}
where
\begin{align}\label{2.8}
\rho(\boldsymbol{r}-\boldsymbol{r}')=\frac{\sigma^3}{\pi^{3/2}}\mathrm{exp}\left[-\sigma^2(\boldsymbol{r}-\boldsymbol{r}')^2\right]
\end{align}
is the smearing function, and $\sigma$ is a smearing parameter. Then, we introduce the momentum dependent factors as follows
\begin{align}\label{2.9}
&\tilde{G}(r)\to\left(1+\frac{p^2}{E_1E_2}\right)^{1/2}\tilde{G}(r)\left(1+\frac{p^2}{E_1E_2}\right)^{1/2},\nonumber\\
&\tilde{V}_i(r)\to \left(\frac{m_1m_2}{E_1E_2}\right)^{1/2+\epsilon_i}\tilde{V}_i(r)
\left(\frac{m_1m_2}{E_1E_2}\right)^{1/2+\epsilon_i}
\end{align}
with $E_{1(2)}=(p^2+m_{1(2)}^2)^{1/2}$, where $\epsilon_i$ correspond to different types of the interactions, and $\tilde{V}_i(r)$ are the effective potentials included in Eqs.~\eqref{2.4}-\eqref{2.6}.

\begin{figure*}[htpb]\centering
\includegraphics[width=0.9\textwidth]{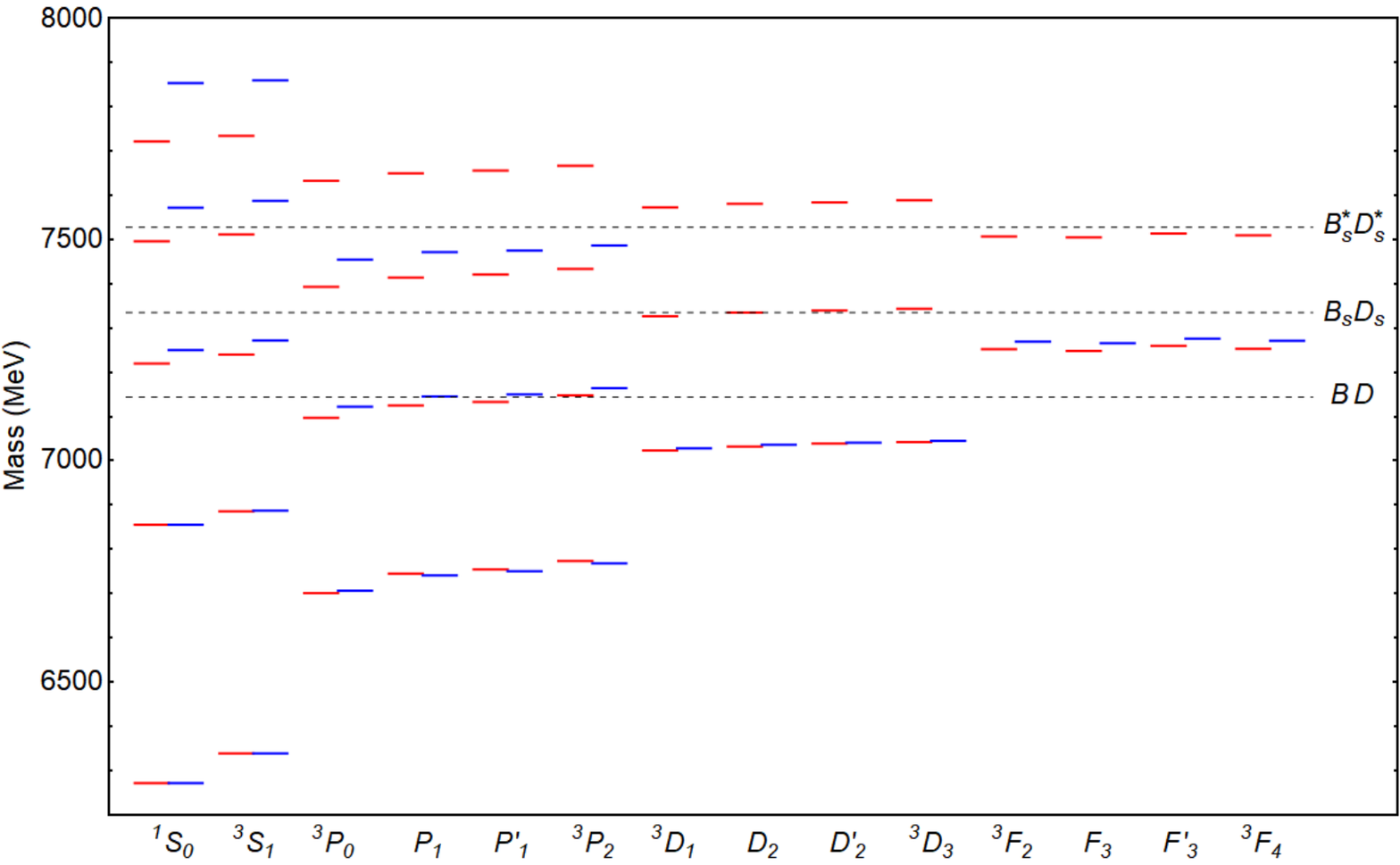}
\caption{\label{fig:spectrum}(Color online) Mass spectrum of the $B_c$ mesons. Here, the red (left) lines and the blue (right) lines are our obtained results and the results from the GI model \cite{Godfrey:2004ya}, respectively, while the short lines denote the thresholds of the $B_{(s)}^{(*)}D_{(s)}^{(*)}$ channels. The masses of the mesons are given in units of MeV.}
\end{figure*}

\begin{table}[htbp]\centering
\caption{The masses (in units of MeV) of the experimental values and our obtained results of the heavy flavor mesons in this work.}
\renewcommand\arraystretch{1.2}
\label{experimental}
\begin{tabular*}{85mm}{c@{\extracolsep{\fill}}cccc}
\toprule[1.5pt]\toprule[0.5pt]
\text{Mesons} & \text{States}   &\text{Experimental values \cite{ParticleDataGroup:2022pth}}  & \text{This work}  \\
\toprule[1pt]
\multirow{2}{*}{$B_c$}       & $1^1S_0$ & $6274.47\pm0.44$ &6271 \\
            & $2^1S_0$ & $6871.2\pm1$ &6855&  \\
      
\hline
\multirow{8}{*}{$c\bar{c}$}  & $1^1S_0$ & $2983.9\pm0.4$ &2969 \\
            & $1^3S_1$ & $3096.9\pm0.006$ &3097& \\
            & $1^3P_0$ & $3414.71\pm0.3$ &3425& \\
            & $1^3P_1$ & $3510.67\pm0.05$ &3497& \\
            & $1^1P_1$ & $3525.38\pm0.11$ &3516& \\
            & $1^3P_2$ & $3556.17\pm0.07$ &3554& \\
            & $2^1S_0$ & $3637.5\pm1.1$ &3616 &\\
            & $2^3S_1$ & $3686.1\pm0.06$ &3668& \\
\hline
\multirow{8}{*}{$b\bar{b}$}  & $1^1S_0$ & $9398.7\pm2.0$ &9415& \\
            & $1^3S_1$ & $9460.3\pm0.26$ &9466& \\
            & $1^3P_0$ & $9859.4\pm0.42\pm0.31$ &9851& \\
            & $1^3P_1$ & $9892.8\pm0.26\pm0.31$ &9880& \\
            & $1^1P_1$ & $9899.3\pm0.8$ &9887& \\
            & $1^3P_2$ & $9912.2\pm0.26\pm0.31$ &9903& \\
            & $2^1S_0$ & $9999\pm6.3$ &9991& \\
            & $2^3S_1$ & $10023.26\pm0.31$ &10012& \\
\hline
\multirow{8}{*}{$D$}         & $1^1S_0$ & $1864.84\pm0.05$ &1862& \\
            & $1^3S_1$  & $2006.85\pm0.05$ &2042& \\
            & $1^3P_0$ & $2343\pm10$ &2291 &\\
            & $1 P_1$  & $2412\pm9$ &2387& \\
            & $1 P'_1$ & $2422.1\pm0.6$ &2467& \\
            & $1^3P_2$ & $2461.1\pm0.7$ &2474& \\
            & $2^1S_0$ & $2549\pm19$ &2541& \\
            & $2^3S_1$ & $2627\pm10$ & 2607 &\\
\hline
\multirow{8}{*}{$D_s$}       & $1^1S_0$ & $1968.35\pm0.07$ &1968& \\
            & $1^3S_1$ & $2112.2\pm0.4$ &2129& \\
            & $1^3P_0$ & $2317.8\pm0.5$ & 2416 &\\
            & $1 P_1$  & $2459.5\pm0.6$ &2518& \\
            & $1 P'_1$ & $2535.11\pm0.06$ &2545& \\
            & $1^3P_2$ & $2569.1\pm0.8$ &2588& \\
            & $2^1S_0$ & $2591\pm13$ &2643&\\
            & $2^3S_1$ & $2714\pm5$ &2704& \\
\hline
\multirow{3}{*}{$B$}         & $1^1S_0$ & $5279.66\pm0.12$ &5308& \\
            & $1^3S_1$ & $5324.71\pm0.21$ &5372& \\
            & $1^3P_2$ & $5737.2\pm0.7$ &5760& \\
\hline
\multirow{3}{*}{$B_s$}       & $1^1S_0$ & $5366.92\pm0.10$ &5391& \\
            & $1^3S_1$ & $5415.4\pm1.8$ &5451& \\
            & $1^3P_2$ & $5839.86\pm0.12$ &5868&\\
\bottomrule[0.5pt]\bottomrule[1.5pt]
\end{tabular*}
\end{table}

\begin{table}[htbp]\centering
\caption{The fitting parameters of the potential model in this work. Besides, the quark masses are chosen as $m_{u}=m_{d}=220\ \text{MeV}$, $m_{s}=419\ \text{MeV}$, $m_{c}=1628\ \text{MeV}$, and $m_{b}=4977\ \text{MeV}$~\cite{Godfrey:2004ya}.}
\renewcommand\arraystretch{1.2}
\label{Parameters}
\begin{tabular*}{85mm}{c@{\extracolsep{\fill}}lccc}
\toprule[1.5pt]\toprule[0.5pt]
\text{Parameters}   &\text{Values}      &\text{Parameters}    &\text{Values}\\
\toprule[1pt]
$b$                   & $0.2053~\text{GeV}^2$ &$\epsilon_t$           &0.5034&\\
$c$                   & $-0.2677$~\text{GeV}    &$\epsilon_{so(V)}$     &$-0.3105$&\\
$\mu$                 & 0.0684~\text{GeV}     &$\epsilon_{so(S)}$     &$-0.3195$&\\
$\epsilon_c$          &$-0.1981$&               &                     &\\
\bottomrule[0.5pt]\bottomrule[1.5pt]
\end{tabular*}
\end{table}

\begin{table}[htbp]\centering
\caption{The obtained masses of the $B_c$ states and comparison with other results and experimental data. Here, the masses are given in units of MeV.}
\renewcommand\arraystretch{1.2}
\footnotesize
\label{Mass spectrum}
\begin{tabular*}{88mm}{c@{\extracolsep{\fill}}ccccccc}
\toprule[1.5pt]\toprule[0.5pt]
\text{States} & \text{Experiments} & \text{This work} & \text{GI \cite{Godfrey:2004ya}} & \text{Ref. \cite{Eichten:2019gig}} & \text{Ref. \cite{Ding:2021dwh}} & \text{Ref. \cite{Wang:2022cxy}} \\
\toprule[1pt]
$1^1S_0$& 6274 & 6271 & 6271 & 6275 & 6275 & 6277\\
$2^1S_0$& 6871 & 6855 & 6855 & 6866 & 6853 & 6867\\
$3^1S_0$&      & 7220 & 7250 & 7253 & 7222 & 7228\\
$4^1S_0$&      & 7496 & 7572 & 7572 & 7484 &     \\
$5^1S_0$&      & 7722 & 7854 &      &      &     \\
$1^3S_1$&      & 6338 & 6338 & 6329 & 6339 & 6332\\
$2^3S_1$&      & 6886 & 6887 & 6897 & 6920 & 6911\\
$3^3S_1$&      & 7240 & 7272 & 7279 & 7283 & 7272\\
$4^3S_1$&      & 7512 & 7588 & 7595 & 7543 &     \\
$5^3S_1$&      & 7735 & 7860 &      &      &     \\
$1^3P_0$&      & 6701 & 6706 &      &      & 6705\\
$1^3P_2$&      & 6773 & 6768 &      &      & 6762\\
$1P_1$  &      & 6745 & 6741 &      &      & 6739\\
$1P'_1$ &       & 6754 & 6750 &      &      & 6748\\
$\theta_{1p}$& & $35.2^{\circ}$ & $22.4^{\circ}$ & & & $32.2^{\circ}$\\
$2^3P_0$&      & 7097 & 7122 & 6692 &      & 7112\\
$2^3P_2$&      & 7148 & 7164 & 6750 &      & 7163\\
$2P_1  $&      & 7125 & 7145 & 6730 &      & 7144\\
$2P'_1 $&      & 7133 & 7150 & 6738 &      & 7149\\
$\theta_{2p}$& & $26.5^{\circ}$ & $18.9^{\circ}$ & $18.7^{\circ}$ & & $30.9^{\circ}$\\
$3^3P_0$&      & 7393 & 7455 & 7104 &      &     \\
$3^3P_2$&      & 7434 & 7487 & 7154 &      &     \\
$3P_1  $&      & 7414 & 7472 & 7135 &      &     \\
$3P'_1 $&      & 7421 & 7475 & 7143 &      &     \\
$\theta_{3p}$& & $23.6^{\circ}$ & $18.9^{\circ}$ & $21.2^{\circ}$ &      & \\
$4^3P_0$&      & 7633 &  &  &  &\\
$4^3P_2$&      & 7667 &  &  &  &\\
$4P_1  $&      & 7650 &  &  &  &\\
$4P'_1 $&      & 7656 &  &  &  &\\
$\theta_{4p}$& & $22.2^{\circ}$ & &  &      & \\
$1^3D_1$&      & 7023 & 7028 &      &      & 7014\\
$1^3D_3$&      & 7042 & 7045 &      &      & 7035\\
$1D_2  $&      & 7032 & 7036 &      &      & 7025\\
$1D'_2 $&      & 7039 & 7041 &      &      & 7029\\
$\theta_{1D}$& & $-53.4^{\circ}$ & $45.5^{\circ}$ &  &  & $38.1^{\circ}$\\
$2^3D_1$&      & 7327 &  &      &      & \\
$2^3D_3$&      & 7344 &  &      &      & \\
$2D_2  $&      & 7335 &  &      &      & \\
$2D'_2 $&      & 7340 &  &      &      & \\
$\theta_{2D}$& & $-48.4^{\circ}$ &  &  &      & \\
$3^3D_1$&      & 7573 &  &      &      & \\
$3^3D_3$&      & 7589 &  &      &      & \\
$3D_2  $&      & 7581 &  &      &      & \\
$3D'_2 $&      & 7584 &  &      &      & \\
$\theta_{3D}$& & $-42.9^{\circ}$ &  &  &      & \\
$1^3F_2$&      & 7252 & 7269 &      &      &     \\
$1^3F_4$&      & 7253 & 7271 &      &      &     \\
$1F_3  $&      & 7248 & 7266 &      &      &     \\
$1F'_3 $&      & 7260 & 7276 &      &      &     \\
$\theta_{1F}$& & $-50.4^{\circ}$ & $41.4^{\circ}$ &      &      &     \\
$2^3F_2$&      & 7507 &  &      &      &     \\
$2^3F_4$&      & 7510 &  &      &      &     \\
$2F_3$  &      & 7505 &  &      &      &     \\
$2F'_3$ &      & 7514 &  &      &      &     \\
$\theta_{2F}$& & $-49.5^{\circ}$ &  &      &      &     \\
\bottomrule[0.5pt]\bottomrule[1.5pt]
\end{tabular*}
\end{table}

For mesons composed of heavy quarks with equal masses, such as the charmonium and bottomonium states, the $L$-$S$ coupling scheme is appropriate and the meson state can be labeled by the notation $n^{2S+1}L_J$. However, for mesons with constituents of different masses, such as the $B_c$ mesons, the spin-dependent terms in the Hamiltonian can mix the spin-singlet and spin-triplet states, and the resulting mixing states can be expressed as
\begin{align}\label{2.10}
L^{\prime}&={ }^{1} L_{J} \cos \theta+{ }^{3} L_{J} \sin \theta, \nonumber\\
L&=-{ }^{1} L_{J} \sin \theta+{ }^{3} L_{J} \cos \theta,
\end{align}
where $\theta$ represents the mixing angle.

To solve the Schr\"{o}dinger equation \cite{Lucha:1991vn,Eichten:1974af}, we use the simple harmonic oscillator (SHO) wave functions as a set of complete bases, i.e.,
\begin{align}\label{2.11}
&\Psi_{n L M_{L}}({\boldsymbol{r}})=R_{n L}(r, \beta) Y_{L, M_{L}}\left(\Omega_{\boldsymbol r}\right), \nonumber\\
&\Psi_{n L M_{L}}({\boldsymbol{p}})=R_{n L}(p, \beta) Y_{L, M_{L}}\left(\Omega_{\boldsymbol p}\right),
\end{align}
where
\begin{align}\label{2.12}
&R_{n L}(r, \beta)=\beta^{\frac{3}{2}} \sqrt{\frac{2 n !}{\Gamma\left(n+L+\frac{3}{2}\right)}}(\beta r)^{L} e^{\frac{-r^{2} \beta^{2}}{2}} L_{n}^{L+\frac{1}{2}}\left(\beta^{2} r^{2}\right), \nonumber\\
&R_{n L}(p, \beta)=\frac{({\!}-{\!}1)^{n}({\!}-{\!}i)^{L}}{\beta^{\frac{3}{2}}} e^{-\frac{p^{2}}{2 \beta^{2}}}{\!} \sqrt{\frac{2 n !}{\Gamma\left(n{\!}+{\!}L{\!}+{\!}\frac{3}{2}\right)}}\left(\frac{p}{\beta}\right)^{L} {\!} L_{n}^{L+\frac{1}{2}}\left(\frac{p^{2}}{\beta^{2}}\right).
\end{align}
Here, the radial wave functions are denoted by $R_{n L}(r, \beta)$ and $R_{n L}(p, \beta)$ in the coordinate and momentum spaces, respectively. $Y_{L, M_{L}}\left(\Omega_{\boldsymbol r}\right)$ and $Y_{L, M_{L}}\left(\Omega_{\boldsymbol p}\right)$ stand for the spherical harmonic functions, and $L_{n}^{L+\frac{1}{2}}(x)$ is the Laguerre polynomial. $\beta$ is a phenomenological parameter in the SHO wave function, and we set it to be $\beta=0.5\ \text{GeV}$ in the calculation.

In Table \ref{experimental}, we list the reported experimental data \cite{ParticleDataGroup:2022pth} of the masses of bottom-charmed, charmonium, bottomonium, charmed, charmed-strange, bottom, and bottom-strange mesons, which are quoted from the PDG \cite{ParticleDataGroup:2022pth} and are used to fit the parameters of the MGI model. Due to the limited availability of experimental data for the $B_c$ mesons, we also include experimental data from other mesons as auxiliary. By fitting the experimental data in Table \ref{experimental}, we can obtain the mass spectrum of the $B_c$ mesons. Here, the accuracy of the fitting data is judged based on the $\chi^{2}$ criterion as follows
\begin{align}
\chi^{2}=\sum_{i} \frac{\left(m^{\text{Th}}_i-m^{\text{Exp}}_i\right)^2}{m^{\text{Er2}}_i},
\end{align}
where $m^{\text{Th}}_i$, $m^{\text{Exp}}_i$, and $m^{\text{Er}}_i$ are the theoretical value, the experimental value, and the error of the $i$-th data, respectively. Here, the errors $m^{\text{Er}}_i=1\ \text{MeV}$ are the uniform values for all considered mesons. 
In the absence of experimental data on the masses of the $B_c$ mesons, we used the experimental data of other heavy flavor mesons listed in Table~\ref{experimental} to fit the parameters of the model. To make the mesons act in the same proportions in our fit, we chose a universe value of 1 MeV as the uncertainty. Thus, in this work we used it only as a mathematical tool to fit the model parameters, rather than a $\chi^{2}$ fit in the usual sense. In this case, we have not given the $\chi^2/\text{d.o.f.}$ value and have not considered the uncertainties of the parameters.

In Table \ref{Parameters}, we present the fitted parameters of the MGI model, and these parameters were determined by selecting the minimum $\chi^{2}$ value. By utilizing the parameters in Table \ref{Parameters}, we can calculate the mass spectrum of the $B_c$ mesons, which are displayed in Table \ref{Mass spectrum}.
In addition to the mass spectrum, the Table \ref{Mass spectrum} also includes the mixing angles of the $P$-wave, $D$-wave, and $F$-wave $B_c$ mesons. Furthermore, we provide a comparison with other theoretical studies \cite{Godfrey:2004ya, Eichten:1994gt, Ding:2021dwh, Wang:2022cxy}.

In Fig. \ref{fig:spectrum}, we present a comparative analysis of the $B_c$ meson mass spectrum, considering both the screening effects and the results obtained by the GI model \cite{Godfrey:2004ya}. Our findings reveal the significant impact of the screening effects on the mass spectrum of the higher excited $B_c$ states, as compared to the predictions based solely on the GI model \cite{Godfrey:2004ya}. Several notable examples are highlighted below:

(i) The $B_c(4 ^1S_0)$ and $B_c(4 ^3S_1)$ states exhibit mass reductions of approximately 76 MeV and 76 MeV, respectively, in our calculations compared to the GI model predictions \cite{Godfrey:2004ya}.


(ii) The impact of the screening effects becomes even more pronounced in the $B_c(5 ^1S_0)$ and $B_c(5 ^3S_1)$ states, with a substantial mass difference of up to 120 MeV observed between our results and those obtained without considering the screening effects \cite{Godfrey:2004ya}. This disparity is particularly evident in the higher excited $B_c$ states, as depicted in Fig. \ref{fig:spectrum}.

Hence, our findings underscore the necessity of accounting for the unquenched effects when investigating the spectroscopy of higher excited $B_c$ mesons. Future experimental studies with more precise data are expected to provide a valuable opportunity for testing and validating our predictions.
\begin{figure*}[htbp]
\centering
\includegraphics[width=17.6cm,keepaspectratio]{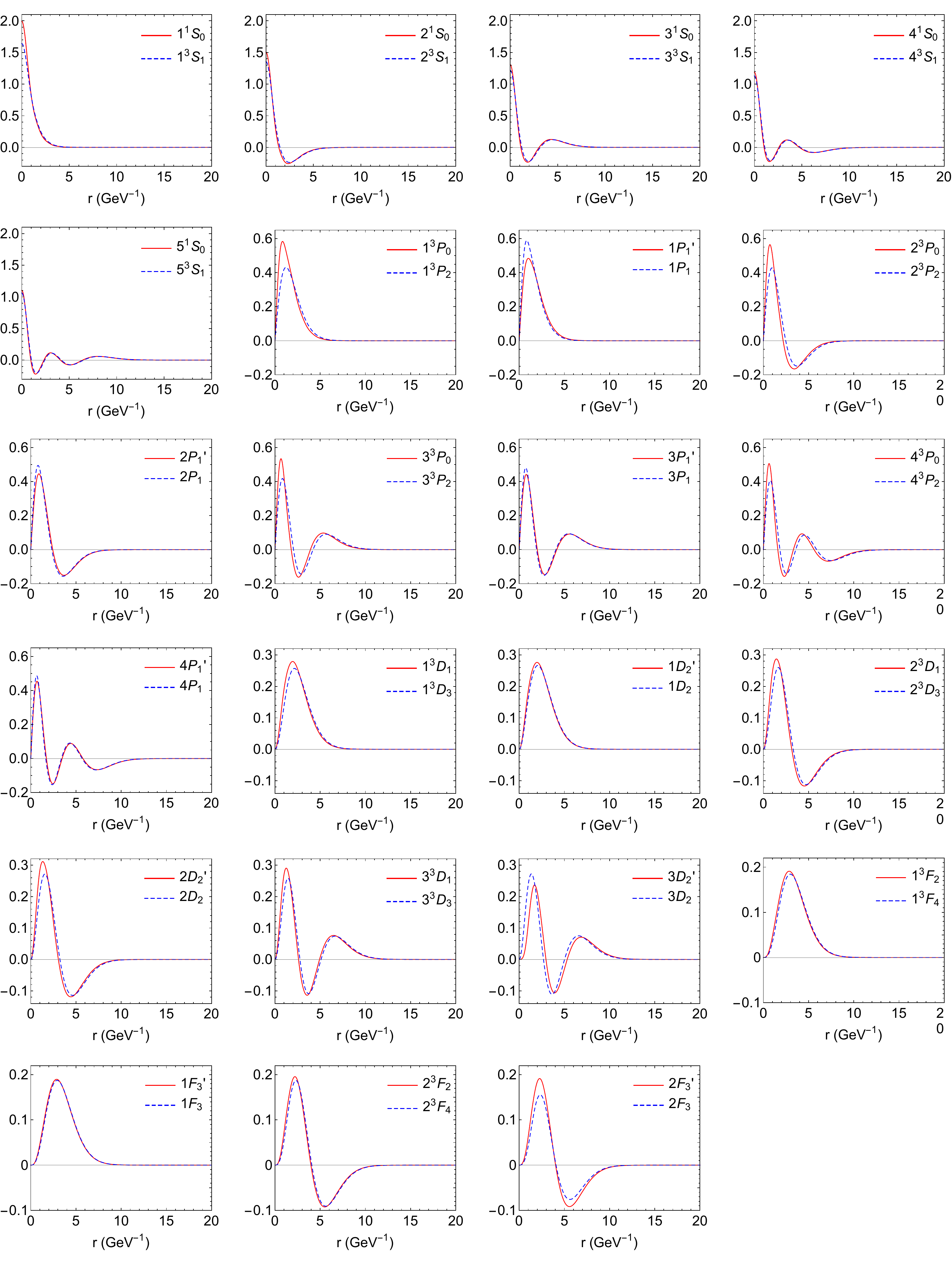}
\caption{The obtained spatial wave functions of the concerned $B_c$ mesons. Here, these $B_c$ mesons are classified by the notation $n^{2S+1}L_J$.}
\label{fig:wavefunctions}
\end{figure*}

In the preceding subsection, we derived the mass spectrum of the $B_c$ mesons. Furthermore, we can extract the corresponding spatial wave functions, which serve as crucial inputs for investigating their decay properties. Consequently, in this subsection, we present these spatial wave functions to explore their characteristics. The figure \ref{fig:wavefunctions} illustrates the $S/P/D/F$-wave spatial wave functions of the $B_c$ mesons. Notably, for the spatial wave functions of the $S$-wave $B_c$ mesons, noticeable node effects emerge when $n$ exceeds 4, such as $n=5$, and so on. These node effects can substantially impact certain physical results, which will be discussed in Section \ref{sec3}.

\section{Two-body OZI-allowed strong decays}\label{sec3}

In this section, our focus lies on the study of two-body OZI-allowed strong decays of the $B_c$ mesons by employing the QPC model in the concrete calculations. The QPC model, also known as the $^3P_0$ model, was initially proposed by Micu in 1968 \cite{Micu:1968mk} and has since been further developed by the Orsay Group \cite{LeYaouanc:1973ldf,LeYaouanc:1974cvx,LeYaouanc:1977fsz}. Over time, the QPC model has been extensively utilized to investigate the two-body OZI-allowed strong decays of various hadrons \cite{LeYaouanc:1972vsx,LeYaouanc:1973ldf,LeYaouanc:1974cvx,LeYaouanc:1977fsz,Ackleh:1996yt,Blundell:1996as}. Here, we provide a brief introduction to the QPC model, with a specific focus on the transition matrix for the $A\to B+C$ process.

In the QPC model, the transition matrix for the $A\to B+C$ decay is defined as follows
\begin{equation}\label{3.6}
\langle BC|\mathcal{T}|A \rangle = \delta ^3(\boldsymbol{P}_B+\boldsymbol{P}_C)\mathcal{M}^{{M}_{J_{A}}M_{J_{B}}M_{J_{C}}}.
\end{equation}
Here, $\mathcal{M}^{{M}_{J_{A}}M_{J_{B}}M_{J_{C}}}$ represents the helicity amplitude, while $\boldsymbol{P}_B$ and $\boldsymbol{P}_C$ denote the momenta of the daughter mesons $B$ and $C$, respectively, in the rest frame of the parent meson $A$. The states $|A \rangle$, $|B \rangle$, and $|C \rangle$ correspond to the mock states associated with mesons $A$, $B$, and $C$, respectively. Furthermore, $\mathcal{T}$ represents the transition operator, which describes the creation of a quark-antiquark pair from the vacuum. In the non-relativistic limit, the transition operator can be expressed as
\begin{align}\label{3.7}
\mathcal{T}& = -3\gamma \sum_{m}\langle 1,m;1,-m|0,0\rangle\int d {\boldsymbol{p}}_3d {\boldsymbol{p}}_4\delta ^3 ({\boldsymbol{p}}_3+{\boldsymbol{p}}_4) \nonumber \\
 & ~
 \quad\times \mathcal{Y}_{1m}\left(\frac{{\boldsymbol{p}}_3-{\boldsymbol{p}}_4}{2}\right)\chi _{1,-m}^{34}\phi _{0}^{34}
\left(\omega_{0}^{34}\right)_{ij}b_{3i}^{\dag}({\boldsymbol{p}}_3)d_{4j}^{\dag}({\boldsymbol{p}}_4),
\end{align}
where the quark and antiquark are represented by indices 3 and 4, respectively. The state $\chi_{1,-m}^{34}$ corresponds to a spin-triplet configuration, while $\phi_0^{34}$ and $\omega_0^{34}$ denote the SU(3) flavor and color singlets, respectively. The parameter $\gamma$ is a dimensionless constant that characterizes the strength of quark-antiquark pair creation from the vacuum, and its value is determined through the fitting of experimental data. The term $\mathcal{Y}_{lm}(\boldsymbol{p}) = |\boldsymbol{p}|^l Y_{lm}(\boldsymbol{p})$ represents the solid harmonic function. According to the Jacobi-Wick formula \cite{Jacob:1959at}, the helicity amplitude can be transformed into the partial wave amplitude, which can be expressed as
\begin{align}\label{3.8}\begin{split}
\mathcal{M}^{J L}({\boldsymbol{P}})=& \frac{\sqrt{4 \pi(2 L+1)}}{2 J_{A}+1} \sum_{M_{J_{B}} M_{J_{C}}}\left\langle L 0 ; J M_{J_{A}} \mid J_{A} M_{J_{A}}\right\rangle \\
& \times\left\langle J_{B} M_{J_{B}} ; J_{C} M_{J_{C}} \mid J_{A} M_{J_{A}}\right\rangle \mathcal{M}^{M_{J_{A}} M_{J_{B}} M_{J_{C}}},
\end{split}\end{align}
where $L$ is the orbital angular momentum between final states $B$ and $C$. The partial width of the $A \to B + C$ process can be given by
\begin{align}\label{3.9}
\Gamma=\frac{\pi}{4} \frac{|\boldsymbol{P}_B|}{m_{A}^{2}} \sum_{J, L}\left|\mathcal{M}^{J L}({\boldsymbol{P}})\right|^{2},
\end{align}
where ${m}_{A}$ is the mass of the parent meson $A$.

In addition, the meson wave function is defined as the mock state \cite{Hayne:1981zy}, i.e.,
\begin{align}\label{3.10}\begin{split}
\left|A\left(n^{2 S+1} L_{J M_{J}}\right)\left({\boldsymbol{p}}_{A}\right)\right\rangle=& \sqrt{2 E} \sum_{M_{S}, M_{L}}\left\langle L M_{L}; S M_{S} \mid J M_{J}\right\rangle \chi_{S M_{S}}^{A} \\
& \times \phi^{A} \omega^{A} \int d {\boldsymbol{p}}_{1} d {\boldsymbol{p}}_{2} \delta^{3}\left({\boldsymbol{p}}_{A}-{\boldsymbol{p}}_{1}-{\boldsymbol{p}}_{2}\right) \\
& \times \Psi_{n L M_{L}}^{A}\left({\boldsymbol{p}}_{1}, {\boldsymbol{p}}_{2}\right)\left|q_{1}\left({\boldsymbol{p}}_{1}\right) \bar{q}_{2}\left({\boldsymbol{p}}_{2}\right)\right\rangle,
\end{split}
\end{align}
where $\chi_{S M_{S}}^{A}$, $\phi^{A}$, $\omega^{A}$, and $\Psi_{n L M_{L}}^{A}\left(\boldsymbol{p}_{1}, \boldsymbol{p}_{2}\right)$ denote the spin, flavor, color, and spatial wave functions of meson $A$, respectively. In our concrete calculations, we utilize the numerical spatial wave functions obtained in Section \ref{sec2} as inputs for these mesons. This approach helps to avoid the dependence on the $\beta$ values by taking a SHO wave function. Furthermore, since experimental width data is lacking, we adopt the value of $\gamma$ for $q\bar q$ as $0.4\sqrt{96\pi}$ from Ref. \cite{Li:2019tbn}, while the creation strength for $s\bar s$ satisfies $\gamma_s=\gamma/\sqrt{3}$.

In Table \ref{strong decay 1}, we present the two-body OZI-allowed strong decay widths and the corresponding branching ratios of the $S$-wave and $P$-wave $B_c$ states with $n=3,4,5$ and $n=3,4$, respectively. These discussed $B_c(n^1S_0)$ states with $n=3,4,5$ can decay into the $B^*D$ channel. As we increase the principal quantum number $n$, such as $n=5$, we observe a significant reduction in the partial width of their decays into the $B^*D$ channel. This reduction is primarily due to the node effects of the corresponding spatial wave functions, as illustrated in Fig. \ref{fig:wavefunctions}. For the $B_c(3^1S_0)$ state, the $B^*D$ is the unique two-body OZI-allowed decay channel around $62~\rm{MeV}$. Thus, the $B^*D$ channel should be the promising channel to observe the $B_c(3^1S_0)$ state. For the $B_c(4^1S_0)$ state, it mainly decays into the $BD^*$, $B^*D$, and $B^*D^*$ channels, and whose corresponding branching ratios are more than 99\%. For the $B_c(5^1S_0)$ state, it dominantly decays into the $BD^*$, $B^*D$, $B^*D^*$, and $BD({}^3P_0)$ channels. 

The node effects are more pronounced for the $B_c(n^3S_1)$ mesons compared to the $B_c(n^1S_0)$ mesons, with a notable example being the $B_c(4^3S_1)$ state. Furthermore, for the $B_c(5^3S_1)$ state, the decay widths of the $B_s^*D_s^*$ and $B^*D(1^3P_0)$ channels are $1.7\times10^{-3}~\rm{MeV}$ and $2.5\times10^{-3}~\rm{MeV}$, respectively, reaching a order of magnitude of $10^{-3}~\rm{MeV}$. Thus, it is evident that the node effects in the $B_c(5^3S_1)$ state are more significant than that in the $B_c(5^1S_0)$ state. In the following, we
summarize several main points:
(i) The total width of the $B_c(3^3S_1)$ state is about 82 MeV, and  the main decay channels of the $B_c(3^3S_1)$ state are $BD$ and $B^*D$; (ii) The $B_c(4^3S_1)$ state mainly decays into the $B^*D$, $BD^*$, and $B^*D^*$ channels, and the $BD$ also has the sizable contribution to the total width; (iii) The main decay modes of the $B_c(5^3S_1)$ state include the $B^*D$, $BD^*$, $B^*D^*$, $BD(1P_1)$, and $BD(1P'_1)$ channels, while the dominant decay channel is the $B^*D^*$ with branching ratio 70.1\%.

For these $3P$ states of the $B_c$ meson, the partial widths of the $B_s D_s$ and $B_s^* D_s$ decay channels are significantly suppressed. Specifically, for the $B_c(3^3P_0)$ state, the $B_s^* D_s$ channel is kinematically forbidden. Similarly, for the $4P$ states of the $B_c$ meson, the decay channel of $B_s^* D_s^*$ is also kinematically forbidden. Furthermore, for the $B_c(3^3P_2)$ and $B_c(4P'_1)$ states, the decay widths of the $B^*D$ and $B(1P_1)D$ channels are $2.6\times10^{-3}~\rm{MeV}$ and $2.3\times10^{-3}~\rm{MeV}$, respectively, reaching the order of magnitude of $10^{-3}~\rm{MeV}$. Our results show that the largest decay width of these discussed $3P$-wave $B_c$ mesons is the $B^*D^*$ channel, which has the estimated branching ratios of 63.1\%, 54.9\%, 49.0\%, and 81.6\% for the $B_c(3^3P_0)$, $B_c(3P_1)$, $B_c(3P_1^{\prime})$, and $B_c(3^3P_2)$ states, respectively. Thus, we suggest the future experiments to search for the $3P$-wave $B_c$ mesons by the $B^*D^*$ channel. For the $4P$ states of the $B_c$ meson, the largest decay width is also the $B^*D^*$ channel. Compared with the $3P$ states of the $B_c$ meson, we find the total widths of the $4P$ states are more smaller than these $3P$ states of the $B_c$ meson. 

In Table \ref{strong decay 2}, we show the two-body OZI-allowed strong decay widths and the corresponding branching ratios for the $D$-wave and $F$-wave $B_c$ states. For these $2D$ states of the $B_c$ meson, no decay channels are significantly suppressed, and the largest decay channels are also different. For $B_c(2^3D_1)$ and $B_c(2D'_2)$ states, the largest decay width is also the $BD^*$ channel, whose branching ratios are 59.8\% and 69.5\%, respectively. However, the largest decay widths of the $B_c(2D_2)$ and $B_c(2^3D_3)$ states are $B^*D$ and $B^*D^*$ channels, respectively. For the $B_c(3D)$ states, the largest decay width is the $B^*D^*$ channel, whose branching ratios are 58.4\%, 37.2\%, 82.1\%, and 64.4\% for the $B_c(3^3D_1)$, $B_c(3D_2)$, $B_c(3D_2^{\prime})$, and $B_c(3^3D_3)$ states, respectively. For the $3D$ states of the $B_c$ meson, the decay widths of some channels are significantly suppressed. Such as the $B_sD_s^*$ channel for the $B_c(3^3D_1)$ state and the $B_sD_s$ channel for the $B_c(3^3D_3)$ state, whose decay widths are $4.7\times10^{-3}~\rm{MeV}$ and $2.3\times10^{-4}~\rm{MeV}$, respectively. 

For the $B_c(1^3F_2)$ and $B_c(1^3F_4)$ states, they can decay into the $BD$ and $B^*D$ channels, and their dominant decay mode is the $BD$ channel. In addition, it is worth noting that the $B_c(1F_3^{\prime})$ and $B_c(1F_3)$ states can decay into the $B^*D$ channel with the branching ratios 100\%, but their decay widths exist obvious difference, which is similar to the decay behaviour of the $B_c(1^3F_2)$ and $B_c(1^3F_4)$ states. The $B^*D^*$ is the dominant decay channel for the $2F$ states of the $B_c$ meson, and the branching ratios are 71.2\%, 48.0\%, 58.6\%, and 51.4\% for the $B_c(2^3F_2)$, $B_c(2F_3)$, $B_c(2F_3^{\prime})$, and $B_c(2^3F_4)$ states, respectively. The $B_sD_s^*$ is a significantly suppressing channel for the $2F$ states of the $B_c$ meson, especially for the $B_c(2F_3)$ and $B_c(2^3F_4)$ states, whose branching ratios are $2.9\times10^{-5}$ and $2.0\times10^{-5}$, respectively.

In general, the information on the concerned two-body OZI-allowed strong decays presented of $B_c$ mesons in Tables \ref{strong decay 1}-\ref{strong decay 2} can provide valuable guidance for further experimental searches for them.  

\begin{table*}[htbp]\centering
\caption{Partial widths and the corresponding branching ratios for these two-body OZI-allowed strong decays of the $S$-wave and $P$-wave $B_c$ mesons. Here, the decay widths of the discussed mesons are given in units of MeV.}
\label{strong decay 1}
\renewcommand\arraystretch{1.2}
\begin{tabular*}{140mm}{c@{\extracolsep{\fill}}lrrcclrr}
\toprule[1.5pt]\toprule[0.5pt]
\text{States}   & \text{Channels} & \text{ Widths} & \text{$\mathcal{B}(\%)$} && \text{States}   & \text{Channels} & \text{ Widths} & \text{$\mathcal{B}(\%)$}\\
\toprule[1pt]
$3^1S_0$  & $B^*D$             & 62.2 & $100.0 $           &&$3P'_1$  & $B^*D$      & 34.7  & 24.6   \\
\cline{2-3}                                                 
         &\text{Total width}& 62.2                                 &&& & $BD^*$      & 36.5  & 25.9  \\
\cline{1-4} 
$4^1S_0$ & $B^*D$             & 26.9 & 30.2                         &&& $B^*D^*$    & 69.1  & 49.0  \\
         & $BD^*$             & 31.4 & 35.2                         &&& $B_s^*D_s$  & 0.8   & 0.6  \\
\cline{7-8}         
         & $B^*D^*$           & 30.2 & 33.8                         &&&\text{Total width}&  141.1 \\
\cline{6-9}         
         & $B_s^*D_s$         & 0.01  & 0.01               &&$3^3P_2$   & $BD$        & 2.1  & 2.1   \\ 
         & $B_sD_s^*$         & 0.7  & 0.8         && & $B^*D$      & $2.6\times10^{-3}$ & $2.7\times10^{-3}$ \\
\cline{2-3}                                                              
         &\text{Total width}& 89.2                                  &&&& $BD^*$      & 14.5 & 15.0 \\
\cline{1-4}
$5^1S_0$ & $B^*D$             & 8.1  & 11.0                         &&& $B^*D^*$    & 79.2 & 81.6  \\
         & $BD^*$             & 19.4 & 26.4                         &&& $B_sD_s$    & 0.6  & 0.7   \\
         & $B^*D^*$           & 42.9 & 58.5                         &&& $B_s^*D_s$  & 0.6  & 0.6 \\
\cline{7-8}         
         & $B_s^*D_s$         & 0.8  & 1.1                          &&& \text{Total width}& 97.1 \\
\cline{6-9}    
         & $B_sD_s^*$         & 0.05  & 0.07               && $4^3P_0$  & $BD$        & 13.2  & 15.7  \\
         & $B_s^*D_s^*$       & 0.08  & 0.1                        &&  & $B^*D^*$    & 67.7  & 80.7   \\
         & $B(^3P_0)D$        & 0.9  & 1.2                          &&& $B_sD_s$    & 2.6   & 3.1   \\
         & $B(^3P_2)D$        & 0.5  & 0.6                          &&& $B_s^*D_s^*$& 0.4   & 0.4   \\
\cline{7-8}    
         & $BD(^3P_0)$        & 0.7  & 0.9                           &&&\text{Total width}& 83.9 \\
\cline{2-3}\cline{6-9}    
         &\text{Total width}& 73.4                       &&&$4P_1$     & $B^*D$      & 4.5   & 5.7  \\
\cline{1-4}
$3^3S_1$ & $BD$               & 25.0 & 30.4                         &&& $BD^*$      & 29.7  & 37.3 \\
         & $B^*D$             & 57.2 & 69.6                         &&& $B^*D^*$    & 34.3  & 43.1  \\
\cline{2-3}    
         &\text{Total width}& 82.2                                  &&&& $B_s^*D_s$  & 1.1   & 1.4 \\
\cline{1-4}
$4^3S_1$ & $BD$               & 1.8  & 2.0                          &&& $B_sD_s^*$  & 0.1   & 0.2  \\
         & $B^*D$             & 11.6 & 12.7                       &&  & $B_s^*D_s^*$& 0.04   & 0.05 \\ 
         & $BD^*$             & 21.9 & 23.9                       &&  & $B(1^3P_0)D$& 0.05   & 0.06  \\  
         & $B^*D^*$           & 54.9 & 60.1                         &&& $B(1P_1)D$  & 0.06   & 0.07  \\
         & $B_sD_s$           & 0.4  & 0.4                          &&& $B(1^3P_2)D$& 9.6   & 12.1 \\
         & $B_s^*D_s$         & 0.1  & 0.2                          &&& $BD(1^3P_0)$& 0.08   & 0.1 \\ 
 \cline{7-8}   
         & $B_sD_s^*$         & 0.6  & 0.6                        &&& \text{Total width}& 79.6   \\  
\cline{2-3}\cline{6-9}   
         &\text{Total width}& 91.4                       & && $4P'_1$  & $B^*D$      & 7.1   & 9.6 \\     
\cline{1-4}
$5^3S_1$ & $BD$               & 0.07  & 0.1                          &&& $BD^*$      & 7.0   & 9.5 \\
         & $B^*D$             & 2.4  & 3.6                          &&& $B^*D^*$    & 39.4  & 53.2 \\
         & $BD^*$             & 10.1 & 15.1                         &&& $B_s^*D_s$  & 1.5   & 2.0 \\
         & $B^*D^*$           & 47.0 & 70.1                         &&& $B_sD_s^*$  & 0.18   & 0.24  \\
         & $B_sD_s$           & 0.3  & 0.4                          &&& $B_s^*D_s^*$& 0.02   & 0.03 \\
         & $B_s^*D_s$         & 0.6  & 0.8                          &&& $B(1^3P_0)D$& 0.2   & 0.28 \\
         & $B_sD_s^*$         & 0.09  & 0.14       &&& $B(1P_1)D$  &$2.3\times10^{-3}$&$3.1\times10^{-3}$ \\  
    & $B_s^*D_s^*$ & $1.7\times10^{-3}$&$2.5\times10^{-3}$          &&& $B(1^3P_2)D$& 17.8  & 24.2  \\
         & $B(1P_1)D$         & 0.2  & 0.3                          &&& $BD(1^3P_0)$& 0.7   & 0.9 \\
\cline{7-8}
         & $B(1P'_1)D$        & 0.7  & 1.0                          &&& \text{Total width}&  73.9 \\
\cline{6-9}
         & $B(1^3P_2)D$       & 0.1  & 0.2                &&$4^3P_2$  & $BD$        & 3.4  & 5.4   \\        
         & $BD(1P_1)$         & 1.9  & 2.9                          &&& $B^*D$      & 1.4  & 2.2  \\
         & $BD(1P'_1)$        & 3.5  & 5.2                          &&& $BD^*$      & 1.3  & 2.0   \\
    & $B^*D(1^3P_0)$ & $2.5\times10^{-3}$ & $3.8\times10^{-3}$      &&& $B^*D^*$    & 46.1 & 72.3  \\  
\cline{2-3}    
         &\text{Total width}&     67.0                              &&&& $B_sD_s$    & 0.1  & 0.2  \\
\cline{1-4}
$3^3P_0$ & $BD$        & 52.5  & 36.9                               &&& $B_s^*D_s$  & 0.4  & 0.6  \\
         & $B^*D^*$    & 89.7  & 63.1                               &&& $B_sD_s^*$  & 0.2  & 0.26 \\
         & $B_sD_s$    & 0.03   & 0.02                                &&& $B_s^*D_s^*$& 0.02  & 0.03 \\
\cline{2-3}
         &\text{Total width}& 142.2                                  &&&& $B(1P_1)D$  & 4.4  & 6.9  \\                
\cline{1-4} 
$3P_1$   & $B^*D$      & 18.2  & 12.3                               &&& $B(1P'_1)D$ & 1.1  & 1.7 \\ 
         & $BD^*$      & 47.5  & 32.2                               &&& $B(1^3P_2)D$& 5.4  & 8.4 \\  
\cline{7-8}
         & $B^*D^*$    & 81.2  & 54.9                              &&&\text{Total width}& 63.7 \\\cline{6-9}
         & $B_s^*D_s$  & 0.8   & 0.5      \\                         
\cline{2-3}   
         &\text{Total width}&   147.7   \\
\bottomrule[0.5pt]\bottomrule[1.5pt]
\end{tabular*}
\end{table*}

\begin{table*}[htbp]\centering
\caption{Partial widths and the corresponding branching ratios for these two-body OZI-allowed strong decays of the $D$-wave and $F$-wave $B_c$ states. Here, the decay width of the meson is given in units of MeV.}
\label{strong decay 2}
\renewcommand\arraystretch{1.2}
\begin{tabular*}{140mm}{c@{\extracolsep{\fill}}lrrcclrr}
\toprule[1.5pt]\toprule[0.5pt]
\text{States}   & \text{Channels} & \text{ Widths} & \text{$\mathcal{B}(\%)$} && \text{States}   & \text{Channels} & \text{ Widths} & \text{$\mathcal{B}(\%)$}\\
\toprule[1pt]
$2^3D_1$ & $BD$        & 4.9  & 10.5             &&$1^3F_2$ & $BD$        & 46.8  & 79.0  \\
         & $B^*D$      & 14.0 & 29.7                      &&& $B^*D$      & 12.5  & 21.0  \\
\cline{7-8}
         & $BD^*$      & 28.2 & 59.8                      &&&\text{Total width}& 59.2    \\
\cline{2-3}    \cline{6-9}
         &\text{Total width}& 47.2               & &&$1F_3$  & $B^*D$      & 0.2 & 100.0  \\
\cline{1-4}\cline{7-8}
$2D_2$   & $B^*D$      & 24.9 & 78.1                        &&&\text{Total width}& 0.2\\
\cline{6-9}
         & $BD^*$      & 3.3  & 10.3            && $1F'_3$  & $B^*D$      & 35.2 & 100.0  \\
\cline{7-8}
         & $B^*D^*$    & 3.7  & 11.6                      &&&\text{Total width}&  35.2     \\ 
\cline{2-3}    \cline{6-9}
         &\text{Total width}& 31.8               &&&$1^3F_4$ & $BD$        & 0.7   & 85.6  \\
\cline{1-4}                                        
$2D'_2$  & $B^*D$      & 34.8 & 26.3                      &&& $B^*D$      & 0.1   & 14.4 \\
\cline{7-8}
         & $BD^*$      & 91.7 & 69.5                      &&&\text{Total width}& 0.8      \\
\cline{6-9}
         & $B^*D^*$    & 5.5  & 4.1              &&$2^3F_2$ & $BD$        & 16.6   & 20.4  \\
\cline{2-3}    
         &\text{Total width}& 132.0                       &&&& $B^*D$     & 4.0    & 4.9   \\
\cline{1-4}
$2^3D_3$ & $BD$        & 19.7   & 23.4                    &&& $BD^*$      & 1.4    & 1.7  \\
         & $B^*D$      & 21.1   & 25.1                    &&& $B^*D^*$    & 58.1   & 71.2  \\
         & $BD^*$      & 2.7    & 3.2                     &&& $B_sD_s$    & 0.5    & 0.6   \\
         & $B^*D^*$    & 40.6   & 48.3                    &&& $B_s^*D_s$  & 0.8    & 1.0   \\ 
\cline{2-3}    
         &\text{Total width}& 84.2                        &&&& $B_sD_s^*$  & 0.1    & 0.2  \\
\cline{1-4}\cline{7-8}
$3^3D_1$ & $BD$        & 0.07   & 0.24                         &&&\text{Total width}& 81.5    \\
\cline{6-9}
         & $B^*D$      & 1.1   & 3.5             &&$2F_3$   & $B^*D$      & 13.8   & 16.3  \\
         & $BD^*$      & 9.0   & 29.5                     &&& $BD^*$      & 28.7   & 33.7  \\
         & $B^*D^*$    & 17.8  & 58.4                     &&& $B^*D^*$    & 40.9   & 48.0  \\
         & $B_sD_s$    & 1.4   & 4.5                      &&& $B_s^*D_s$  & 1.7    & 2.0  \\
         & $B_s^*D_s$  & 0.7   & 2.2                      &&& $B_sD_s^*$  & 0.2    & $2.9\times 10^{-3}$  \\
\cline{7-8}
         & $B_sD_s^*$  & $4.7\times10^{-3}$ &0.01            &&&\text{Total width}& 85.2 \\
\cline{6-9}
         & $B_s^*D_s^*$& 0.5   & 1.6             &&$2F'_3$  & $B^*D$      & 27.1   & 37.6   \\
\cline{2-3}    
         &\text{Total width}& 30.5                    &    &&& $BD^*$      & 1.7    & 2.4  \\
\cline{1-4} 
$3D_2$   & $B^*D$      & 15.0  & 27.7                     &&& $B^*D^*$    & 42.2   & 58.6  \\
         & $BD^*$      & 18.1  & 33.4                     &&& $B_s^*D_s$  & 0.5    & 0.7  \\
         & $B^*D^*$    & 20.1  & 37.2                     &&& $B_sD_s^*$  & 0.5    & 0.7  \\
\cline{7-8}
         & $B_s^*D_s$  & 0.4   & 0.7                      &&&\text{Total width}& 72.0   \\
\cline{6-9}
         & $B_sD_s^*$  & 0.08   & 0.16             &&$2^3F_4$ & $BD$        & 8.2    & 10.2  \\
         & $B_s^*D_s^*$& 0.5   & 0.9                      &&& $B^*D$        & 14.2   & 17.8   \\
\cline{2-3}    
         &\text{Total width}& 54.17                      &  &&& $BD^*$      & 15.7   & 19.7  \\
\cline{1-4}
$3D'_2$  & $B^*D$      & 0.6   & 2.3                      &&& $B^*D^*$    & 40.9   & 51.4  \\
         & $BD^*$      & 1.2   & 4.5                      &&& $B_sD_s$    & 0.4    & 0.5   \\
         & $B^*D^*$    & 21.0  & 82.1                     &&& $B_s^*D_s$  & 0.2    & 0.3   \\
         & $B_s^*D_s$  & 1.8   & 7.0             &&& $B_sD_s^*$  & $1.6\times10^{-3}$ & $2.0\times10^{-3}$ \\
\cline{7-8}
         & $B_sD_s^*$  & 0.5   & 2.1                        &&&\text{Total width}& 79.7      \\
         \cline{6-9}
         & $B_s^*D_s^*$& 0.5   & 1.9  \\
\cline{2-3}    
         &\text{Total width}& 25.6   \\ 
\cline{1-4}
$3^3D_3$ & $BD$        & 10.3   & 16.1  \\
         & $B^*D$      & 10.2   & 15.9   \\
         & $BD^*$      & 1.4    & 2.2   \\
         & $B^*D^*$    & 41.2   & 64.4  \\
         & $B_sD_s$    & $2.3\times10^{-4}$ & $3.6\times10^{-4}$   \\
         & $B_s^*D_s$  & 0.1    & 0.2   \\
         & $B_sD_s^*$  & 0.3    & 0.5   \\
         & $B_s^*D_s^*$& 0.5    & 0.8   \\
\cline{2-3}    
         &\text{Total width}& 64.0      \\
\bottomrule[0.5pt]\bottomrule[1.5pt]
\end{tabular*}
\end{table*}

\section{Dipion transitions between $B_c$ mesons}
\label{sec4}

In this section, we investigate the dipion transitions using the QCDME approach. 
The QCDME has been widely used to study the dipion or $\eta$ hadronic transitions between the low-lying heavy quarkonium \cite{Yan:1980uh,Kuang:1981se,Zhou:1990ik,Kuang:2006me,Wang:2015xsa,Godfrey:2015dia,Segovia:2016xqb,Wang:2018rjg}. The idea of this approach is that the QZI-suppressed hadronic transition is represented by
the parent heavy quarkonium first emitting a gluon to form an intermediate hybrid state, and then recombining itself to the daughter heavy quarkonium and the light meson(s) (like a $\eta$ or a pair of pions) with emitting another gluon via the hadronization process.

Following Ref. \cite{Kuang:1981se}, one can calculate the decay width of the dipion transition as depicted by the following expression:
\begin{widetext}
\begin{equation}
    \begin{split}
        \Gamma(A\to B+\pi^{+}\pi^{-})=&\delta_{l_{i}l_{f}}\delta_{J_{i}J_{f}}(G\vert{c_{1}}\vert^{2}-\frac{2}{3}H\vert{c_{2}}\vert^{2})
        \Big{\vert}\sum_{l}(2l+1)f_{if}^{l}
        \left(\begin{array}{ccc}l_{i} &1 &l \\ 0 &0 &0\end{array}\right)
        \left(\begin{array}{ccc}l     &1 &l_{i} \\ 0 &0 &0\end{array}\right)
        \Big{\vert}^{2}\\
        &+(2l_{i}+1)(2l_{f}+1)(2J_{f}+1)\sum_{k}(2k+1)[1+(-1)^{k}]
        \left\{\begin{array}{ccc}s     &l_{f} &J_{f} \\ k &J_{i} &l_{i}\end{array}\right\}^{2}
        H\vert{c_{2}}\vert^{2}\\
        &\times \Big{\vert}\sum_{l}(2l+1)f_{if}^{l}
        \left(\begin{array}{ccc}l_{f} &1 &l \\ 0 &0 &0\end{array}\right)
        \left(\begin{array}{ccc}l     &1 &l_{i} \\ 0 &0 &0\end{array}\right)
        \left\{\begin{array}{ccc}l_{i} &l &1 \\ 1 &k &l_{f}\end{array}\right\}
        \Big{\vert}^{2},
    \end{split}
\label{eq:HT}
\end{equation}
\end{widetext}
where $c_{1}$ and $c_{2}$ are undetermined parameters, while $l_{i(f)}$ and $J_{i(f)}$ are the orbital and total angular momenta of meson $A$($B$), respectively. Both mesons $A$ and $B$ possess an identical spin value denoted as $s$. Furthermore, the phase-space factors, denoted as $G$ and $H$ in this context, are defined as follows:
\begin{equation}
G=\frac{3}{4}\frac{m_{f}}{m_{i}}\pi^{3}\int dm_{\pi^{+}\pi^{-}}^{2} K \Big{(}1-\frac{4m_{\pi}^{2}}{m_{\pi^{+}\pi^{-}}^{2}}\Big{)}^{1/2}(m_{\pi^{+}\pi^{-}}^{2}-2m_{\pi}^{2})^{2},
\end{equation}
\begin{equation}
    \begin{split}
        H=&\frac{1}{20}\frac{m_{f}}{m_{i}}\pi^{3}
        \int dm_{\pi^{+}\pi^{-}}^{2} K \Bigg{(}1-\frac{4m_{\pi}^{2}}{m_{\pi^{+}\pi^{-}}^{2}}\Bigg{)}^{1/2}\Big{[}\big{(}m_{\pi^{+}\pi^{-}}^{2}-4m_{\pi}^{2}\big{)}^{2}\\
        &\times\Bigg{(}1+\frac{2}{3}\frac{K^{2}}{m_{\pi^{+}\pi^{-}}^{2}}\Bigg{)}+\frac{8K^{4}}{15m_{\pi^{+}\pi^{-}}^{4}}(m_{\pi^{+}\pi^{-}}^{4}+2m_{\pi}^{2}m_{\pi^{+}\pi^{-}}^{2}+6m_{\pi}^{4})
        \Big{]},
    \end{split}
\end{equation}
respectively, with
\begin{equation}
K=\frac{\sqrt{ \big{(}(m_{i}+m_{f})^{2}-m_{\pi^{+}\pi^{-}}^{2}\big{)} \big{(}(m_{i}-m_{f})^{2}-m_{\pi^{+}\pi^{-}}^{2}\big{)}} }{2m_{i}}.
\end{equation}
Here, $m_i$ and $m_f$ are the masses of the initial and final $B_{c}$ mesons, respectively, while $m_{\pi}$ is the mass of the pion. Additionally, the dynamical-associated part $f_{if}^{l}$ is written as
\begin{equation}
\begin{split}
f_{if}^{l}=&\sum_{k}\frac{1}{m_{i}-m_{kl}} \int dr r^{3}\mathcal{R}_{f}(r)\mathcal{R}_{kl}(r)\\
&\times \int dr^{\prime} r^{\prime3} \mathcal{R}_{kl}(r^{\prime})\mathcal{R}_{i}(r^{\prime}),
\end{split}
\end{equation}
where $\mathcal{R}_{i}(r)$ and $\mathcal{R}_{f}(r)$ are the radial wave functions of the parent and daughter $B_{c}$ mesons, respectively, while  $\mathcal{R}_{kl}(r)$ denotes the intermediate vibrational state. $m_{kl}$ is the mass of the intermediate vibrational state with the radial quantum number $k$ and the orbital angular momentum $l$.

The spatial wave functions associated with the relevant $B_{c}$ mesons can be given by utilizing the MGI model as mentioned above. However, for the intermediate vibrational states, the appropriate approach involves resorting to the quark confining string (QCS) model, as established by Refs. \cite{Giles:1977mp,Tye:1975fz,Buchmuller:1979gy}. The potential of a hybrid state adopted in our calculations is \cite{Wang:2018rjg}
\begin{equation}
V_{\text{hyb}}(r)=V_{G}(r)+V_{S}(r)+\big{[} V_{n}(r)-\sigma(r) r\big{]},
\end{equation}
where $V_{G}(r)=-4\alpha_{s}(r)/(3r)$ is the one-gluon exchange potential and $V_{S}(r)=\sigma(r)r+c$ is the color confining potential with
\begin{equation}
\begin{split}
\alpha_{s}(r)=&\sum_{k=1}^{3}\alpha_{k}\frac{2}{\sqrt{\pi}}\int_{0}^{\gamma_{k}r}e^{-x^{2}}dx,\\
\sigma(r)=&\frac{b(1-e^{-\mu r})}{\mu r}.
\end{split}
\end{equation}
Here, we adopt $\{\alpha_1,\alpha_2,\alpha_3\}=\{0.25,0.15,0.20\}$ and $\{\gamma_1,\gamma_2,\gamma_3\}=\{1/2,\sqrt{10}/2,\sqrt{1000}/2\}$ \cite{Godfrey:1985xj,Capstick:1985xss}.

The effective vibrational potential $V_{n}(r)$ is chosen as \cite{Giles:1977mp,Wang:2015xsa}
\begin{equation}
V_{n}(r)=\sigma(r)\,r\,\Bigg{(}1+\frac{2n\pi}{\sigma(r)\big{[}(r-2d)^{2}+4d^{2}\big{]}}\Bigg{)}^{1/2}
\end{equation}
with $d=\sigma(r)r^{2}\alpha_{n}/\big{(}4[m_{b}+m_{c}+\sigma(r)r\alpha_{n}]\big{)}$. In the calculations, we only consider the lowest string excitation, i.e., $n=1$. And then, we choose $\alpha_{1}=\sqrt{1.5}$ \cite{Wang:2018rjg}.

Before performing the concrete calculations, we should determine the parameters $c_{1}$ and $c_{2}$ in Eq.~\eqref{eq:HT}. Since there have no measurement of the dipion decays between different $B_{c}$ states, we would fit these parameters from the bottomomium segment, which is the same as in Ref. \cite{Martin-Gonzalez:2022qwd}. Numerically, we take $\vert c_{1}\vert^{2}=61.8\times10^{-6}$ and $\vert c_{2}\vert^{2}=1.93\times10^{-6}$ \cite{Kuang:1981se} in the calculations.

With the above preparations, we calculate the decay rates of dipion transitions between $B_c$ states. In Table \ref{Dipion transitions}, we present our results of the decay rates of dipion transitions $B_c(2S)\to B_c(1S)\pi^{+}\pi^{-}$, $B_c(2P)\to B_c(1P)\pi^{+}\pi^{-}$, and $B_c(1D)\to B_c(1S)\pi^{+}\pi^{-}$. We notice the discrepancies between the results from this work and Refs. \cite{Godfrey:2004ya,Martin-Gonzalez:2022qwd}. In fact, 
there are also notable differences between the results from the GI model \cite{Godfrey:2004ya} and Ref. \cite{Martin-Gonzalez:2022qwd}. But, we still find some consistent conclusions for the decay rates, i.e.,
(i) For the $B_{c}(2S)\to B_{c}(1S)\pi^{+}\pi^{-}$ transitions, although our results are about a half of those of Refs. \cite{Godfrey:2004ya,Martin-Gonzalez:2022qwd}, but our result is comparable to those of the other two theoretical groups. Thus, we come to the same conclusion about the decay rates for these transitions, which are sizable.
(ii) For the $B_{c}(2P)\to B_{c}(1P)\pi^{+}\pi^{-}$ transitions, our results show that $B_c(2^3P_0)\to B_c(1^3P_0)\pi^+\pi^-$,
$B_c(2^3P_2)\to B_c(1^3P_2)\pi^+\pi^-$, 
$B_c(2P_1)\to B_c(1P_1)\pi^+\pi^-$, $B_c(2P_1)\to B_c(1P_1^\prime)\pi^+\pi^-$, $B_c(2P_1^\prime)\to B_c(1P_1)\pi^+\pi^-$, and $B_c(2P_1^\prime)\to B_c(1P_1^\prime)\pi^+\pi^-$ have significant decay rates and other transitions are suppressed. In general, this conclusion was supported by the GI model \cite{Godfrey:2004ya} and Ref. \cite{Martin-Gonzalez:2022qwd}.  
(iii) Regarding the $B_{c}(1D)\to B_{c}(1S)\pi^{+}\pi^{-}$ transitions, it is noteworthy that our result is an order of magnitude smaller than the corresponding result from the GI model \cite{Godfrey:2004ya}, but comparable to that of Ref.  \cite{Martin-Gonzalez:2022qwd},  taking $B_c(1^3D_1)\to B_c(1^3S_1)\pi^+\pi^-$ and $B_c(1^3D_3)\to B_c(1^3S_1)\pi^+\pi^-$ as examples. Obviously, the study of these dipion transitions of $B_c$ mesons should be further pursued both theoretically and experimentally. 

\begin{table}[htbp]\centering
\caption{Decay rates of dipion transitions between $B_c$ states. Here, the decay rates are given in units of keV.}
\renewcommand\arraystretch{1.2}
\label{Dipion transitions}
\begin{tabular*}{85mm}{c@{\extracolsep{\fill}}lccc}
\toprule[1.5pt]\toprule[0.5pt]
\text{Initial~states} & \text{Final~states}  & \text{This~work} & \text{GI \cite{Godfrey:2004ya}} &\text{Ref. \cite{Martin-Gonzalez:2022qwd}} \\
\toprule[1pt]
$2^1S_0$    & $1^1S_0 + \pi^{+}\pi^{-}$ & 25 & 57 & 42 \\\hline
$2^3S_1$    & $1^3S_1 + \pi^{+}\pi^{-}$ & 21 & 57 & 41 \\
\hline
\multirow{4}*{$2^3P_0$}    & $1^3P_0 + \pi^{+}\pi^{-}$ & 2.8 & 0.97 & 12 \\
            & $1^3P_2 + \pi^{+}\pi^{-}$ & $1.2\times10^{-4}$ & 0.055 & $5.5\times10^{-3}$ \\
            & $1 P_1 + \pi^{+}\pi^{-}$  & 0 & 0 & 0 \\
            & $1 P'_1 + \pi^{+}\pi^{-}$ & 0 & 0 & 0 \\\hline
            
\multirow{4}*{$2^3P_2$}    & $1^3P_0 + \pi^{+}\pi^{-}$ & $5.7\times10^{-3}$ & 0.011 & 0.018 \\
            & $1^3P_2 + \pi^{+}\pi^{-}$ & 3.0 & 1.0 & 11 \\
            & $1 P_1 +  \pi^{+}\pi^{-}$  & $2.7\times10^{-3}$ & 0.021 & 0.02 \\
            & $1 P'_1 + \pi^{+}\pi^{-}$ & $9.7\times10^{-4}$ & $4.0\times10^{-3}$ \\\hline
            
\multirow{4}*{$2 P_1$}     & $1^3P_0 + \pi^{+}\pi^{-}$ & 0 & 0 & 0\\
            & $1^3P_2 + \pi^{+}\pi^{-}$ & $6.3\times10^{-4}$ & 0.037 & 0.012\\
            & $1 P_1 +  \pi^{+}\pi^{-}$  & 1.5 & 2.7 & 11\\
            & $1 P'_1 + \pi^{+}\pi^{-}$ & 0.77 & 0.020 \\
            \hline
\multirow{4}*{$2 P'_1$}    & $1^3P_0 + \pi^{+}\pi^{-}$ & 0 & 0 & 0 \\
            & $1^3P_2 + \pi^{+}\pi^{-}$ & $2.7\times10^{-4}$ & $4.0\times10^{-3}$ \\
            & $1 P_1 +  \pi^{+}\pi^{-}$  & 1.4 & 0.10 \\
            & $1 P'_1 + \pi^{+}\pi^{-}$ & 1.6 & 1.2 & 11\\
\hline
$1^3D_1$    & $1^3S_1 + \pi^{+}\pi^{-}$ & 0.15 & 4.3 & 0.75 \\\hline
$1^3D_3$    & $1^3S_1 + \pi^{+}\pi^{-}$ & 0.23 & 4.3 & 0.84 \\\hline
\multirow{2}*{$1 D_2$}     & $1^1S_0 + \pi^{+}\pi^{-}$ & 0.20 & 2.1 \\
            & $1^3S_1 + \pi^{+}\pi^{-}$ & 0.066 & 2.2 \\\hline
\multirow{2}*{$1 D'_2$}    & $1^1S_0 + \pi^{+}\pi^{-}$ & 0.12 & 2.2 \\
            & $1^3S_1 + \pi^{+}\pi^{-}$ & 0.13 & 2.1 \\
\bottomrule[0.5pt]\bottomrule[1.5pt]
\end{tabular*}
\end{table}

In addition to focusing on the decay rates of these dipion transitions of $B_c$ mesons, we should also pay more attention to their dipion mass distributions, which are also accessible in experiment. Taking some dipion transitions of heavy quarkonia as an example, we find that the experiment already reported the dipion mass distributions of $\psi(3686)\to J/\psi \pi^+\pi^-$ \cite{CLEO:2008kwj}, $\psi(3770)\to J/\psi\pi^+\pi^-$ \cite{CLEO:2005zky}, $\Upsilon(2S,3S,4S)\to \Upsilon(1S)\pi^+\pi^-$ \cite{CLEO:1993fsd,CLEO:1998fxo,Belle:2017vat}, and so on. The QCDME approach can reproduce well the line shape of the dipion mass distribution of the dipion transitions of some low-lying heavy quarkonia such as $\psi(3686)\to J/\psi \pi^+\pi^-$ and $\Upsilon(2S)\to \Upsilon(1S)\pi^+\pi^-$ \cite{Yan:1980uh}. However, for higher heavy quarkonia near or above the threshold of the OZI-allowed hadronic channel, it is difficult to directly apply the QCDME approach to describe the corresponding dipion mass distribution \cite{Wang:2015xsa,Bai:2022cfz,Chen:2011jp,Chen:2011qx}, since the QCDME approach 
is a typical quenched quark model. Usually, for higher heavy quarkonia, the unquenched effects become important \cite{Moxhay:1988ri,Zhou:1990ik,Wang:2015xsa,Chen:2011jp,Bai:2022cfz,Chen:2011qx,Meng:2007tk,Meng:2008dd,Danilkin:2011sh,Chen:2011zv,Wang:2016qmz,Huang:2017kkg,Huang:2018cco,Huang:2018pmk,Li:2021jjt,Li:2022leg,Chen:2011pu,Chen:2012nva,Chen:2014sra,Chen:2014ccr}. 

\begin{figure*}[htpb]\centering
\includegraphics[width=1\textwidth]{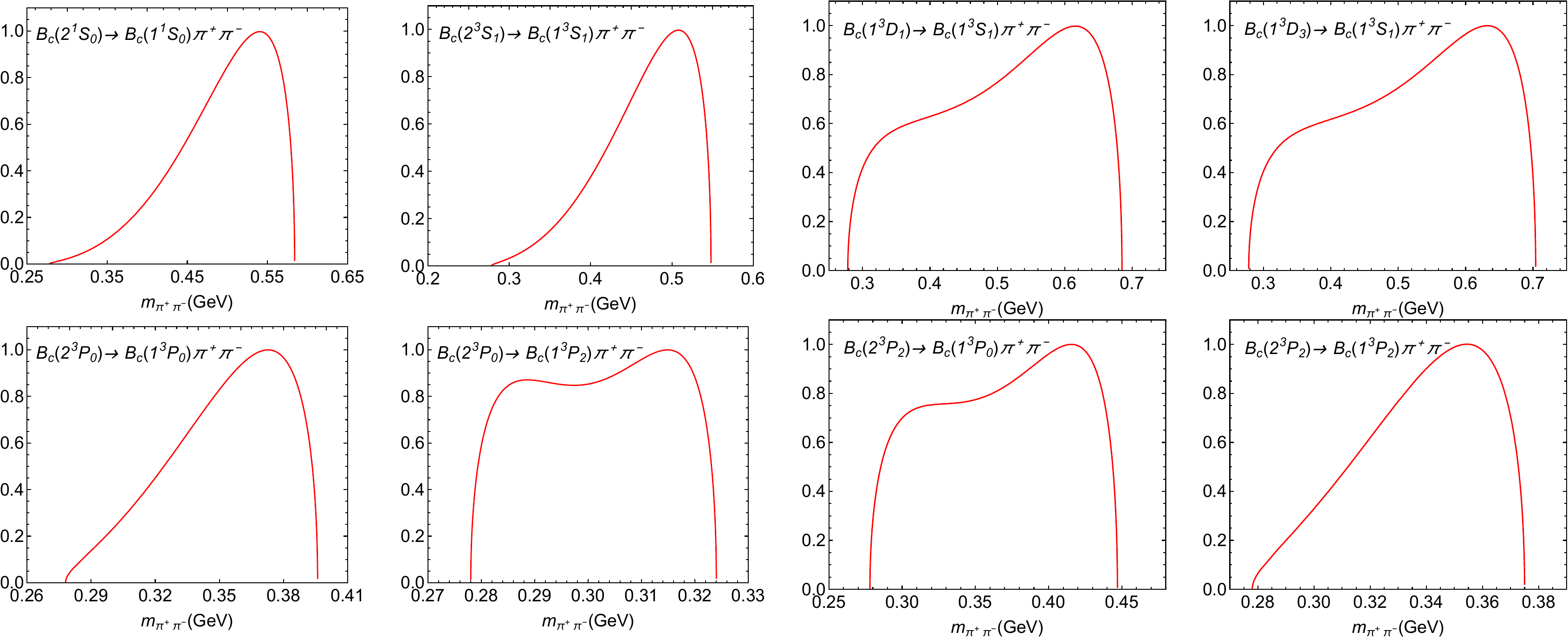}
\caption{\label{fig:dipion distribution}The dipion invariant mass spectrum distributions of $B_{c}(2^{1}S_{0})\to B_{c}(1^{1}S_{0})\pi^{+}\pi^{-}$ and $B_{c}(2^{3}S_{1})\to B_{c}(1^{3}S_{1})\pi^{+}\pi^{-}$, and $B_{c}(2^{3}P_{0})\to B_{c}(1^{3}P_{0(2)})\pi^{+}\pi^{-}$ and $B_{c}(2^{3}P_{2})\to B_{c}(1^{3}P_{0(2)})\pi^{+}\pi^{-}$, and $B_{c}(1^{3}D_{1(3)})\to B_{c}(1^{3}S_{1})\pi^{+}\pi^{-}$ processes. Here, we have normalized the maxima of the dipion distributions to 1.}
\end{figure*}

In this work, we hope to provide the dipion mass distributions of some discussed dipion transitions of $B_c$ mesons, which are shown in Fig. \ref{fig:dipion distribution}. We want to emphasize that the maxima of dipion mass distributions are all normalized to 1, since we only concern on their line shapes. Here, we not only select some decays of low-lying $B_c$ mesons far below the $BD$ threshold, i.e., $B_{c}(2^{3}S_{1})\to B_{c}(1^{3}S_{1})\pi^{+}\pi^{-}$ and $B_{c}(2^{1}S_{0})\to B_{c}(1^{1}S_{0})\pi^{+}\pi^{-}$, but also take some decays of $B_c$ mesons above (or near) the $BD$ threshold, like $B_{c}(2^{3}P_{0})\to B_{c}(1^{3}P_{0(2)})\pi^{+}\pi^{-}$, $B_{c}(2^{3}P_{2})\to B_{c}(1^{3}P_{0(2)})\pi^{+}\pi^{-}$, and $B_{c}(1^{3}D_{1(3)})\to B_{c}(1^{3}S_{1})\pi^{+}\pi^{-}$. At the present, this experimental information is still missing. Therefore, we strongly suggest our experimental colleagues to carry out the measurement of the dipion mass distributions of the $B_c$ dipion transitions, which is sensitive to reflect the difference in the result under the quenched and unquenched pictures.

\section{Radiative decays and Magnetic moments}\label{sec5}

In this section, we will delve into the electric dipole (E1) and magnetic dipole (M1) radiative decay behaviors and the magnetic moments exhibited by the relevant $B_c$ states. These observations hold significant implications for unraveling their intricate internal structures.

\subsection{The E1 transitions}

To commence our explorations, we will delve into the E1 transition decays of the aforementioned $B_c$ states. The partial widths associated with the E1 radiative transition in the nonrelativistic quark model can be obtained using the following expression \cite{Brodsky:1968ea,Kwong:1988ae}
\begin{align}
\Gamma(i \to f+\gamma)=\frac{4}{3}\left\langle e_{Q}\right\rangle^{2} \alpha \omega^{3} C_{f i} \delta_{S S^{\prime}}|\langle f|r| i\rangle|^{2}
\end{align}
with
\begin{align}
C_{f i}=\operatorname{Max}\left(L, L^{\prime}\right)\left(2 J^{\prime}+1\right)\left\{\begin{array}{ccc}
L^{\prime} & J^{\prime} & S \\
J & L & 1
\end{array}\right\}^{2},
\end{align}
\begin{align}
\left\langle e_{Q}\right\rangle=\frac{m_{b} e_{c}-m_{c} e_{\bar{b}}}{m_{b}+m_{c}}.
\end{align}
Additionally, we denote $m_c$ and $m_b$ as the masses of the charm and bottom quarks, respectively. The corresponding charges of the charm and bottom quarks are $e_{c}=2/3$ and $e_{b}=-1/3$, respectively. The energy of the photon, denoted as $\omega$, can be obtained through the conservation of energy and momentum, expressed as follows:
\begin{align}
M_{i}=\sqrt{M_{f}^{2}+\omega^{2}}+\omega,
\end{align}
where $M_{i}$ and $M_{f}$ are the masses of the parent and daughter mesons, respectively.

\begin{table*}[htbp]\centering
\caption{Widths of the E1 transitions for the $1P$-wave, $1D$-wave, $1F$-wave, $2S$-wave, $2P$-wave, and $3S$-wave $B_c$ states compared with those from other theoretical work. Here, the width is in units of keV.}
\renewcommand\arraystretch{1.2}
\label{Electric compare}
\begin{tabular*}{177mm}{c@{\extracolsep{\fill}}cccccc|cccccc}
\toprule[1.5pt]\toprule[0.5pt]
\text{Initial~states} & \text{Final~states}  & \text{This~work} & \text{GI \cite{Godfrey:2004ya}} & \text{Ref. \cite{Ebert:2002pp}} & \text{Ref. \cite{Gershtein:1994dxw}} && \text{Initial~states} & \text{Final~states}  & \text{This~work} & \text{GI \cite{Godfrey:2004ya}} & \text{Ref. \cite{Ebert:2002pp}} & \text{Ref. \cite{Gershtein:1994dxw}}\\
\toprule[1pt]
$1^3P_0$    & $1^3S_1\gamma$ & 52.3  & 55 & 67 & 65       && \multirow{3}*{$2^3P_0$}   & $1^3S_1\gamma$ & 0.4 & 1.0 &      & 16.1 \\
\cline{1-6}
$1^3P_2$    & $1^3S_1\gamma$ & 85.1  & 83 & 107  & 103                &&& $2^3S_1\gamma$ & 33.6 & 42 & 29 & 26 \\
\cline{1-6}
\multirow{2}*{$1P'_1$}     & $1^1S_0\gamma$ & 64.0 & 80 & 132  & 11                  &&& $1^3D_1\gamma$ & 2.4 & 4.2 & 0.036 & 3.2 \\
\cline{8-13}
            & $1^3S_1\gamma$ & 25.6 & 11 & 14 & 8.1       &&\multirow{6}*{$2^3P_2$}    & $1^3S_1\gamma$ & 15.5  & 14 &      & 19 \\
\cline{1-6}
\multirow{2}*{$1P_1$}      & $1^1S_0\gamma$ & 30.1  & 13 & 18 & 12                   &&& $2^3S_1\gamma$ & 48.9  & 55 & 57 & 49 \\
            & $1^3S_1\gamma$ & 47.8 & 60 & 79 & 78                    &&& $1^3D_1\gamma$ & 0.08& 0.10 & 0.035 & 0.1 \\ 
\cline{1-6}
\multirow{4}*{$1^3D_1$}    & $1^3P_0\gamma$ & 53.8 & 55 & 128  & 80                  &&& $1^3D_3\gamma$ & 5.4 & 6.8 & 1.6 & 11 \\
            & $1^3P_2\gamma$ & 1.7  & 1.8 & 5.5 & 2.2                 &&& $1D'_2\gamma$  & 0.6 & 0.70 & 0.11 & 0.5 \\
            & $1P'_1\gamma$  & 9.1 & 4.4& 7.7 & 3.3                   &&& $1D_2\gamma$   & 0.4 & 0.60 & 0.27 & 1.5 \\
\cline{8-13}
            & $1P_1\gamma$   & 21.2  & 28 & 74 & 39      && \multirow{7}*{$2P'_1$}     & $1^1S_0\gamma$ & 14.6  & 19  \\
\cline{1-6}
$1^3D_3$    & $1^3P_2\gamma$ & 74.6 & 78 & 102 & 75                   &&& $2^1S_0\gamma$ & 42.4  & 52  \\
\cline{1-6}
\multirow{3}*{$1D'_2$}     & $1^3P_2\gamma$ & 11.7 & 8.8 & 13 & 6.8                  &&& $1^3S_1\gamma$ & 1.0  & 0.6  \\
            & $1P'_1\gamma$  & 74.2  & 63 & 116  & 46                 &&& $2^3S_1\gamma$ & 9.6  & 5.5 \\
            & $1P_1\gamma$   & 0.8 & 7  & 7.3 & 25                    &&& $1^3D_1\gamma$ & 0.3 & 0.2  \\              
\cline{1-6}
\multirow{3}*{$1D_2$}      & $1^3P_2\gamma$ & 6.0 & 9.6& 28 & 12                     &&& $1D'_2\gamma$ & 3.3 & 5.5  \\
            & $1P'_1\gamma$  & 0.1 & 15 & 14 & 18                     &&& $1D_2\gamma$ & 1.3 & 1.3  \\
\cline{8-13}
            & $1P_1\gamma$   & 67.6  & 64 & 112  & 45     &&\multirow{7}*{$2P_1$}      & $1^1S_0\gamma$  & 3.5  & 2.1  \\
\cline{1-6}
\multirow{4}*{$1^3F_2$}    & $1^3D_1\gamma$ & 68.2 & 75&      &                      &&& $2^1S_0\gamma$ & 9.7  & 5.7  \\
            & $1^3D_3\gamma$ & 0.3  & 0.4  &      &                   &&& $1^3S_1\gamma$ & 3.8  & 5.4  \\
            & $1D'_2\gamma$  & 6.9 & 6.3 &      &                     &&& $2^3S_1\gamma$ & 35.1  & 45  \\
            & $1D_2\gamma$   & 4.2 & 6.5 &      &                     &&& $1^3D_1\gamma$ & 1.1& 1.6  \\              
\cline{1-6}
$1^3F_4$    & $1^3D_3\gamma$ & 70.6  & 81 &      &                    &&& $1D'_2\gamma$  & 0.5 & 0.8  \\
\cline{1-6}
\multirow{3}*{$1F'_3$}     & $1^3D_3\gamma$ & 5.1 & 3.7&      &                      &&& $1D_2\gamma$   & 3.0 & 3.6  \\           
\cline{8-13}
            & $1D'_2\gamma$ & 71.6  & 78 &      &         &&\multirow{4}*{$3^1S_0$}    & $1P'_1\gamma$  & 3.0 &  \\
            & $1D_2\gamma$ & 0.6  & 0.5 &      &                      &&& $1P_1\gamma$   & 1.6  &  \\
\cline{1-6}
\multirow{3}*{$1F_3$}      & $1^3D_3\gamma$ & 3.0 & 5.4 &      &                     &&& $2P'_1\gamma$  & 9.0 &  \\
            & $1D'_2\gamma$  & 0.03  & 0.04 &      &                  &&& $2P_1\gamma$   & 2.9  &  \\
\cline{8-13}
            & $1D_2\gamma$   & 69.1 & 82 &      &         &&\multirow{8}*{$3^3S_1$}    & $1^3P_0\gamma$ & 0.04 & \\
\cline{1-6}
\multirow{2}*{$2^1S_0$}    & $1P'_1\gamma$ & 4.4 & 6.1& 3.7 & 16                     &&& $1^3P_2\gamma$ & 2.2  &  \\
            & $1P_1\gamma$ & 2.8 & 1.3& 1.0 & 1.9                     &&& $1P'_1\gamma$  & 0.04 &  \\
\cline{1-6}
\multirow{4}*{$2^3S_1$}    & $1^3P_0\gamma$ & 3.0 & 2.9 & 3.8 & 7.7                  &&& $1P_1\gamma$   & 0.3  &  \\
            & $1^3P_2\gamma$ & 5.3 & 5.7 & 5.2 & 15                   &&& $2^3P_0\gamma$ & 3.8 &  \\
            & $1P'_1\gamma$ & 1.3 & 0.7 & 0.63 & 1.0                  &&& $2^3P_2\gamma$ & 7.6  &  \\
            & $1P_1\gamma$ & 3.6 & 4.7 & 5.1 & 13                     &&& $2P'_1\gamma$  & 1.2 &  \\
            &&&&&                                                     &&& $2P_1\gamma$   & 5.8  &  \\
\bottomrule[0.5pt]\bottomrule[1.5pt]
\end{tabular*}
\end{table*}

In Table \ref{Electric compare}, we present the widths of electric dipole decays for the discussed $B_c$ states and compare them with those obtained from other models. Upon scrutinizing the results in Table \ref{Electric compare}, we observe that the overall magnitude of our calculations remains below 0.1 MeV, which is several orders of magnitude smaller than the two-body OZI-allowed strong decays calculated earlier. Given the limited availability of experimental data, we can only discuss our findings in relation to other theoretical predictions. When comparing our results with those obtained from the GI model \cite{Godfrey:2004ya}, we find substantial agreement in most cases, with only a few discrepancies. These variations could potentially arise from the introduction of the screening effects, such as the transitions $B_{c}({1D'_2})\to B_{c}(1P_1)\gamma$, $B_{c}(1D_2)\to B_{c}(1P'_1)\gamma$, and so on.

Now let's delve into a detailed analysis of the obtained results presented in Table \ref{Electric compare}. Regarding the calculations for the initial state of $1P$-wave, our findings exhibit consistency with the GI model \cite{Godfrey:2004ya}, as well as with Refs. \cite{Ebert:2002pp,Gershtein:1994dxw}, and no instances of underestimation are observed. However, for the cases involving initial states of $1D$-wave and $1F$-wave, we have noticed a systematic tendency toward underestimation in the results. We notice that these results for the processes $B_{c}(1^3D_1)\to B_{c}(1^3P_2)\gamma$, $B_{c}(1^3F_2)\to B_{c}(1^3D_3)\gamma$, and $B_{c}(1D'_2)\to B_{c}(1P_1)\gamma$, as well as  $B_{c}(1F'_3)\to B_{c}(1D_2)\gamma$, $B_{c}(1D_2)\to B_{c}(1P'_1)\gamma$, and $B_{c}(1F_3)\to B_{c}(1D'_2)\gamma$, are underestimated. Upon closer examination of the transition processes listed above, we can discern a fascinating pattern: the underestimation occurs consistently across similar transition processes when transitioning from a higher-order excited state to a lower-order excited state. Truly, this phenomenon is a remarkable finding.

\subsection{The M1 transitions}

We perform calculations and explore the M1 transition decays of the aforementioned $B_c$ states. The M1 transition is frequently employed to investigate the radiative transition behavior of quarkonia. Since the $B_c$ meson comprises an antibottom quark and a charm quark, its properties lie intermediate to those of charmonium and bottomonium. Consequently, we can leverage the formulas previously discussed for quarkonia to analyze the behavior of $B_c$ mesons. The magnetic dipole transition width between quarkonium states is given by the following expression \cite{Brodsky:1968ea,Novikov:1977dq}
\begin{align}
\Gamma(i \to f+\gamma)=\frac{\alpha}{3} \mu^{2} \omega^{3}\left(2 J_{f}+1\right)\left|\left\langle f\left|j_{0}\left(\frac{k {r}}{2}\right)\right| i\right\rangle\right|^{2}
\end{align}
under the nonrelativistic approximation, where
\begin{align}
\mu=\frac{e_{c}}{m_{c}}-\frac{e_{\bar{b}}}{m_{\bar{b}}},
\end{align}
and $j_{0}(x)$ is the spherical Bessel function, i.e.,
\begin{align}
j_{0}(x)=\frac{\sin x}{x}.
\end{align}

\begin{table*}[htbp]\centering
\caption{Partial widths of the M1 transitions for the $S$-wave $B_c$ states compared with other theoretical works. Here, the width is in units of eV.}
\renewcommand\arraystretch{1.2}
\label{Magnetic compare}
\begin{tabular*}{120mm}{c@{\extracolsep{\fill}}ccccccc}
\toprule[1.5pt]\toprule[0.5pt]
\text{Initial~states} & \text{Final~states}  & \text{This~work} & \text{GI \cite{Godfrey:2004ya}} & \text{Ref. \cite{Eichten:1994gt}} & \text{Ref. \cite{Ebert:2002pp}} & \text{Ref. \cite{Fulcher:1998ka}} \\
\toprule[1pt]
$1^3S_1$    & $1^1S_0\gamma$ & 83.6   & 80  & 135 & 33  & 59 \\
\hline
\multirow{2}*{$2^3S_1$}    & $2^1S_0\gamma$ & 8.3    & 10  & 29 & 17  & 12 \\
            & $1^1S_0\gamma$ & 559.3  & 600 & 123 & 428 & 122 \\
\hline
\multirow{3}*{$3^3S_1$}    & $3^1S_0\gamma$ & 2.2    & 3   &    &     &  \\
            & $2^1S_0\gamma$ & 139.3   & 200 &    &     &  \\
            & $1^1S_0\gamma$ & 503.7  & 600 &    &     &  \\
\hline
\multirow{4}*{$4^3S_1$}    & $4^1S_0\gamma$ & 1.1     \\
            & $3^1S_0\gamma$ & 62.2     \\
            & $2^1S_0\gamma$ & 151.2   \\
            & $1^1S_0\gamma$ & 443.0   \\
\hline
\multirow{5}*{$5^3S_1$}    & $5^1S_0\gamma$ & 0.6     \\
            & $4^1S_0\gamma$ & 34.8    \\
            & $3^1S_0\gamma$ & 73.1     \\
            & $2^1S_0\gamma$ & 146.5    \\
            & $1^1S_0\gamma$ & 390.0    \\
\hline
$2^1S_0$    & $1^3S_1\gamma$ & 320.6  & 300 & 93 & 488 & 59 \\
\hline
\multirow{2}*{$3^1S_0$}    & $2^3S_1\gamma$ & 53.0   & 60  &    &     &  \\
            & $1^3S_1\gamma$ & 443.5  & 4200&    &     &  \\
\hline
\multirow{3}*{$4^1S_0$}    & $3^3S_1\gamma$ & 15.6    \\
            & $2^3S_1\gamma$ & 114.0    \\
            & $1^3S_1\gamma$ & 452.2    \\
\hline
\multirow{4}*{$5^1S_0$}    & $4^3S_1\gamma$ & 6.0     \\
            & $3^3S_1\gamma$ & 45.6     \\
            & $2^3S_1\gamma$ & 138.5    \\
            & $1^3S_1\gamma$ & 431.5    \\
\bottomrule[0.5pt]\bottomrule[1.5pt]
\end{tabular*}
\end{table*}

In Table \ref{Magnetic compare}, we present the magnetic dipole decay widths of the aforementioned $B_c$ states. Upon examining the results, we observe that for a given initial state, the decay width increases as the principal quantum number $n$ decreases. This finding aligns with the conclusions drawn by previous studies \cite{Godfrey:2004ya}. Conversely, for a fixed final state, the decay width decreases as the initial state's principal quantum number $n$ increases. This intriguing phenomenon warrants further investigations. We have observed two transitions that exhibit significant suppression, which are $B_{c}(4^3S_1)\to B_{c}(4^1S_0)\gamma$ and $B_{c}(5^3S_1)\to B_{c}(5^1S_0)\gamma$, with corresponding values of 1.1 eV and 0.6 eV, respectively. Although experimental progress regarding the radiative transition behavior of $B_c$ mesons remains elusive, we 
expect that our discussions on this topic will offer theoretical insights for future experimental explorations of $B_c$ mesons and will prove valuable in this regard.

\subsection{Magnetic moments}

We delve into the magnetic moments of the discussed $B_c$ states, which serve as essential and significant physical observables of hadrons. Magnetic moments are fundamental quantities that have garnered considerable attention and sparked extensive discussions over the past decades, particularly in relation to the magnetic moments of the decuplet and octet baryons. Numerous theoretical models and approaches have been employed to investigate the magnetic moments of hadronic states, including the constituent quark model, the Bag model, the lattice QCD simulations, the chiral perturbation theory, the QCD sum rule, and others \cite{Meng:2022ozq}.

The magnetic moments of the $B_c$ states can be calculated by the following expectation values \cite{Liu:2003ab,Huang:2004tn,Zhu:2004xa,Haghpayma:2006hu,Wang:2016dzu,Deng:2021gnb,Gao:2021hmv,Zhou:2022gra,Wang:2022tib,Li:2021ryu,Schlumpf:1992vq,Schlumpf:1993rm,Cheng:1997kr,Ha:1998gf,Ramalho:2009gk,Girdhar:2015gsa,
Menapara:2022ksj,Mutuk:2021epz,Menapara:2021vug,Menapara:2021dzi,Gandhi:2018lez,Dahiya:2018ahb,Kaur:2016kan,Thakkar:2016sog,Shah:2016vmd,Dhir:2013nka,Sharma:2012jqz,Majethiya:2011ry,Sharma:2010vv,Dhir:2009ax,Simonis:2018rld,
Ghalenovi:2014swa,Kumar:2005ei,Rahmani:2020pol,Hazra:2021lpa,Gandhi:2019bju,Majethiya:2009vx,Shah:2016nxi,Shah:2018bnr,Ghalenovi:2018fxh,Wang:2022ugk,Mohan:2022sxm,An:2022qpt,Kakadiya:2022pin,Wu:2022gie,Wang:2023bek}
\begin{eqnarray}
\mu_{{B_c}}=\left\langle J_{B_c},J_{B_c} \left|\hat{\mu}_{z}\right| J_{B_c},J_{B_c} \right\rangle.
\end{eqnarray}
In the case of the $B_c$ states, the spatial wave function fulfills the normalization condition, and the color wave function is unity due to the color confinement. Consequently, the magnetic moments of the $B_c$ states are determined by their flavor and spin-orbit wave functions.

In this study, we employ the constituent quark model to explore the magnetic moments of the discussed $B_c$ states. This approach is analogous to the investigations of the magnetic moments of the decuplet and octet baryons conducted in previous studies \cite{Schlumpf:1993rm,Kumar:2005ei,Ramalho:2009gk}. Within the framework of the constituent quark model, the total magnetic moment of the $B_c$ state is composed of two distinct components: the spin magnetic moment and the orbital magnetic moment. The $z$-component of the spin magnetic moment operator, denoted as $\hat{\mu}_{z}^{\rm spin}$, and the orbital magnetic moment operator, denoted as $\hat{\mu}_{z}^{\rm orbital}$, can be expressed as \cite{Liu:2003ab,Huang:2004tn,Zhu:2004xa,Haghpayma:2006hu,Wang:2016dzu,Deng:2021gnb,Gao:2021hmv,Zhou:2022gra,Wang:2022tib,Li:2021ryu,Schlumpf:1992vq,Schlumpf:1993rm,Cheng:1997kr,Ha:1998gf,Ramalho:2009gk,Girdhar:2015gsa,
Menapara:2022ksj,Mutuk:2021epz,Menapara:2021vug,Menapara:2021dzi,Gandhi:2018lez,Dahiya:2018ahb,Kaur:2016kan,Thakkar:2016sog,Shah:2016vmd,Dhir:2013nka,Sharma:2012jqz,Majethiya:2011ry,Sharma:2010vv,Dhir:2009ax,Simonis:2018rld,
Ghalenovi:2014swa,Kumar:2005ei,Rahmani:2020pol,Hazra:2021lpa,Gandhi:2019bju,Majethiya:2009vx,Shah:2016nxi,Shah:2018bnr,Ghalenovi:2018fxh,Wang:2022ugk,Mohan:2022sxm,An:2022qpt,Kakadiya:2022pin,Wu:2022gie,Wang:2023bek}
\begin{eqnarray}
\hat{\mu}_{z}^{\rm spin}&=&\sum_{i=c,\bar b}\mu_i\hat{\sigma}_{iz},\\
\hat{\mu}_{z}^{\rm orbital}&=&\frac{m_{c}\mu_{\bar{b}}}{m_{c}+m_{\bar{b}}} \hat{L}_z +\frac{m_{b}\mu_{{c}}}{m_{c}+m_{\bar{b}}} \hat{L}_z,
\end{eqnarray}
respectively. Here, $\mu_{i}=q_{i}/(2m_{i})$ represents the magnetic moment of the $i$-th quark, where $q_i$ and $m_i$ denote the charge and mass of the $i$-th quark, respectively. Moreover, $\hat{\sigma}_{iz}$ corresponds to the Pauli spin operator associated with the $i$-th quark. In the context of the $B_c$ meson, the subscripts $c$ and $\bar{b}$ refer to the charm quark and the antibottom quark, respectively. Additionally, $\hat{L}_z$ represents the $z$-component of the orbital angular momentum operator between the $c$ and $\bar{b}$ quarks.

In the case of the $S$-wave $B_c$ states, the contribution from the orbital magnetic moment is zero due to the absence of the orbital angular momentum ($L_z = 0$). Therefore, we only need to consider the contribution from the spin magnetic moment for the $S$-wave $B_c$ states. On the other hand, for the orbital excited $B_c$ states, the total magnetic moments comprise both the spin magnetic moment and the orbital magnetic moment. The spin-orbit wave function $|^{2S+1}L_J \rangle$ for the orbital excited $B_c$ states can be expanded by coupling the orbital wave function $Y_{L,m_L}$ with the spin wave function $\chi_{S,m_S}$, resulting in the following expression \cite{Zhou:2022gra}:
\begin{align}
|^{2S+1}L_J \rangle=\sum_{m_L,m_S} C_{Lm_L, Sm_S}^{JM} Y_{L,m_L} \chi_{S,m_S},
\end{align}
where $C_{Lm_L,Sm_S}^{JM}$ is the Clebsch-Gordan coefficient. Here, we need to point out that the orbital magnetic moment and the spin magnetic moment can be obtained by sandwiching the orbital and spin magnetic moment operators between the relevant spin-orbit and flavor wave functions.

To illustrate the calculations of magnetic moments for the orbital excited $B_c$ states, let us consider the $nP$-wave $B_c$ states, which include the $B_c({n }^{3} P_{0})$, $B_c({n }^{3} P_{1})$, $B_c({ n}^{1} P_{1})$, and $B_c({n }^{3} P_{2})$ states. In this case, due to the mixture of the spin-singlet and spin-triplet states, it is necessary to expand their spin-orbital wave functions $|^{2S+1}L_J \rangle$ as follows:
\begin{eqnarray}
\left|{n }^{3} P_{0}\right\rangle&=&\frac{1}{\sqrt{3}}Y_{1, -1}\chi_{1,1}-\frac{1}{\sqrt{3}}Y_{1, 0}\chi_{1,0}+\frac{1}{\sqrt{3}}Y_{1, 1}\chi_{1,-1},\nonumber\\
\left|{ n}^{3} P_{1}\right\rangle&=&\frac{1}{\sqrt{2}}Y_{1, 1}\chi_{1,0}-\frac{1}{\sqrt{2}}Y_{1, 0}\chi_{1,1},\nonumber\\
\left|{n }^{1} P_{1}\right\rangle&=&Y_{1, 1}\chi_{0,0},\nonumber\\
\left|{n }^{3} P_{2}\right\rangle&=&Y_{1, 1}\chi_{1,1}.
\end{eqnarray}
Using the previously expanded spin-orbital wave functions, we can proceed to calculate the magnetic moments and the transition magnetic moments of the $nP$-wave $B_c$ states. The explicit expressions for these quantities are
\begin{eqnarray}
\mu_{B_c\left({ n}^{3} P_{0}\right)}&=&0,\nonumber\\
\mu_{B_c\left({ n}^{3} P_{1}\right)}&=&\frac{1}{2}\mu_{c}+\frac{1}{2}\mu_{\bar b}+\frac{1}{2}\mu_{c \bar b}^L,\nonumber\\
\mu_{B_c\left({n }^{1} P_{1}\right)}&=&\mu_{c \bar b}^L,\\
\mu_{B_c\left({n }^{3} P_{2}\right)}&=&\mu_{c}+\mu_{\bar b}+\mu_{c \bar b}^L,\nonumber\\
\mu_{B_c\left({ n}^{3} P_{1}\right) \to B_c\left({n }^{1} P_{1}\right)}&=&\frac{1}{\sqrt 2}\mu_{c}-\frac{1}{\sqrt 2}\mu_{\bar b},\nonumber
\end{eqnarray}
where we introduce the notation $\mu_{c \bar b}^L=\frac{m_{c}\mu_{\bar{b}}}{m_{c}+m_{\bar{b}}} +\frac{m_{b}\mu_{{c}}}{m_{c}+m_{\bar{b}}}$ to simplify the magnetic moment calculations. It is important to note that the calculation method for the transition magnetic moments between the $nP$-wave $B_c$ states is similar to that of the magnetic moments of the $nP$-wave $B_c$ states, with the only difference being the wave functions of the initial and final states. For a more detailed calculation of the transition magnetic moments, please refer to Refs. \cite{Li:2021ryu,Zhou:2022gra,Wang:2022tib,Wang:2022ugk,Majethiya:2009vx,Majethiya:2011ry,Shah:2016nxi,Gandhi:2018lez,Simonis:2018rld,Ghalenovi:2018fxh,Gandhi:2019bju,Rahmani:2020pol,Hazra:2021lpa,Menapara:2021dzi,Menapara:2022ksj,Kakadiya:2022pin,Wang:2023bek}.

For the $nP$-wave $B_c$ states, it is important to consider the mixing of the $B_c({n}^{1} P_{1})$ and $B_c({n}^{3} P_{1})$ states, as discussed in Section \ref{sec2}. Referring to Eq. (\ref{2.10}), we can obtain the following expressions:
\begin{equation}
\begin{split}
\left|nP_1^{\prime}\right\rangle=&\left|n{}^1P_1\right\rangle{\rm cos}\theta_{nP}+\left|n{}^3P_1\right\rangle{\rm sin}\theta_{nP},\\
\left|nP_1\right\rangle=&-\left|n{}^1P_1\right\rangle{\rm sin}\theta_{nP}+\left|n{}^3P_1\right\rangle{\rm cos}\theta_{nP}.
\end{split}
\end{equation}
Here, $\theta_{nP}$ represents the mixing angle between the $B_c({n}^{1} P_{1})$ and $B_c({n}^{3} P_{1})$ states, as provided in Table \ref{Mass spectrum}. With these preparations, we can proceed to calculate the magnetic moments of the mixed states of the $B_c({n}^{1} P_{1})$ and $B_c({n}^{3} P_{1})$ states. The magnetic moments of these mixed states can be calculated using the following relations:
\begin{equation}
\begin{split}
\mu_{B_c(nP_1^{\prime})}=&\mu_{\left|{n}^{1} P_{1}\right\rangle}{\rm cos}^2\theta_{nP}+\mu_{\left|{n }^{3} P_{1}\right\rangle \to \left|{n }^{1} P_{1}\right\rangle}{\rm sin}2\theta_{nP}\\
&+\mu_{\left|{n}^{3} P_{1}\right\rangle}{\rm sin}^2\theta_{nP},\\
\mu_{B_c(nP_1)}=&\mu_{\left|{n }^{1} P_{1}\right\rangle}{\rm sin}^2\theta_{nP}-\mu_{\left|{
n}^{3} P_{1}\right\rangle \to \left|{ n}^{1} P_{1}\right\rangle}{\rm sin}2\theta_{nP}\\
&+\mu_{\left|{n}^{3} P_{1}\right\rangle}{\rm cos}^2\theta_{nP}.
\end{split}
\end{equation}
Therefore, the magnetic moments of the mixed states of the $B_c({n}^{1} P_{1})$ and $B_c({n}^{3} P_{1})$ states are determined not only by the magnetic moments and the transition magnetic moments of the $B_c({n}^{1} P_{1})$ and $B_c({n}^{3} P_{1})$ states but also by the mixing angles of the $B_c({n}^{1} P_{1})$ and $B_c({n}^{3} P_{1})$ states.

Within the framework of the constituent quark model, the masses of the involved quarks play a crucial role in the study of their magnetic moment properties. In our calculations, we adopt the quark masses $m_c=1.660$ GeV and $m_b=4.730$ GeV, which have been widely used to describe the hadronic magnetic moments quantitatively \cite{Lichtenberg:1976fi,Li:2017cfz,Meng:2017dni,Li:2017pxa,Wang:2019mhm,Gao:2021hmv,Wang:2023bek}. Table \ref{magnetic moment} presents the magnetic moments of the $S$/$P$/$D$/$F$-wave $B_c$ mesons obtained from our calculations. Furthermore, we compare our results with those from other theoretical works and find that our findings are in good agreement with the theoretical predictions in Ref. \cite{Wang:2023bek}. It is important to note that the investigations of the magnetic moments of $B_c$ states have not received much attention thus far. Therefore, we hope that our work will stimulate further theoretical and experimental efforts to explore the magnetic moments of the $B_c$ states.

\begin{table}[htbp]\centering
\caption{Magnetic moments of the $S/P/D/F$-wave $B_c$ mesons. Here, the magnetic moment of the hadron is in units of the nuclear magneton $\mu_N=e/2m_p$.}
\renewcommand\arraystretch{1.05}
\label{magnetic moment}
\begin{tabular*}{85mm}{c@{\extracolsep{\fill}}ccccc}
\toprule[1.5pt]\toprule[0.5pt]
\text{States} & \text{Expressions}  & \text{Results} & \text{Ref. \cite{Wang:2023bek}} & \text{Ref. \cite{Simonis:2016pnh}} \\
\toprule[1pt]
$^1S_0$    & 0 & 0 & 0 & 0 &\\
$^3S_1$    & $\mu_{c}+\mu_{\bar{b}}$ & 0.443 &  & 0.350 & \\
\hline
$^3P_0$    & 0 & 0 & 0 &  & \\
$^3P_2$    & $\mu_c+\mu_{\bar{b}}+\mu^L_{c\bar{b}}$ & 0.739 & 0.739 &  & \\
$1P'_1$    &  & 0.527 &  &  &  \\
$1P_1$     &  & 0.138 &  &  &  \\
$2P'_1$    &  & 0.486 & 0.437 &  &  \\
$2P_1$     &  & 0.179 & 0.229 &  &  \\
$3P'_1$    &  & 0.469 & 0.454 &  &  \\
$3P_1$     &  & 0.197 &  &  &  \\
$4P'_1$    &  & 0.460 \\
$4P_1$     &  & 0.205 \\
\hline
$^3D_1$    & $-\frac{1}{2}(\mu_c+\mu_{\bar{b}})+\frac{3}{2}\mu^{L}_{c\bar{b}}$ & 0.223 &  \\
$^3D_3$    & $\mu_c+\mu_{\bar{b}}+2\mu^{L}_{c\bar{b}}$ & 1.035 &  \\
$1D'_2$    &  & 0.381 &  \\
$1D_2$     &  & 0.852 &  \\
$2D'_2$    &  & 0.368 &  \\
$2D_2$     &  & 0.865 &  \\
$2D'_2$    &  & 0.362 &  \\
$3D_2$     &  & 0.871 &  \\
\hline
$^3F_2$      & $-\frac{2}{3}(\mu_c+\mu_{\bar{b}})+\frac{8}{3}\mu^{L}_{c\bar{b}}$ & 0.494 &  \\
$^3F_4$      & $\mu_c+\mu_{\bar{b}}+3\mu^{L}_{c\bar{b}}$ & 1.331 &  \\
$1F'_3$      &  & 0.646 &  \\
$1F_3$       &  & 1.167 &  \\
$2F'_3$      &  & 0.644 &  \\
$2F_3$       &  & 1.169 &  \\
\bottomrule[0.5pt]\bottomrule[1.5pt]
\end{tabular*}
\end{table}

As widely acknowledged, phenomena in the higher mass region are highly intricate. This domain encompasses various conventional $B_c$ states, which possess identical quantum numbers and similar masses, as well as predicted $B_c$-like molecular states \cite{Sun:2012sy}. Distinguishing between these states poses a critical challenge for both theoretical and experimental aspects. For instance, consider the conventional $B_c(2 P_{1}^{\prime})$ state and the $DB^*$ molecular state with $I(J^P)=0(1^+)$, which have closely aligned masses. However, it has been observed that their magnetic moments exhibit evident differences, i.e.,
$\mu_{B_c(2 P_{1}^{\prime})}=0.486\mu_N$, $\mu_{DB^*[0(1^+)]}=0.532\mu_N$ \cite{Wang:2023bek}.
There are many such examples. Consequently, investigating the magnetic moment properties provides a means to differentiate between states sharing identical quantum numbers and similar masses.

\section{Some typical weak decays of the $B_c(1^3S_0)$ meson}\label{sec6}

As the lowest bottom-charmed meson, the $B_{c}(1^3S_0)$ meson can only decay via the weak process. For simplicity, we use $B_{c}$ to denote the $B_{c}(1^3S_0)$ in this section. In experiments, a series weak processes have been observed, while the absolute branching ratios are deficiency for most of them \cite{ParticleDataGroup:2022pth}. Previous theoretical studies on $B_{c}$ weak decays have been conducted using lattice QCD (LQCD) \cite{Lytle:2016ixw,Harrison:2020gvo}, perturbative QCD \cite{Hu:2019qcn,Liu:2023kxr}, QCD sum rule \cite{Kiselev:2002vz,Leljak:2019eyw}, light-cone sum rule \cite{Huang:2007kb}, various quark models \cite{Scora:1995ty,Colangelo:1999zn,Ivanov:2000aj,Ebert:2003cn,Ivanov:2006ni,Wang:2007sxa,Wang:2008xt,Wang:2009mi,Ke:2013yka,Shi:2016gqt,Chen:2021ywv,Zhang:2023ypl,Sun:2023iis}, and other methods \cite{Yao:2021pyf}. With the updated high luminosity of the LHC, the experimental measurements of these weak decays become feasible, and this will demand more theoretical studies.

In theoretical aspects, the light-front quark model is a powerful phenomenological model to calculate the weak transition matrix elements of mesons and baryons decays. Specially, the $B_{c}$ weak decays have been studied by the CLFQM in Refs. \cite{Wang:2007sxa,Wang:2008xt,Wang:2009mi,Ke:2013yka,Shi:2016gqt,Chen:2021ywv,Zhang:2023ypl,Sun:2023iis}. However, in the concrete calculations, the important input, i.e., the spatial wave function, is usually adopted as the Gaussian-like form, and may  bring large uncertainty due to the phenomenological parameter $\beta$. In this work, we also employ the CLFQM to revisit the $B_{c}$ weak decays. In our calculations, we adopt the numerical spatial wave functions of the involved mesons, which benefit from the MGI model introduced in Section \ref{sec2}, as opposed to SHO wave functions with a phenomenological parameter $\beta$ used in previous works based on the CLFQM \cite{Wang:2007sxa,Wang:2008xt,Wang:2009mi,Ke:2013yka,Shi:2016gqt,Chen:2021ywv,Zhang:2023ypl,Sun:2023iis}. This approach reduces the dependence on phenomenological parameters in determining the form factors. 

In order to make a comprehensive discussion, we first discuss the $B_{c}\to M$ weak transition form factors, where $M$ represents a pseudoscalar meson ($P$), vector meson ($V$), scalar meson ($S$), or axial meson ($A$). These transitions involve the quark-level transitions $b\to c(u)$ and $c\to s(d)$. Additionally, we employ these form factors as inputs to investigate various weak decays, including semileptonic decays and typical two-body nonleptonic decays.

Generally, the $B_{c}\to M$ transitions induced by the $V-A$ current can be expressed as \cite{Wirbel:1985ji}
\begin{widetext}
\begin{equation}
\begin{split}
\langle P(p^{\prime\prime})\vert V_{\mu} \vert B_{c}(p^{\prime})\rangle=&\Bigg{(}P_{\mu}-\frac{m_{B_{c}}^{2}-m_{P}^{2}}{q^{2}}q_{\mu}\Bigg{)}F_{1}^{B_cP}(q^{2})+\frac{m_{B_{c}}^{2}-m_{P}^{2}}{q^{2}}q_{\mu}F_{0}^{B_{c}P}(q^{2}),\\
\langle V(p^{\prime\prime})\vert V_{\mu} \vert B_{c}(p^{\prime})\rangle=&-\frac{1}{m_{B_{c}}+m_{V}}\epsilon_{\mu\nu\alpha\beta}\varepsilon_{V}^{*\nu}P^{\alpha}q^{\beta}V^{B_{c}V}(q^{2}),\\
\langle V(p^{\prime\prime})\vert A_{\mu} \vert B_{c}(p^{\prime})\rangle=&i\Bigg{\{}(m_{B_{c}}+m_{V})\varepsilon_{V\mu}^{*}A_{1}^{B_{c}V}(q^{2})-\frac{\varepsilon_{V}^{*}\cdot P}{m_{B_{c}}+m_{V}}P_{\mu}A_{2}^{B_{c}V}(q^{2})-2m_{V}\frac{\varepsilon_{V}^{*}\cdot P}{q^{2}}q_{\mu}\Big{[}A_{3}^{B_{c}V}(q^{2})-A_{0}^{B_{c}V}(q^{2})\Big{]}\Bigg{\}},\\
\langle A(p^{\prime\prime})\vert V_{\mu} \vert B_{c}(p^{\prime})\rangle=&-i\Bigg{\{}(m_{B_{c}}-m_{A})\varepsilon_{A\mu}^{*}V_{1}^{B_{c}A}(q^{2})-\frac{\varepsilon_{A}^{*}\cdot P}{m_{B_{c}}-m_{A}}P_{\mu}V_{2}^{B_{c}A}(q^{2})-2m_{A}\frac{\varepsilon_{A}^{*}\cdot P}{q^{2}}q_{\mu}\Big{[}V_{3}^{B_{c}A}(q^{2})-V_{0}^{B_{c}A}(q^{2})\Big{]}\Bigg{\}},\\
\langle A(p^{\prime\prime})\vert A_{\mu} \vert B_{c}(p^{\prime})\rangle=&-\frac{1}{m_{B_{c}}-m_{A}}\epsilon_{\mu\nu\alpha\beta}\varepsilon_{A}^{*\nu}P^{\alpha}q^{\beta}A^{B_{c}A}(q^{2}),\\
\langle S(p^{\prime\prime})\vert A_{\mu} \vert B_{c}(p^{\prime})\rangle=&\Bigg{(}P_{\mu}-\frac{m_{B_{c}}^{2}-m_{S}^{2}}{q^{2}}q_{\mu}\Bigg{)}F_{1}^{B_{c}S}(q^{2})+\frac{m_{B_{c}}^{2}-m_{S}^{2}}{q^{2}}q_{\mu}F_{0}^{B_{c}S}(q^{2}),
\end{split}
\end{equation}
\end{widetext}
where $p^{\prime}$ and $p^{\prime\prime}$ are the momenta of the initial state meson $B_{c}$ and the final state meson $P/V/A/S$, respectively. Besides, we define $P_{\mu}=p_{\mu}^{\prime}+p_{\mu}^{\prime\prime}$ and $q_{\mu}=p_{\mu}^{\prime}-p_{\mu}^{\prime\prime}$, while the convention $\epsilon_{0123}=+1$ is used.

In the frame of the CLFQM, the constituent (anti)quark inside a meson system are off-shell. The parent and daughter mesons have the four momenta $P^{\prime}=p_{1}^{\prime}+p_{2}$ and $P^{\prime\prime}=p_{1}^{\prime\prime}+p_{2}$, where $p_{1}^{\prime(\prime\prime)}$ and $p_{2}$ are the four momenta of the quark and the antiquark, respectively. These momenta can be expressed in terms of the following internal variables ($x_{i},\vec{k}_{\bot}^{\prime}$) ($i=1,2$):
\begin{equation}
p_{1}^{\prime+}=x_{1}P^{\prime+},~~p_{1}^{+}=x_{2}P^{\prime+},~~\vec{p}_{1\bot}^{\prime}=x_{1}\vec{P}_{\bot}^{\prime}+\vec{k}_{\bot}^{\prime},
\end{equation}
where they must satisfy the relation $x_{1}+x_{2}=1$.

The $B_{c}\to M$ weak transition form factors have been widely studied by CLFQM in Refs. \cite{Wang:2007sxa,Wang:2008xt,Wang:2009mi,Ke:2013yka,Shi:2016gqt,Chen:2021ywv,Zhang:2023ypl}. As derived in Refs. \cite{Jaus:1999zv,Cheng:2003sm,Wang:2008xt,Shi:2016gqt,Zhang:2023ypl,Verma:2011yw}, the $B_{c}\to P$ weak transition form factors are
\begin{equation}
\begin{split}
F_{1}^{B_{c}P}(q^{2})=&\frac{N_{c}}{16\pi^{3}}\int dx_{2}d^{2}\vec{k}_{\bot}^{\prime}
\frac{h_{B_{c}}^{\prime}h_{P}^{\prime\prime}}{x_{2}\hat{N}_{1}^{\prime}\hat{N}_{1}^{\prime\prime}}
\Big{[}x_{1}(M_{0}^{\prime2}+M_{0}^{\prime\prime2})+x_{2}q^{2}\\
&-x_{2}(m_{1}^{\prime}-m_{1}^{\prime\prime})^{2}-x_{1}(m_{1}^{\prime}-m_{2})^{2}-x_{1}(m_{1}^{\prime\prime}-m_{2})^{2}\Big{]},
\label{eq:formfactorPF1}
\end{split}
\end{equation}
\begin{equation}
\begin{split}
F_{0}^{B_{c}P}(q^{2})=&F_{1}^{B_{c}P}(q^{2})+\frac{q^{2}}{P\cdot q}\frac{N_{c}}{16\pi^{3}}\int dx_{2}d^{2}\vec{k}_{\bot}^{\prime}
\frac{2h_{B_{c}}^{\prime}h_{P}^{\prime\prime}}{x_{2}\hat{N}_{1}^{\prime}\hat{N}_{1}^{\prime\prime}}\\
&\times\Big{\{}-x_{1}x_{2}M^{\prime2}-k_{\bot}^{\prime2}-m_{1}^{\prime}m_{2}+(m_{1}^{\prime\prime}-m_{2})\\
&\times(x_{2}m_{1}^{\prime}+x_{1}m_{2})+2\frac{P\cdot q}{q^{2}}\Big{(}k_{\bot}^{\prime2}+2\frac{(\vec{k}_{\bot}^{\prime}\cdot \vec{q}_{\bot})^{2}}{q^{2}}\Big{)}\\
&+2\frac{(\vec{k}_{\bot}^{\prime}\cdot \vec{q}_{\bot})^{2}}{q^{2}}-\frac{\vec{k}_{\bot}^{\prime}\cdot \vec{q}_{\bot}}{q^{2}}\Big{[}M^{\prime\prime2}-x_{2}(q^{2}+P\cdot q)\\
&-{\!}(x_{2}{\!}-{\!}x_{1})M^{\prime2}{\!}+{\!}2x_{1}M_{0}^{\prime2}{\!}-{\!}2(m_{1}^{\prime}{\!}-{\!}m_{2})(m_{1}^{\prime}{\!}+{\!}m_{1}^{\prime\prime})\Big{]}\Big{\}},
\label{eq:formfactorPF0}
\end{split}
\end{equation}
where
\begin{equation}
\begin{split}
h_{B_{c}}^{\prime}=&(M^{\prime2}-M_{0}^{\prime2})\sqrt{\frac{x_{1}x_{2}}{N_{c}}}
\frac{1}{\sqrt{2}\tilde{M}_{0}^{\prime}}\phi_{s}(x_{2},\vec{k}_{\bot}^{\prime}),\\
h_{P}^{\prime\prime}=&(M^{\prime\prime2}-M_{0}^{\prime\prime2})\sqrt{\frac{x_{1}x_{2}}{N_{c}}}
\frac{1}{\sqrt{2}\tilde{M}_{0}^{\prime\prime}}\phi_{s}(x_{2},\vec{k}_{\bot}^{\prime\prime})
\end{split}
\end{equation}
with
\begin{equation}
\begin{split}
M_{0}^{\prime(\prime\prime)2}&=\frac{\vec{k}_{\bot}^{\prime(\prime\prime)2}+m_{1}^{\prime(\prime\prime)2}}{x_{1}}+\frac{\vec{k}_{\bot}^{\prime(\prime\prime)2}+m_{2}^{\prime(\prime\prime)2}}{x_{2}},\\
\tilde{M}_{0}^{\prime(\prime\prime)}&=\sqrt{M_{0}^{\prime(\prime\prime)2}-(m_{1}^{\prime(\prime\prime)}-m_{2})^{2}},\\
\vec{k}_{\bot}^{\prime\prime}&=\vec{k}_{\bot}^{\prime}-x_{2}\vec{q}_{\bot}.
\end{split}
\end{equation}

Analogously, the $B_{c}\to S$ weak transition form factors are written as \cite{Cheng:2003sm,Wang:2009mi,Shi:2016gqt,Zhang:2023ypl,Verma:2011yw}
\begin{equation}
\begin{split}
F_{1}^{B_{c}S}(q^{2})=&\frac{N_{c}}{16\pi^{3}}\int dx_{2}d^{2}\vec{k}_{\bot}^{\prime}
\frac{h_{B_{c}}^{\prime}h_{S}^{\prime\prime}}{x_{2}\hat{N}_{1}^{\prime}\hat{N}_{1}^{\prime\prime}}
\Big{[}x_{1}(M_{0}^{\prime2}+M_{0}^{\prime\prime2})+x_{2}q^{2}\\
&-x_{2}(m_{1}^{\prime}+m_{1}^{\prime\prime})^{2}-x_{1}(m_{1}^{\prime}-m_{2})^{2}-x_{1}(m_{1}^{\prime\prime}+m_{2})^{2}\Big{]},
\label{eq:formfactorSF1}
\end{split}
\end{equation}
\begin{equation}
\begin{split}
F_{0}^{B_{c}S}(q^{2})=&F_{1}^{B_{c}S}(q^{2})+\frac{q^{2}}{P\cdot q}\frac{N_{c}}{16\pi^{3}}\int dx_{2}d^{2}\vec{k}_{\bot}^{\prime}
\frac{2h_{B_{c}}^{\prime}h_{S}^{\prime\prime}}{x_{2}\hat{N}_{1}^{\prime}\hat{N}_{1}^{\prime\prime}}\\
&\times\Big{\{}-x_{1}x_{2}M^{\prime2}-\vec{k}_{\bot}^{\prime2}-m_{1}^{\prime}m_{2}-(m_{1}^{\prime\prime}+m_{2})\\
&\times(x_{2}m_{1}^{\prime}+x_{1}m_{2})+2\frac{P\cdot q}{q^{2}}\Big{(}\vec{k}_{\bot}^{\prime2}+2\frac{(\vec{k}_{\bot}^{\prime}\cdot \vec{q}_{\bot})^{2}}{q^{2}}\Big{)}\\
&+2\frac{(\vec{k}_{\bot}^{\prime}\cdot \vec{q}_{\bot})^{2}}{q^{2}}-\frac{\vec{k}_{\bot}^{\prime}\cdot \vec{q}_{\bot}}{q^{2}}\Big{[}M^{\prime\prime2}-x_{2}(q^{2}+P\cdot q)\\
&-{\!}(x_{2}-x_{1})M^{\prime2}{\!}+{\!}2x_{1}M_{0}^{\prime2}{\!}-{\!}2(m_{1}^{\prime}-m_{2})(m_{1}^{\prime}-m_{1}^{\prime\prime})\Big{]}\Big{\}},
\label{eq:formfactorSF0}
\end{split}
\end{equation}
where
\begin{equation}
h_{S}^{\prime\prime}=(M^{\prime\prime2}-M_{0}^{\prime\prime2})\sqrt{\frac{x_{1}x_{2}}{N_{c}}}\frac{1}{\sqrt{2}\tilde{M}_{0}^{\prime\prime}}
\frac{\tilde{M}_{0}^{\prime\prime}}{2\sqrt{3}M_{0}^{\prime\prime}}\phi_{p}(x_{2},\vec{k}_{\bot}^{\prime\prime}).
\end{equation}

The $B_{c}\to V$ transition form factors can be evaluated by
\begin{equation}
\begin{split}
V^{B_{c}V}{\!}=&-(m_{B_{c}}+m_{V})g,\\
A_{1}^{B_{c}V}{\!}=&-f/(m_{B_{c}}+m_{V}),~~~A_{2}^{B_{c}V}{\!}=(m_{B_{c}}+m_{V})a_{+},\\
A_{0}^{B_{c}V}{\!}=&\frac{m_{B_{c}}{\!}+{\!}m_{V}}{2m_{V}}A_{1}^{B_{c}V}{\!}-{\!}\frac{m_{B_{c}}{\!}-{\!}m_{V}}{2m_{V}}A_{2}^{B_{c}V}{\!}-{\!}\frac{q^{2}}{2m_{V}}a_{-},
\label{eq:formfactorV}
\end{split}
\end{equation}
where $g$, $f$, $a_{+}$, and $a_{-}$ are the scalar functions of $q^{2}$. Their expressions in CLFQM are \cite{Jaus:1999zv,Cheng:2003sm,Wang:2007sxa,Wang:2008xt,Verma:2011yw,Shi:2016gqt,Zhang:2023ypl}
\begin{widetext}
\begin{equation}
\begin{split}
g(q^{2})=&-\frac{N_{c}}{16\pi^{3}}\int dx_{2}d^{2}\vec{k}_{2\bot}\frac{2h_{B_{c}}^{\prime}h_{V}^{\prime\prime}}{x_{2}\hat{N}_{1}^{\prime}\hat{N}_{1}^{\prime\prime}}
\Big{\{}x_{2}m_{1}^{\prime}+x_{1}m_{2}+(m_{1}^{\prime}-m_{1}^{\prime\prime})\frac{\vec{k}_{\bot}^{\prime}\cdot\vec{q}_{\bot}}{q^{2}}+\frac{2}{\omega_{V}^{\prime\prime}}\Big{[}\vec{k}_{\bot}^{\prime2}+\frac{(\vec{k}_{\bot}^{\prime}\cdot\vec{q}_{\bot})^{2}}{q^{2}}\Big{]}\Big{\}},
\end{split}
\label{eq:formfactorVg}
\end{equation}
\begin{equation}
\begin{split}
f(q^{2})=&\frac{N_{c}}{16\pi^{3}}\int dx_{2}d^{2}\vec{k}_{\bot}^{\prime}\frac{h_{B_{c}}^{\prime}h_{V}^{\prime\prime}}{x_{2}\hat{N}_{1}^{\prime}\hat{N}_{1}^{\prime\prime}}
\Big{\{}2x_{1}(m_{2}-m_{1}^{\prime})(M_{0}^{\prime2}+M_{0}^{\prime\prime2})-4x_{1}m_{1}^{\prime\prime}M_{0}^{\prime2}+2x_{2}m_{1}^{\prime}P\cdot q+2m_{2}q^{2}\\
&-2x_{1}m_{2}(M^{\prime2}+M^{\prime\prime2})+2(m_{1}^{\prime}-m_{2})(m_{1}^{\prime}+m_{1}^{\prime\prime})^{2}
+8(m_{1}^{\prime}-m_{2})\Big{[}\vec{k}_{\bot}^{\prime2}+\frac{(\vec{k}_{\bot}^{\prime}\cdot\vec{q}_{\bot})^{2}}{q^{2}}\Big{]}\\
&+2(m_{1}^{\prime}+m_{1}^{\prime\prime})(q^{2}+P\cdot q)\frac{\vec{k}_{\bot}^{\prime}\cdot\vec{q}_{\bot}}{q^{2}}
-4\frac{q^{2}\vec{k}_{\bot}^{\prime2}+(\vec{k}_{\bot}^{\prime}\cdot\vec{q}_{\bot})^{2}}{q^{2}\omega_{V}^{\prime\prime}}
\Big{[}2x_{1}(M^{\prime2}+M_{0}^{\prime2})-q^{2}-P\cdot q\\
&-2(q^{2}+P\cdot q)\frac{\vec{k}_{\bot}^{\prime}\cdot\vec{q}_{\bot}}{q^{2}}-2(m_{1}^{\prime}-m_{1}^{\prime\prime})(m_{1}^{\prime}-m_{2})\Big{]}\Big{\}},
\label{eq:formfactorVf}
\end{split}
\end{equation}
\begin{equation}
\begin{split}
a_{+}(q^{2})=&\frac{N_{c}}{16\pi^{3}}\int dx_{2}d^{2}\vec{k}_{\bot}^{\prime}
\frac{2h_{B_{c}}^{\prime}h_{V}^{\prime\prime}}{x_{2}\hat{N}_{1}^{\prime}\hat{N}_{1}^{\prime\prime}}
\Big{\{}(x_{1}-x_{2})(x_{2}m_{1}^{\prime}+x_{1}m_{2})
-\big{[}2x_{1}m_{2}+m_{1}^{\prime\prime}+(x_{2}-x_{1})m_{1}^{\prime}\big{]}\frac{\vec{k}_{\bot}^{\prime}\cdot\vec{q}_{\bot}}{q^{2}}\\
&-2\frac{x_{2}q^{2}+\vec{k}_{\bot}^{\prime}\cdot\vec{q}_{\bot}}{x_{2}q^{2}\omega_{V}^{\prime\prime}}
\big{[}\vec{k}_{\bot}^{\prime}\cdot\vec{k}_{\bot}^{\prime\prime}+(x_{1}m_{2}+x_{2}m_{1}^{\prime})(x_{1}m_{2}-x_{2}m_{1}^{\prime\prime})\big{]}\Big{\}},
\label{eq:formfactorVaplus}
\end{split}
\end{equation}
\begin{equation}
\begin{split}
a_{-}(q^{2})=&\frac{N_{c}}{16\pi^{3}}\int dx_{2}d^{2}\vec{k}_{\bot}^{\prime}
\frac{h_{B_{c}}^{\prime}h_{V}^{\prime\prime}}{x_{2}\hat{N}_{1}^{\prime}\hat{N}_{1}^{\prime\prime}}
\Big{\{}2(2x_{1}-3)(x_{2}m_{1}^{\prime}+x_{1}m_{2})-8(m_{1}^{\prime}-m_{2})\Big{[}\frac{\vec{k}_{\bot}^{\prime2}}{q^{2}}+2\frac{(\vec{k}_{\bot}^{\prime}\cdot\vec{q}_{\bot})^{2}}{q^{4}}\Big{]}\\
&-\big{[}(14-12x_{1})m_{1}^{\prime}-2m_{1}^{\prime\prime}-(8-12x_{1})m_{2}\big{]}\frac{\vec{k}_{\bot}^{\prime}\cdot\vec{q}_{\bot}}{q^{2}}
+\frac{4}{\omega_{V}^{\prime\prime}}\Big{(}
\big{[}M^{\prime2}+M^{\prime\prime2}-q^{2}+2(m_{1}^{\prime}-m_{2})(m_{1}^{\prime\prime}+m_{2})\big{]}\\
&\times(A_{3}^{2}+A_{4}^{(2)}-A_{2}^{1})+Z_{2}(3A_{2}^{(1)}-2A_{4}^{(2)}-1)
+\frac{1}{2}P\cdot q(A_{1}^{(1)}+A_{2}^{(1)}-1)\big{[}x_{1}(q^{2}+P\cdot q)-2M^{\prime2}-2\vec{k}_{\bot}^{\prime}\cdot\vec{q}_{\bot}\\
&-2m_{1}^{\prime}(m_{1}^{\prime\prime}+m_{2}-2m_{2}(m_{1}^{\prime}-m_{2}))\big{]}
\Big{[}\frac{\vec{k}_{\bot}^{\prime2}}{q^{2}}+\frac{(\vec{k}_{\bot}^{\prime}\cdot\vec{q}_{\bot})^{2}}{q^{4}}\Big{]}(4A_{2}^{(1)}-3)\Big{)}\Big{\}},
\label{eq:formfactorVamin}
\end{split}
\end{equation}
where
\begin{equation}
\begin{split}
h_{V}^{\prime\prime}&=(M^{\prime\prime2}-M_{0}^{\prime\prime2})\sqrt{\frac{x_{1}x_{2}}{N_{c}}}
\frac{1}{\sqrt{2}\tilde{M}_{0}^{\prime\prime}}\phi_{s}(x_{2},\vec{k}_{\bot}^{\prime\prime}),\\
\omega_{V}^{\prime\prime}&=M_{0}^{\prime\prime}+m_{1}^{\prime\prime}+m_{2}.
\end{split}
\end{equation}

For the $B_{c}\to{^{3}A(^{1}A)}$ transition, the form factors can be evaluated by the relations:
\begin{equation}
\begin{split}
A^{B_{c}{^{3}A(^{1}A)}}=&-(m_{B_{c}}-m_{A})q^{{^{3}A(^{1}A)}},\\
V_{1}^{B_{c}{^{3}A(^{1}A)}}=&-l^{{^{3}A(^{1}A)}}/(m_{B_{c}}-m_{A}),~~~
V_{2}^{B_{c}{^{3}A(^{1}A)}}=(m_{B_{c}}-m_{A})c^{{^{3}A(^{1}A)}}_{+},\\
V_{0}^{B_{c}{^{3}A(^{1}A)}}=&\frac{m_{B_{c}}{\!}-{\!}m_{A}}{2m_{A}}V_{1}^{B_{c}{^{3}A(^{1}A)}}
-\frac{m_{B_{c}}{\!}+{\!}m_{A}}{2m_{A}}V_{2}^{B_{c}{^{3}A(^{1}A)}}
-\frac{q^{2}}{2m_{A}}c_{-}^{{^{3}A(^{1}A)}},
\label{eq:formfactorA}
\end{split}
\end{equation}
where $q^{^{3}A(^{1}A)}$, $l^{^{3}A(^{1}A)}$, and $c_{\pm}^{^{3}A(^{1}A)}$ are functions of $q^{2}$, with the concrete expressions in the CLFQM as \cite{Cheng:2003sm,Wang:2007sxa,Zhang:2023ypl}
\begin{equation}
\begin{split}
q^{^{3}A{(^{1}A)}}(q^{2})=&-\frac{N_{c}}{16\pi^{3}}\int dx_{2}d^{2}\vec{k}_{2\bot}
\frac{2h_{B_{c}}^{\prime}h_{^{3}A{(^{1}A)}}^{\prime\prime}}{x_{2}\hat{N}_{1}^{\prime}\hat{N}_{1}^{\prime\prime}}
\Bigg{\{}\frac{2}{\omega_{^{3}A{(^{1}A)}}^{\prime\prime}}\Bigg{[}\vec{k}_{\bot}^{\prime2}+\frac{(\vec{k}_{\bot}^{\prime}\cdot\vec{q}_{\bot})^{2}}{2^{2}}\Bigg{]}
\Bigg{(}+x_{2}m_{1}^{\prime}+x_{1}m_{2}+(m_{1}^{\prime}+m_{1}^{\prime\prime})\frac{\vec{k}_{\bot}^{\prime}\cdot\vec{q}_{\bot}}{q^{2}}\Bigg{)}\Bigg{\}},
\end{split}
\label{eq:formfactorAg}
\end{equation}
\begin{equation}
\begin{split}
l^{^{3}A{(^{1}A)}}(q^{2})=&\frac{N_{c}}{16\pi^{3}}\int dx_{2}d^{2}\vec{k}_{\bot}^{\prime}
\frac{h_{B_{c}}^{\prime}h_{^{3}A{(^{1}A)}}^{\prime\prime}}{x_{2}\hat{N}_{1}^{\prime}\hat{N}_{1}^{\prime\prime}}
\Bigg{\{}-4\frac{q^{2}\vec{k}_{\bot}^{\prime2}+(\vec{k}_{\bot}^{\prime}\cdot\vec{q}_{\bot})^{2}}{q^{2}\omega_{^{3}A{(^{1}A)}}^{\prime\prime}}
\Bigg{[}2x_{1}(M^{\prime2}+M_{0}^{\prime2})-q^{2}-P\cdot q-2(q^{2}+P\cdot q)\frac{\vec{k}_{\bot}^{\prime}\cdot\vec{q}_{\bot}}{q^{2}}\\
&-2(m_{1}^{\prime}+m_{1}^{\prime\prime})(m_{1}^{\prime}-m_{2})\Bigg{]}
\Bigg{(}+2x_{1}(m_{2}-m_{1}^{\prime})(M_{0}^{\prime2}+M_{0}^{\prime\prime2})+4x_{1}m_{1}^{\prime\prime}M_{0}^{\prime2}
+2x_{2}m_{1}^{\prime}P\cdot q+2m_{2}q^{2}-2x_{1}m_{2}\\
&\times(M^{\prime2}+M^{\prime\prime2})
+2(m_{1}^{\prime}-m_{2})(m_{1}^{\prime}-m_{1}^{\prime\prime})^{2}
+8(m_{1}^{\prime}-m_{2})\Big{[}\vec{k}_{\bot}^{\prime2}+\frac{(\vec{k}_{\bot}^{\prime}\cdot\vec{q}_{\bot})^{2}}{q^{2}}\Big{]}
+2(m_{1}^{\prime}-m_{1}^{\prime\prime})(q^{2}+P\cdot q)\frac{\vec{k}_{\bot}^{\prime}\cdot\vec{q}_{\bot}}{q^{2}}\Bigg{)}
\Bigg{\}},
\label{eq:formfactorAf}
\end{split}
\end{equation}
\begin{equation}
\begin{split}
c^{^{3}A{(^{1}A)}}_{+}(q^{2})=&\frac{N_{c}}{16\pi^{3}}\int dx_{2}d^{2}\vec{k}_{\bot}^{\prime}
\frac{2h_{B_{c}}^{\prime}h_{^{3}A{(^{1}A)}}^{\prime\prime}}{x_{2}\hat{N}_{1}^{\prime}\hat{N}_{1}^{\prime\prime}}
\Bigg{\{}-2\frac{x_{2}q^{2}+\vec{k}_{\bot}^{\prime}\cdot\vec{q}_{\bot}}{x_{2}q^{2}\omega_{^{3}A{(^{1}A)}}^{\prime\prime}}
\big{[}\vec{k}_{\bot}^{\prime}\cdot\vec{k}_{\bot}^{\prime\prime}+(x_{1}m_{2}+x_{2}m_{1}^{\prime})(x_{1}m_{2}+x_{2}m_{1}^{\prime\prime})\big{]}\\
&{\Bigg{(}+(x_{1}-x_{2})(x_{2}m_{1}^{\prime}+x_{1}m_{2})
-\big{[}2x_{1}m_{2}-m_{1}^{\prime\prime}+(x_{2}-x_{1})m_{1}^{\prime}\big{]}\frac{\vec{k}_{\bot}^{\prime}\cdot\vec{q}_{\bot}}{q^{2}}\Bigg{)}}\Bigg{\}},
\label{eq:formfactorAaplus}
\end{split}
\end{equation}
\begin{equation}
\begin{split}
c^{^{3}A{(^{1}A)}}_{-}(q^{2})=&\frac{N_{c}}{16\pi^{3}}\int dx_{2}d^{2}\vec{k}_{\bot}^{\prime}
\frac{h_{B_{c}}^{\prime}h_{^{3}A{(^{1}A)}}^{\prime\prime}}{x_{2}\hat{N}_{1}^{\prime}\hat{N}_{1}^{\prime\prime}}
\Bigg{(}\frac{4}{\omega_{^{3}A{(^{1}A)}}^{\prime\prime}}\Bigg{(}\big{[}M^{\prime2}+M^{\prime\prime2}-q^{2}+2(m_{1}^{\prime}-m_{2})(-m_{1}^{\prime\prime}+m_{2})\big{]}(A_{3}^{2}+A_{4}^{(2)}-A_{2}^{1})\\
&+Z_{2}(3A_{2}^{(1)}-2A_{4}^{(2)}-1)+\frac{1}{2}P\cdot q(A_{1}^{(1)}+A_{2}^{(1)}-1)\big{[}x_{1}(q^{2}+P\cdot q)-2M^{\prime2}-2\vec{k}_{\bot}^{\prime}\cdot\vec{q}_{\bot}-2m_{1}^{\prime}(-m_{1}^{\prime\prime}+m_{2}\\
&-2m_{2}(m_{1}^{\prime}-m_{2}))\big{]}\Big{[}\frac{\vec{k}_{\bot}^{\prime2}}{q^{2}}+
\frac{(\vec{k}_{\bot}^{\prime}\cdot\vec{q}_{\bot})^{2}}{q^{4}}\Big{]}(4A_{2}^{(1)}-3)\Bigg{)}
\Bigg{(}+2(2x_{1}-3)(x_{2}m_{1}^{\prime}+x_{1}m_{2})-8(m_{1}^{\prime}-m_{2})\\
&\times\Big{[}\frac{\vec{k}_{\bot}^{\prime2}}{q^{2}}
+2\frac{(\vec{k}_{\bot}^{\prime}\cdot\vec{q}_{\bot})^{2}}{q^{4}}\Big{]}
-\big{[}(14-12x_{1})m_{1}^{\prime}+2m_{1}^{\prime\prime}-(8-12x_{1})m_{2}\big{]}
\frac{\vec{k}_{\bot}^{\prime}\cdot\vec{q}_{\bot}}{q^{2}}
\Bigg{)}\Bigg{)},
\label{eq:formfactorAamin}
\end{split}
\end{equation}
\end{widetext}
where
\begin{equation}
\begin{split}
h_{^{3}A}^{\prime\prime}=&(M^{\prime\prime2}-M_{0}^{\prime\prime2})\sqrt{\frac{x_{1}x_{2}}{N_{c}}}
\frac{1}{\sqrt{2}\tilde{M}_{0}^{\prime\prime}}\frac{\tilde{M}_{0}^{\prime\prime}}{2\sqrt{2}M_{0}^{\prime\prime}}\phi_{p}(x_{2},\vec{k}_{\bot}^{\prime\prime}),\\
h_{^{1}A}^{\prime\prime}=&(M^{\prime\prime2}-M_{0}^{\prime\prime2})\sqrt{\frac{x_{1}x_{2}}{N_{c}}}
\frac{1}{\sqrt{2}\tilde{M}_{0}^{\prime\prime}}\phi_{p}(x_{2},\vec{k}_{\bot}^{\prime\prime}),
\end{split}
\end{equation}
$\omega_{{}^{3}A}^{\prime\prime}=\frac{\tilde{M}_{0}^{\prime\prime2}}{m_{1}^{\prime\prime}-m_{2}}$, and $\omega_{{}^{1}A}^{\prime\prime}=2$.

In the previous theoretical works \cite{Wang:2007sxa,Wang:2008xt,Wang:2009mi,Ke:2013yka,Shi:2016gqt,Chen:2021ywv,Zhang:2023ypl}, the phenomenological Gaussian-type wave functions
\begin{equation}
\begin{split}
\phi_{s}(x_{2},\vec{k}_{\bot}^{\prime(\prime\prime)})&=4\Big{(}\frac{\pi}{\beta^{\prime(\prime\prime)2}}\Big{)}^{3/4}{\!}
\sqrt{\frac{e_{1}^{\prime(\prime\prime)}e_{2}}{x_{1}x_{2}M_{0}^{\prime(\prime\prime)}}}
\exp\Big{(}{\!}-{\!}\frac{\vec{k}_{\bot}^{\prime(\prime\prime)2}+k_{z}^{\prime(\prime\prime)2}}{2\beta^{\prime(\prime\prime)2}}\Big{)},\\
\phi_{p}(x_{2},\vec{k}_{\bot}^{\prime\prime})&=4\Big{(}\frac{\pi}{\beta^{\prime\prime2}}\Big{)}^{3/4}{\!}
\sqrt{\frac{2}{\beta^{\prime\prime2}}}
\sqrt{\frac{e_{1}^{\prime\prime}e_{2}}{x_{1}x_{2}M_{0}^{\prime\prime}}}
\exp\Big{(}{\!}-{\!}\frac{\vec{k}_{\bot}^{\prime\prime2}+k_{z}^{\prime\prime2}}{2\beta^{\prime\prime2}}\Big{)}
\label{eq:wavefunctionbeta}
\end{split}
\end{equation}
with
\begin{equation}
\begin{split}
k_{z}^{\prime(\prime\prime)}&=\frac{x_{2}M_{0}^{\prime(\prime\prime)}}{2}-\frac{m_{2}^{2}+\vec{k}_{\bot}^{\prime(\prime\prime)2}}{2x_{2}M_{0}^{\prime(\prime\prime)}},\\
e_{1}^{\prime(\prime\prime)}&=\sqrt{m_{1}^{\prime(\prime\prime)2}+\vec{k}_{\bot}^{\prime(\prime\prime)2}+k_{z}^{\prime(\prime\prime)2}},\\
e_{2}&=\sqrt{m_{2}^{2}+\vec{k}_{\bot}^{\prime2}+k_{z}^{\prime2}},
\end{split}
\end{equation}
are widely used. However, this treatment unavoidably results in the dependence of decay width on the parameter $\beta$, which is a parameter within the utilized SHO wave function. In this study, we capitalize on the knowledge acquired from the discussion on meson spectrum employing the MGI model in Section \ref{sec2} to acquire the numerical spatial wave functions of the mesons under consideration. To accomplish this, we replace the form provided in Eq.~\eqref{eq:wavefunctionbeta} with a refined expression
\begin{equation}
\begin{split}
\phi_{l}(x_{2},\vec{k}_{\bot}^{\prime(\prime\prime)})&=\sqrt{4}\pi\sum_{n=1}^{N_{\text{max}}}c_{n}\sqrt{\frac{e_{1}^{\prime(\prime\prime)}e_{2}}{x_{1}x_{2}M_{0}^{\prime(\prime\prime)}}}
R_{nl}\Big{(}\sqrt{\vec{k}_{\bot}^{\prime(\prime\prime)2}+k_{z}^{\prime(\prime\prime)2}}\Big{)},\\
\phi_{s}(x_{2},\vec{k}_{\bot}^{\prime(\prime\prime)})&\equiv\phi_{l=0}(x_{2},\vec{k}_{\bot}^{\prime(\prime\prime)}),\\
\phi_{p}(x_{2},\vec{k}_{\bot}^{\prime})&\equiv\phi_{l=1}(x_{2},\vec{k}_{\bot}^{\prime}),
\label{eq:wavefunctionSHO}
\end{split}
\end{equation}
where the expansion coefficients $c_n$ represent the values of the corresponding eigenvectors, while $l$ denotes the orbital angular momentum of the meson. By incorporating these modifications, we can effectively eliminate the associated uncertainties. To ensure proper normalization, the inclusion of the factor $\sqrt{4}\pi$ is necessary, i.e., 
\begin{equation}
\int\frac{dx_{2}d\vec{k}_{\bot}}{2(2\pi)^{3}}\phi_{l}^{*}(x_{2},\vec{k}_{\bot})\phi_{l}(x_{2},\vec{k}_{\bot})=1.
\end{equation}
Besides, $R_{nl}(\vert p \vert)$ is the SHO wave function as
\begin{equation}
R_{nl}(\vert p \vert){\!}={\!}\frac{(-1)^{n-1}}{\beta^{3/2}}{\!}\sqrt{\frac{2(n-1)!}{\Gamma(n{\!}+{\!}l{\!}+{\!}1/2)}}
\Bigg{(}\frac{p}{\beta}\Bigg{)}^{l}\exp\Bigg{(}{\!}-{\!}\frac{p^{2}}{2\beta^{2}}\Bigg{)}L_{n{\!}-{\!}1}^{l}\Bigg{(}\frac{p^{2}}{\beta^{2}}\Bigg{)}.
\label{eq:SHO}
\end{equation}
The parameter $\beta=0.5\ \text{GeV}$ used in the above equation is consistent with Section \ref{sec2}. In Eq. \eqref{eq:SHO}, we neglect the factor $(-i)^{l}$ since it does not affect the final results, but it does introduce a common factor of $i$ to the $P$-wave final state weak transition form factors, which makes the form factors less concise.

Following the approach described in Refs. \cite{Jaus:1999zv,Cheng:2003sm}, we adopt the condition $q^{+}=0$. This implies that our form factor calculations are performed in the space-like region ($q^{2}<0$), and therefore we need to extrapolate them to the time-like region ($q^{2}>0$). To perform the analytical continuations, we utilize the $z$-series parametrization \cite{Bourrely:2008za}, which has the form as \cite{Chen:2017vgi}\footnote{We employ a uniform representation, denoted as $\mathcal{F}(q^{2})$, to encompass all the relevant form factors.}
\begin{equation}
\begin{split}
\mathcal{F}(q^{2}){\!}=&{\!}\frac{1}{1{\!}-{\!}q^{2}/m_{\text{pole}}^{2}}\Big{[}a_{0}{\!}
+{\!}a_{1}\Big{(}z(q^{2}){\!}-{\!}z(0){\!}-{\!}\frac{1}{3}\big{(}z(q^{2})^{2}{\!}-{\!}z(0)^{2}\big{)}\Big{)}\\
&+a_{2}\Big{(}z(q^{2}){\!}-{\!}z(0){\!}+{\!}\frac{2}{3}\big{(}z(q^{2})^{2}{\!}-{\!}z(0)^{2}\big{)}\Big{)}\Big{]},
\end{split}
\label{eq:fitting1}
\end{equation}
where $a_{0}$, $a_{1}$, and $a_{2}$ are free parameters needed to be fitted in $q^{2}<0$ region, and $z(q^{2})$ is written as \cite{Bourrely:2008za,Khodjamirian:2010vf,Cheng:2017smj}
\begin{equation}
z(q^{2})=\frac{\sqrt{t_{+}-q^{2}}-\sqrt{t_{+}-t_{0}}}{\sqrt{t_{+}-q^{2}}+\sqrt{t_{+}-t_{0}}}
\end{equation}
with $t_{0}=t_{+}\Big{(}1-\sqrt{1-t_{-}/t_{+}}\Big{)}$ and $t_{\pm}=(m_{B_{c}}\pm m_{f})^{2}$. $m_{B_{c}}$ and $m_{f}$ are the masses of the $B_{c}$ meson and daughter meson, respectively. This parametrization are more convenient to reflect the character $F_{1}^{B_{c}P(S)}(q^{2}=0)=F_{0}^{B_{c}P(S)}(q^{2}=0)$. Moreover, we have $\mathcal{F}(0)=a_{0}$.

\begin{figure}[htbp]\centering
  \begin{tabular}{cc}
  \includegraphics[width=42mm]{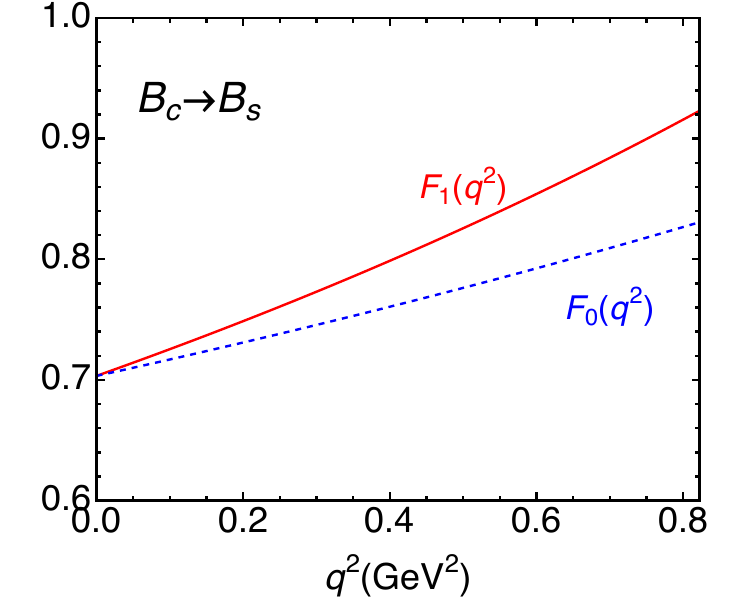}
  \includegraphics[width=42mm]{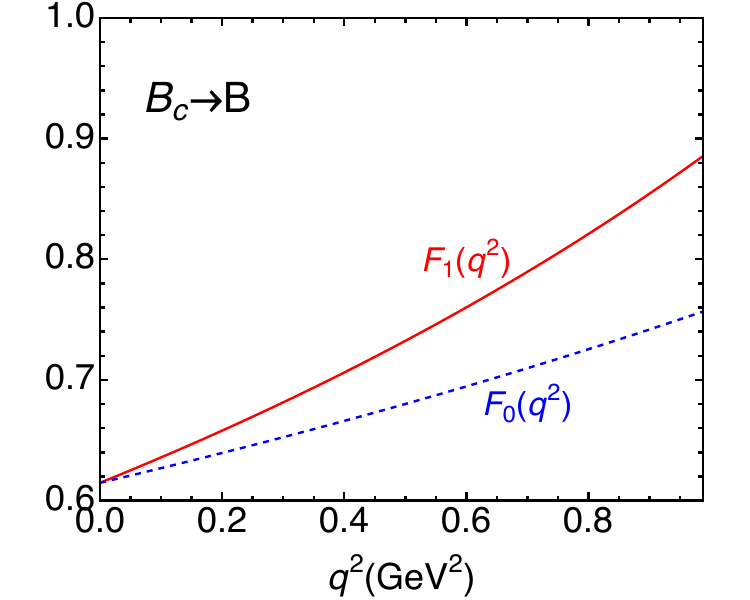}\\
  \includegraphics[width=42mm]{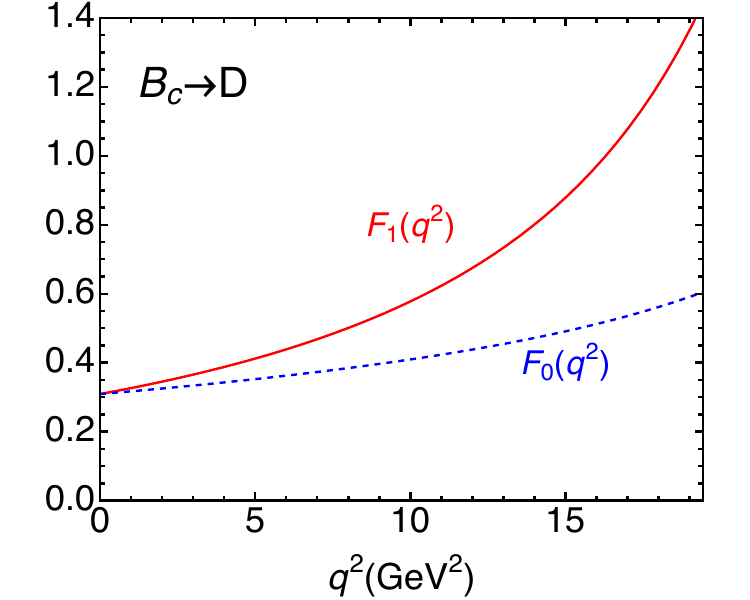}
  \includegraphics[width=42mm]{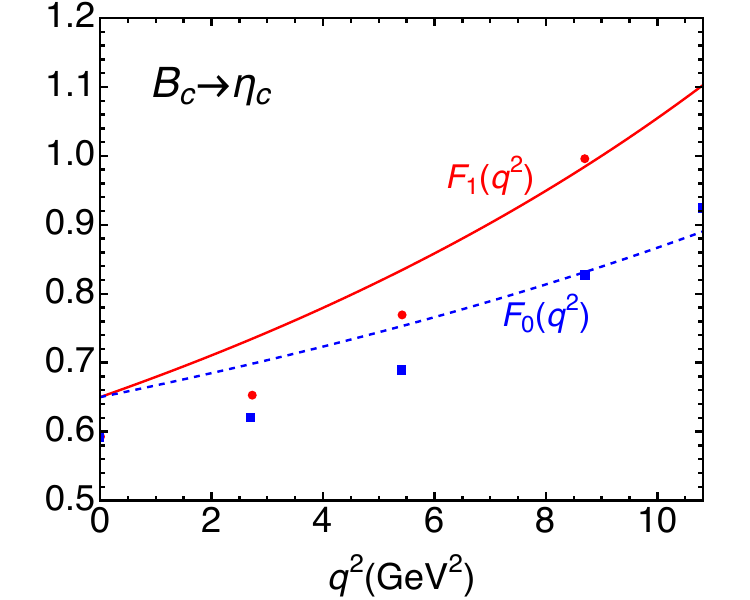}
  \end{tabular}
  \caption{The $q^{2}$ dependence of the weak transition form factors of the $B_{c}\to B_{s},B,D,\eta_{c}$ processes. In the $B_c \to \eta_c$ transition, the form factors $F_1$ and $F_0$ are also depicted as the results from LQCD, represented by a red circle and a blue square, respectively, as shown in Figure of Ref.  \cite{Lytle:2016ixw}.}
\label{fig:ffsP}
\end{figure}

\begin{figure}[htbp]\centering
  \begin{tabular}{lc}
  \includegraphics[width=42mm]{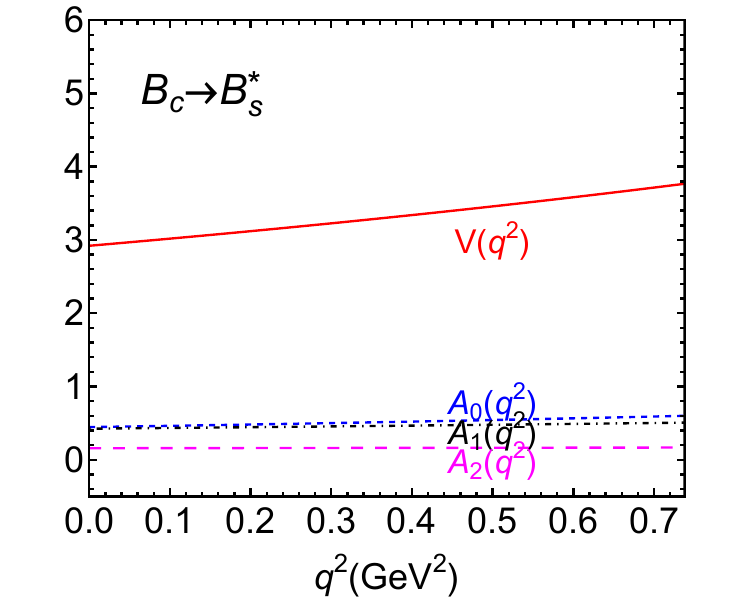}
  \includegraphics[width=42mm]{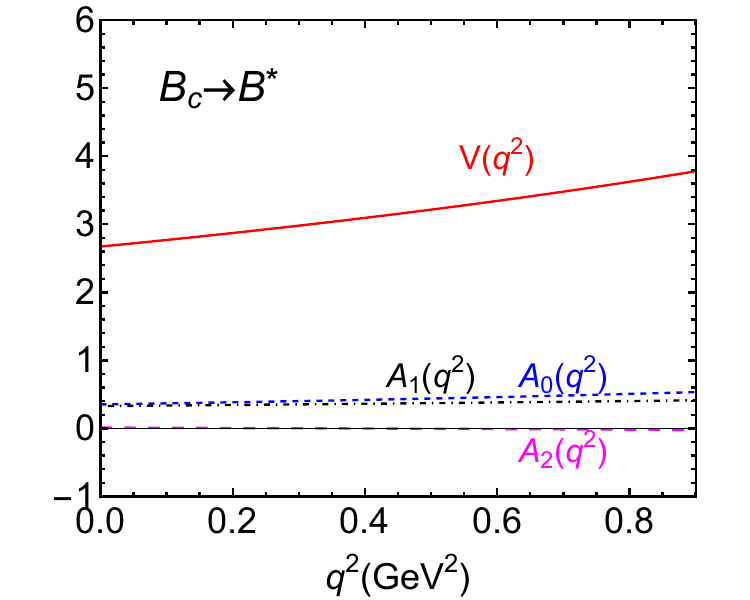}\\
  \includegraphics[width=42mm]{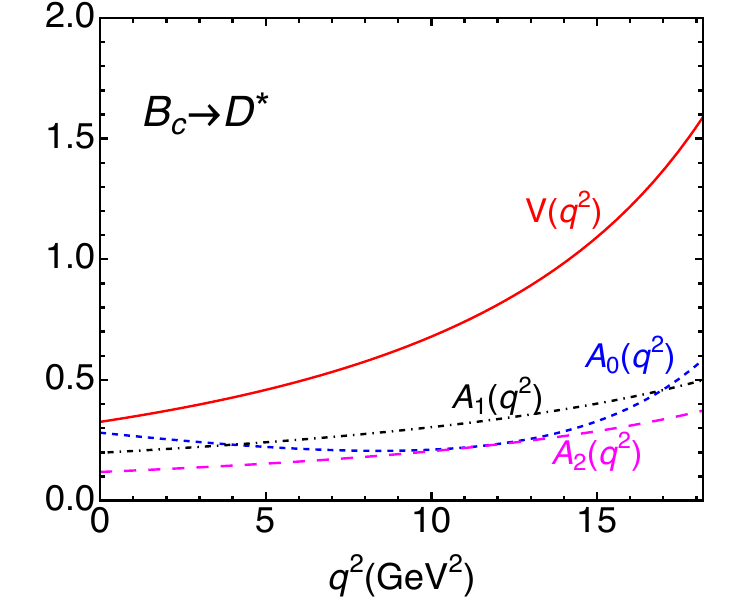}
  \includegraphics[width=42mm]{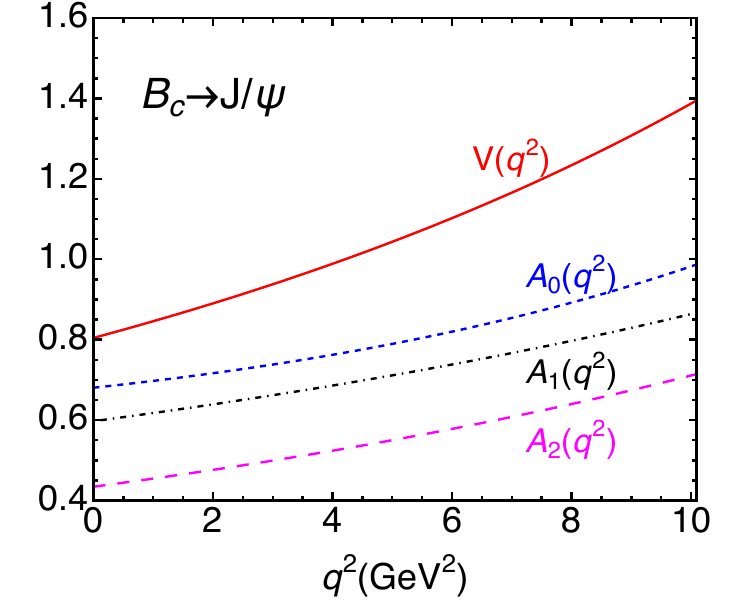}\\
  \includegraphics[width=42mm]{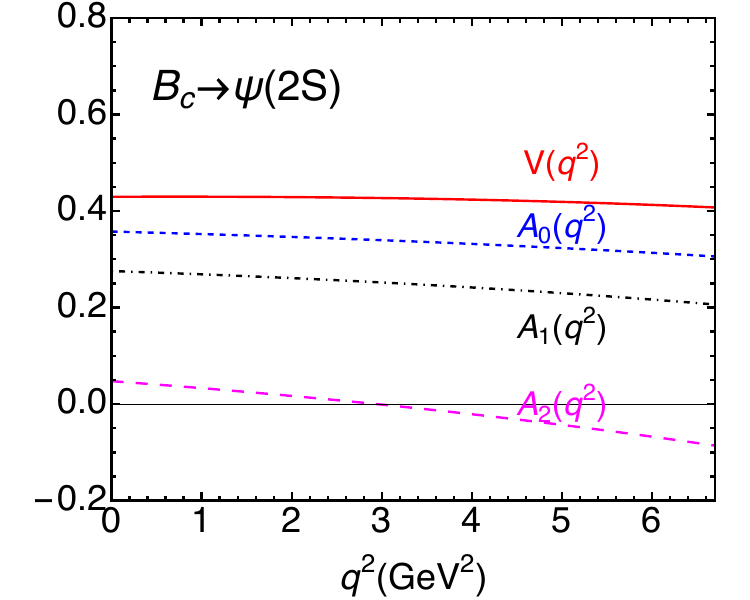}\\
  \end{tabular}
  \caption{The $q^{2}$ dependence of the weak transition form factors of the $B_{c}\to B_{s}^{*},B^{*},D^{*},J/\psi,\psi(2S)$ processes.}
\label{fig:ffsV}
\end{figure}

\begin{figure}[htbp]\centering
  \begin{tabular}{cc}
  \includegraphics[width=42mm]{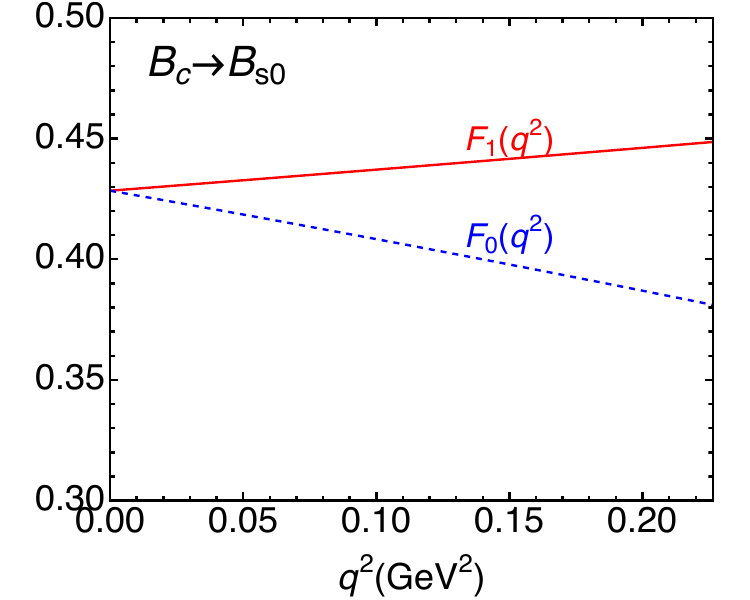}
  \includegraphics[width=42mm]{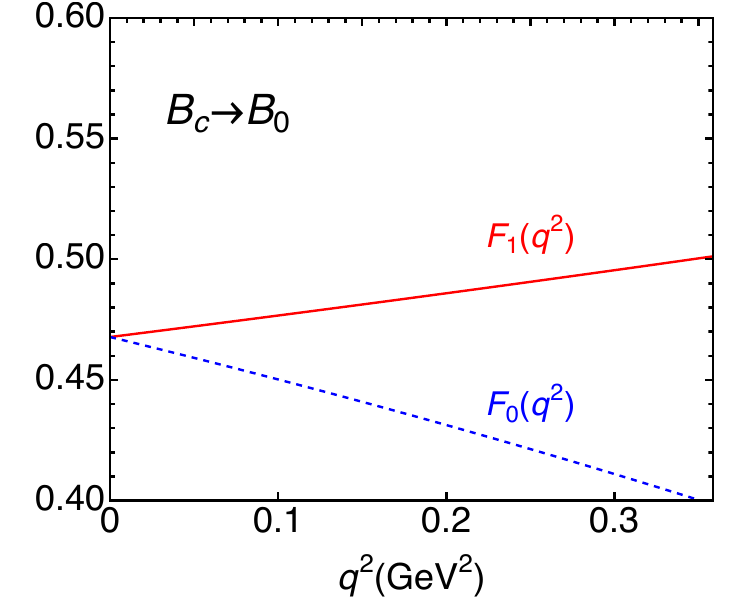}\\
  \includegraphics[width=42mm]{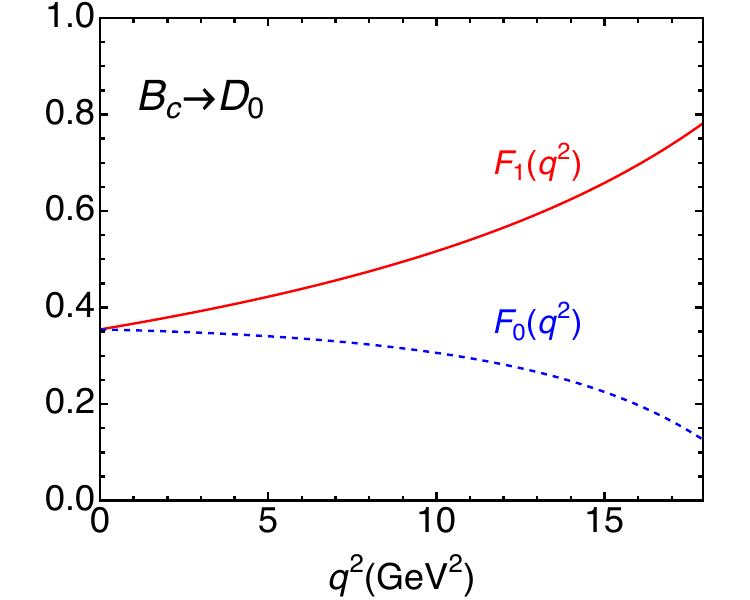}
  \includegraphics[width=42mm]{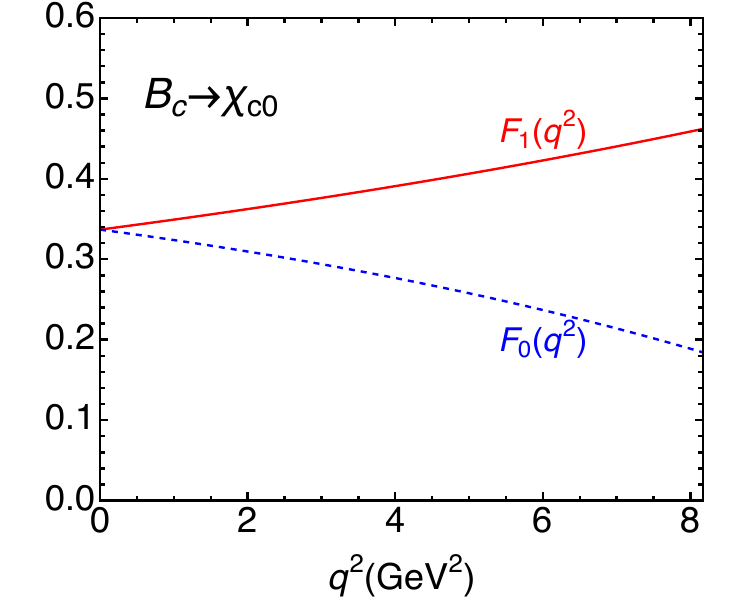}
  \end{tabular}
  \caption{The $q^{2}$ dependence of the weak transition form factors of the $B_{c}\to B_{s0},B_{0},D_{0},\chi_{c0}$ processes.}
\label{fig:ffsS}
\end{figure}

\begin{figure*}[htbp]\centering
  \begin{tabular}{lccr}
  \includegraphics[width=42mm]{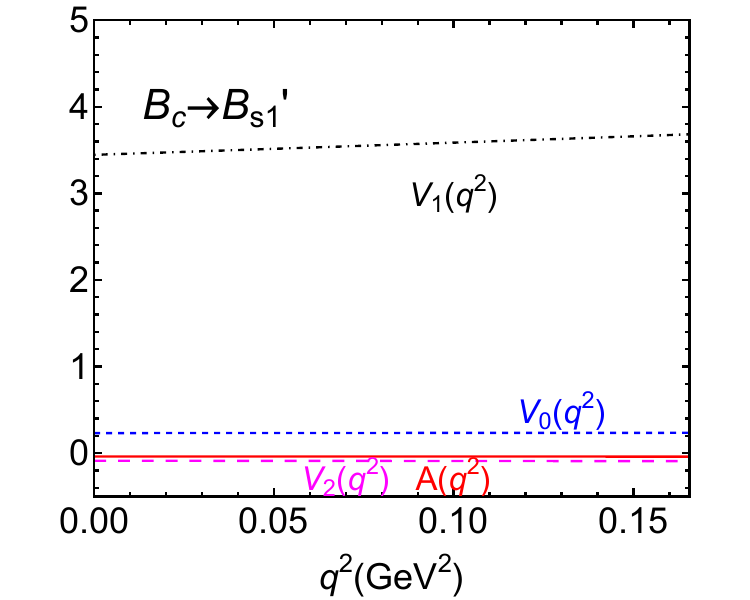}
  \includegraphics[width=42mm]{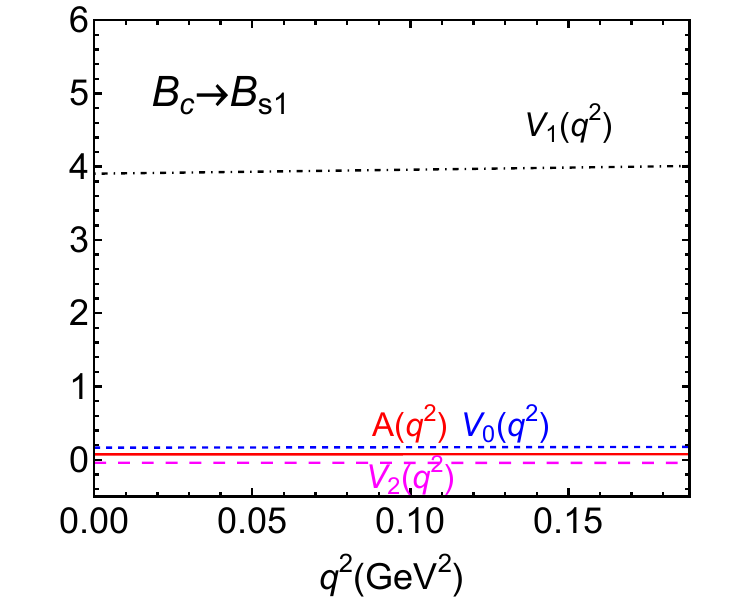}
  \includegraphics[width=42mm]{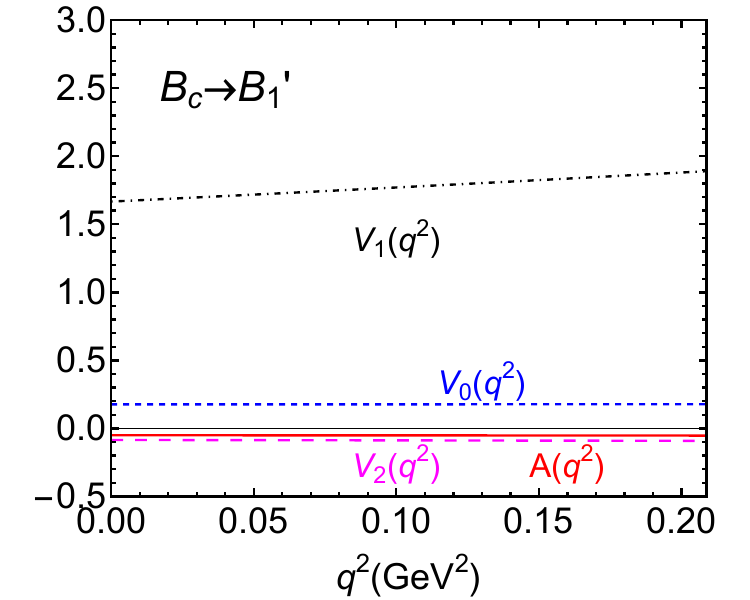}
  \includegraphics[width=42mm]{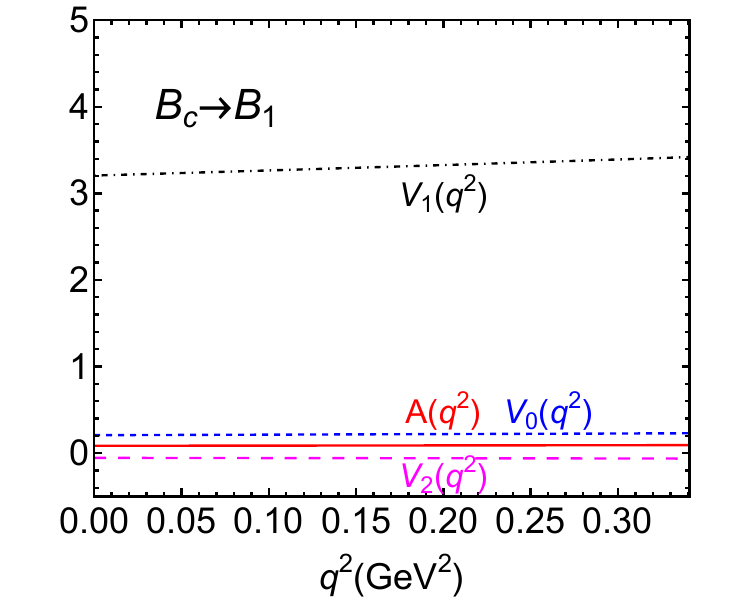}\\
  \includegraphics[width=42mm]{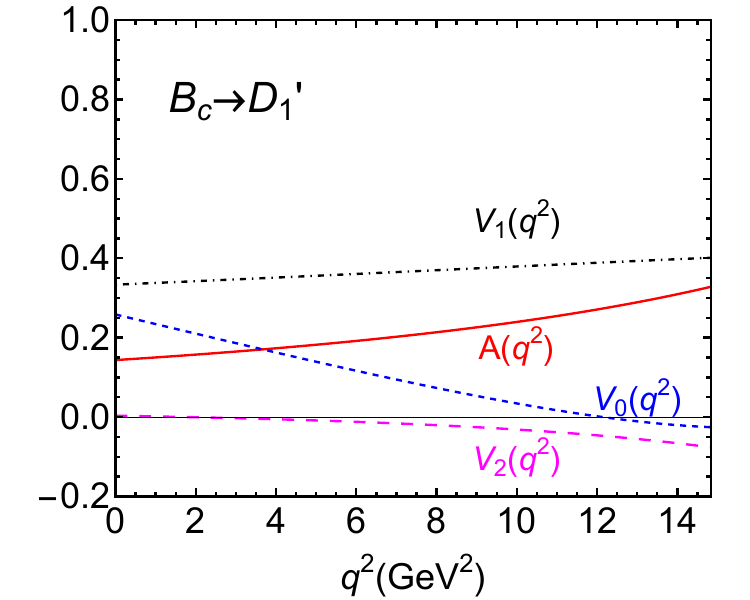}
  \includegraphics[width=42mm]{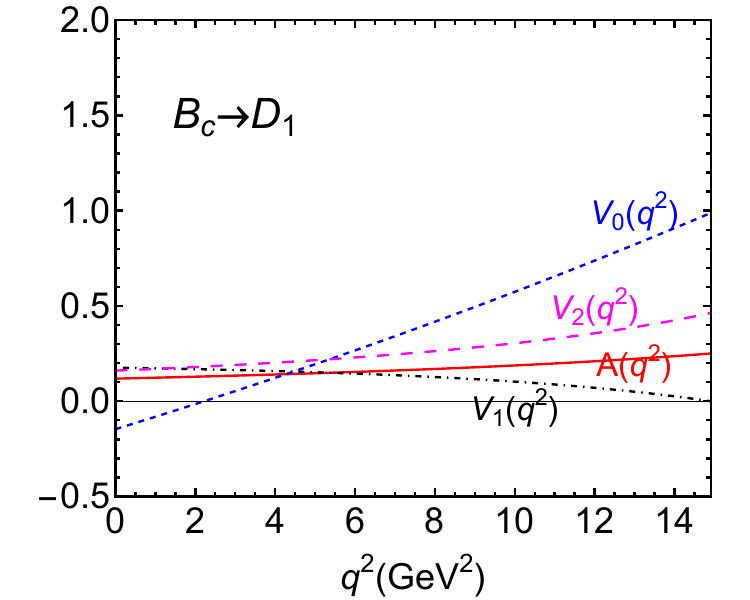}
  \includegraphics[width=42mm]{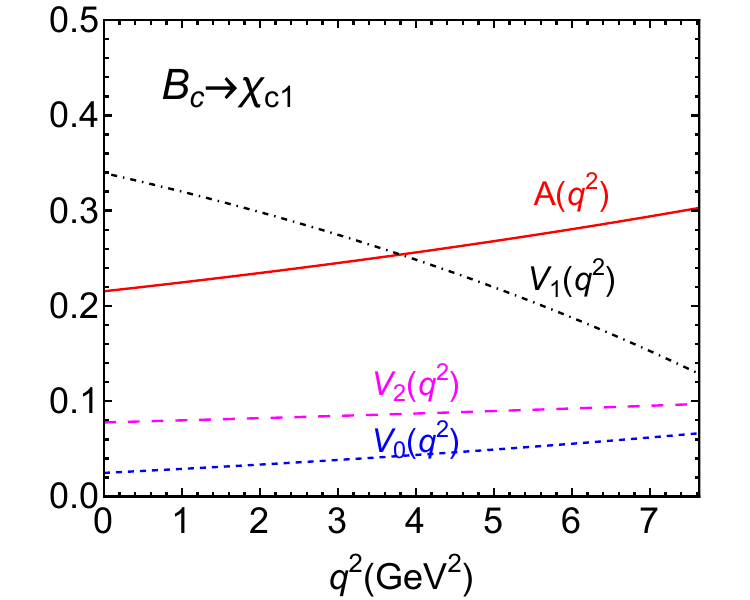}
  \includegraphics[width=42mm]{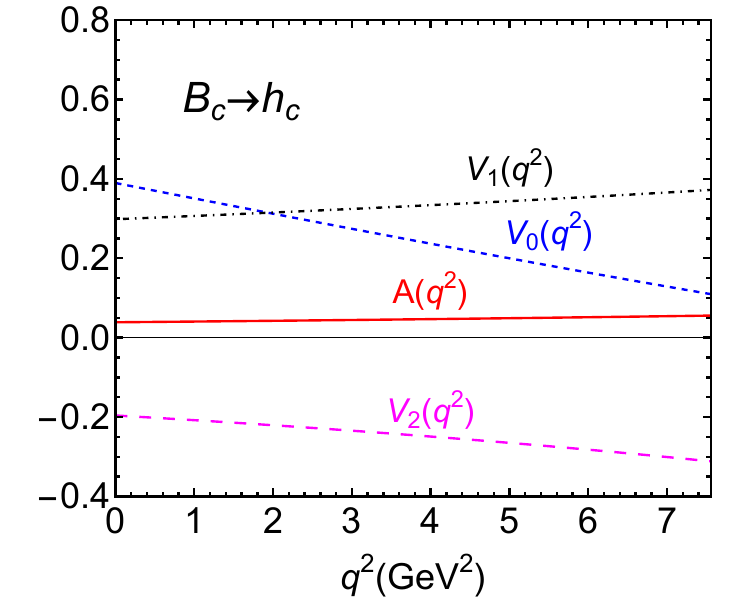}
  \end{tabular}
  \caption{The $q^{2}$ dependence of the weak transition form factors of the $B_{c}\to B_{s1}^{(\prime)},B_{1}^{(\prime)},D_{1}^{(\prime)},\chi_{c1},h_{c}$ processes.}
\label{fig:ffsA}
\end{figure*}

To determine the values of the free parameters $a_{0}$, $a_{1}$, and $a_{2}$\footnote{For $B_{c}\to P(S)$ transitions, only the parameters $a_{1}$ and $a_{2}$ need to be fitted, as $a_{0}$ can be obtained from $\mathcal{F}(0)$ using Eq. \eqref{eq:fitting1}, and $\mathcal{F}(0)$ can be accurately calculated using Eq. \eqref{eq:formfactorPF1} and Eq. \eqref{eq:formfactorSF1}.}, as stated in Eq. \eqref{eq:fitting1}, we perform numerical calculations at 200 equally spaced points for each form factor, ranging from $-20\ \text{GeV}^{2}$ to $-0.1\ \text{GeV}^{2}$, utilizing Eqs. \eqref{eq:formfactorPF1}-\eqref{eq:formfactorAamin}. Subsequently, we fit the calculated points using Eq. \eqref{eq:fitting1}. The fitted values of the free parameters, as well as $\mathcal{F}(0)$, $\mathcal{F}(q_{\text{max}}^{2})$, and the pole masses $m_{\text{pole}}$, are compiled in Table~\ref{tab:ffsP} (\ref{tab:ffsV}, \ref{tab:ffsS}, \ref{tab:ffsA}) for $B_{c}\to P(V,S,A)$ transitions. Additionally, the $q^{2}$ dependence of the $B_{c}\to P(V,S,A)$ transition form factors is illustrated in Fig.~\ref{fig:ffsP} (\ref{fig:ffsV}, \ref{fig:ffsS}, \ref{fig:ffsA}). Particularly, in the $B_c \to \eta_c$ panel, we also present the LQCD's result \cite{Lytle:2016ixw}. Our result is slightly larger than the LQCD in low $q^{2}$ region, but is anastomotic in large $q^{2}$ region.

\begin{table}[htbp]\centering
\caption{The form factors of the $B_{c}\to B_{s},B,D,\eta_{c}$ transitions in CLFQM.}
\label{tab:ffsP}
\renewcommand\arraystretch{1.2}
\begin{tabular*}{86mm}{c@{\extracolsep{\fill}}ccccc}
\toprule[1pt]
\toprule[0.5pt]
                        &$\mathcal{F}(0)=a_{0}$   &$\mathcal{F}(q_{\text{max}}^{2})$   &$m_{\text{pole}}$ (GeV)   &$a_{1}$  &$a_{2}$\\
\midrule[0.5pt]
$F_1^{B_c\to B_s}$      &$0.703$       &$0.923$       &$2.112$        &$-32.550$     &$515.53$  \\
$F_0^{B_c\to B_s}$      &$0.703$       &$0.831$       &$2.460$        &$-9.520$     &$-13.871$   \\
\specialrule{0em}{2pt}{2pt}
$F_1^{B_c\to B}$        &$0.615$       &$0.886$       &$2.007$        &$-28.916$     &$447.924$\\
$F_0^{B_c\to B}$        &$0.615$       &$0.757$       &$2.412$        &$-7.279$      &$-69.149$\\
\specialrule{0em}{2pt}{2pt}
$F_1^{B_c\to D}$        &$0.309$       &$1.446$       &$5.325$        &$-1.682$      &$2.017$\\
$F_0^{B_c\to D}$        &$0.309$       &$0.605$       &$5.675$        &$0.802$       &$-3.070$\\
\specialrule{0em}{2pt}{2pt}
$F_1^{B_c\to \eta_c}$   &$0.650$       &$1.103$       &$6.336$        &$-4.612$      &$12.277$\\
$F_0^{B_c\to \eta_c}$   &$0.650$       &$0.891$       &$6.706$        &$-0.777$      &$1.108$\\
\bottomrule[0.5pt]
\bottomrule[1pt]
\end{tabular*}
\end{table}

\begin{table}[htbp]\centering
\caption{The form factors of the $B_{c}\to B_{s}^{*},B^{*},D^{*},J/\psi,\psi(2S)$ transitions in CLFQM.}
\label{tab:ffsV}
\renewcommand\arraystretch{1.2}
\begin{tabular*}{86mm}{c@{\extracolsep{\fill}}ccccc}
\toprule[1pt]
\toprule[0.5pt]
                        &$\mathcal{F}(0)=a_{0}$   &$\mathcal{F}(q_{\text{max}}^{2})$   &$m_{\text{pole}}$   &$a_{1}$   &$a_{2}$\\
\midrule[0.5pt]
$V^{B_c\to B_s^{*}}$       &$2.921$       &$3.767$       &$2.112$        &$-164.601$    &$2805.64$\\
$A_0^{B_c\to B_s^{*}}$     &$0.447$       &$0.601$       &$1.968$        &$-29.193$    &$615.132$\\
$A_1^{B_c\to B_s^{*}}$     &$0.425$       &$0.507$       &$2.535$        &$-17.847$    &$278.218$\\
$A_2^{B_c\to B_s^{*}}$     &$0.158$       &$0.168$       &$2.535$        &$7.108$    &$-254.943$\\
\specialrule{0em}{3pt}{3pt}
$V^{B_c\to B^{*}}$         &$2.672$       &$3.780$       &$2.007$        &$-156.204$    &$2657.84$\\
$A_0^{B_c\to B^{*}}$       &$0.353$       &$0.534$       &$1.865$        &$-25.298$    &$550.667$\\
$A_1^{B_c\to B^{*}}$       &$0.328$       &$0.413$       &$2.422$        &$-12.978$    &$188.302$\\
$A_2^{B_c\to B^{*}}$       &$0.012$       &$-0.022$       &$2.422$        &$18.505$    &$-479.418$\\
\specialrule{0em}{3pt}{3pt}
$V^{B_c\to D^{*}}$         &$0.326$       &$1.592$       &$5.325$        &$-3.167$    &$7.189$\\
$A_0^{B_c\to D^{*}}$       &$0.281$       &$0.585$       &$5.280$        &$0.944$    &$71.415$\\
$A_1^{B_c\to D^{*}}$       &$0.198$       &$0.499$       &$5.726$        &$-0.312$    &$-0.704$\\
$A_2^{B_c\to D^{*}}$       &$0.119$       &$0.373$       &$5.726$        &$-0.615$    &$0.440$\\
\specialrule{0em}{3pt}{3pt}
$V^{B_c\to J/\psi}$        &$0.804$       &$1.395$       &$6.336$        &$-7.864$    &$26.235$\\
$A_0^{B_c\to J/\psi}$      &$0.681$       &$0.987$       &$6.274$        &$-1.761$    &$72.843$\\
$A_1^{B_c\to J/\psi}$      &$0.597$       &$0.867$       &$6.749$        &$-2.543$    &$5.721$\\
$A_2^{B_c\to J/\psi}$      &$0.434$       &$0.714$       &$6.749$        &$-3.994$    &$13.044$\\
\specialrule{0em}{3pt}{3pt}
$V^{B_c\to \psi(2S)}$        &$0.429$      &$0.407$       &$6.336$       &$5.159$    &$-67.300$\\
$A_0^{B_c\to \psi(2S)}$      &$0.357$       &$0.306$       &$6.274$        &$5.929$    &$-23.284$\\
$A_1^{B_c\to \psi(2S)}$      &$0.276$       &$0.206$       &$6.749$        &$5.704$    &$-46.34$\\
$A_2^{B_c\to \psi(2S)}$      &$0.047$       &$-0.086$       &$6.749$        &$6.922$    &$-61.037$\\
\bottomrule[0.5pt]
\bottomrule[1pt]
\end{tabular*}
\end{table}

\begin{table}[htbp]\centering
\caption{The form factors of the $B_{c}\to B_{s0},B_{0},D_{0}^{*},\chi_{c0}$ transitions in CLFQM.}
\label{tab:ffsS}
\renewcommand\arraystretch{1.2}
\begin{tabular*}{86mm}{c@{\extracolsep{\fill}}ccccc}
\toprule[1pt]
\toprule[0.5pt]
                           &$\mathcal{F}(0)=a_{0}$   &$\mathcal{F}(q_{\text{max}}^{2})$   &$m_{\text{pole}}$   &$a_{1}$    &$a_{2}$\\
\midrule[0.5pt]
$F_1^{B_c\to B_{s0}}$      &$0.428$       &$0.449$       &$2.535$        &$-11.305$    &$126.279$   \\
$F_0^{B_c\to B_{s0}}$      &$0.428$       &$0.381$       &$1.968$        &$179.073$   &$-3346.19$   \\
\specialrule{0em}{2pt}{2pt}
$F_1^{B_c\to B_0}$         &$0.468$       &$0.501$       &$2.422$        &$-4.449$    &$-65.925$   \\
$F_0^{B_c\to B_0}$         &$0.468$       &$0.398$       &$1.865$        &$175.706$   &$-2991.27$   \\
\specialrule{0em}{2pt}{2pt}
$F_1^{B_c\to D_0}$         &$0.355$       &$0.782$       &$5.726$        &$0.007$    &$-3.225$   \\
$F_0^{B_c\to D_0}$         &$0.355$       &$0.126$      &$5.280$         &$4.130$      &$-2.009$    \\
\specialrule{0em}{2pt}{2pt}
$F_1^{B_c\to \chi_{c0}}$   &$0.337$       &$0.462$       &$6.749$        &$-1.861$     &$4.133$   \\
$F_0^{B_c\to \chi_{c0}}$   &$0.337$       &$0.184$       &$6.274$        &$8.389$      &$-23.161$   \\
\bottomrule[0.5pt]
\bottomrule[1pt]
\end{tabular*}
\end{table}

\begin{table}[htbp]\centering
\caption{The form factors of the $B_{c}\to B_{s1}^{(\prime)},B_{1}^{(\prime)},D_{1}^{(\prime)},h_{c},\chi_{c1}$ transitions in CLFQM.}
\label{tab:ffsA}
\renewcommand\arraystretch{1.2}
\begin{tabular*}{86mm}{c@{\extracolsep{\fill}}ccccc}
\toprule[1pt]
\toprule[0.5pt]
                        &$\mathcal{F}(0)=a_{0}$   &$\mathcal{F}(q_{\text{max}}^{2})$   &$m_{\text{pole}}$   &$a_{1}$   &$a_{2}$\\
\midrule[0.5pt]
$A^{B_c\to B_{s1}^{\prime}}$       &$-0.038$      &$-0.039$      &$2.535$      &$1.605$    &$-24.722$\\
$V_0^{B_c\to B_{s1}^{\prime}}$     &$0.232$       &$0.235$       &$2.460$      &$11.187$     &$239.345$\\
$V_1^{B_c\to B_{s1}^{\prime}}$     &$3.444$       &$3.682$       &$2.112$      &$-357.075$     &$6870.02$\\
$V_2^{B_c\to B_{s1}^{\prime}}$     &$-0.087$      &$-0.092$      &$2.112$      &$5.528$    &$-107.399$\\
\specialrule{0em}{3pt}{3pt}
$A^{B_c\to B_{s1}}$                &$0.074$       &$0.077$      &$2.535$     &$-3.468$    &$57.551$\\
$V_0^{B_c\to B_{s1}}$              &$0.166$       &$0.176$      &$2.460$     &$-14.452$    &$378.227$\\
$V_1^{B_c\to B_{s1}}$              &$3.904$       &$4.010$      &$2.112$     &$199.722$    &$-5484.34$\\
$V_2^{B_c\to B_{s1}}$              &$-0.039$      &$-0.042$     &$2.112$     &$3.266$   &$-66.969$\\
\specialrule{0em}{3pt}{3pt}
$A^{B_c\to B_{1}^{\prime}}$         &$-0.050$      &$-0.052$     &$2.422$    &$1.640$    &$-19.246$\\
$V_0^{B_c\to B_{1}^{\prime}}$       &$0.177$       &$0.178$      &$2.412$    &$15.181$     &$150.109$\\
$V_1^{B_c\to B_{1}^{\prime}}$       &$1.667$       &$1.890$      &$2.007$    &$-351.092$     &$7009.86$\\
$V_2^{B_c\to B_{1}^{\prime}}$       &$-0.085$      &$-0.091$     &$2.007$    &$5.153$    &$-98.674$\\
\specialrule{0em}{3pt}{3pt}
$A^{B_c\to B_{1}}$                  &$0.085$       &$0.093$      &$2.422$    &$-3.474$    &$50.953$\\
$V_0^{B_c\to B_{1}}$                &$0.208$       &$0.230$      &$2.412$    &$-14.503$    &$390.001$\\
$V_1^{B_c\to B_{1}}$                &$3.205$       &$3.419$      &$2.007$    &$127.525$    &$-3507.11$\\
$V_2^{B_c\to B_{1}}$                &$-0.054$      &$-0.062$     &$2.007$    &$4.546$   &$-91.094$\\
\specialrule{0em}{3pt}{3pt}
$A^{B_c\to D_{1}^{\prime}}$         &$0.144$      &$0.328$       &$5.726$      &$-0.662$     &$0.696$\\
$V_0^{B_c\to D_{1}^{\prime}}$       &$0.258$      &$-0.025$       &$5.675$      &$4.944$     &$85.952$\\
$V_1^{B_c\to D_{1}^{\prime}}$       &$0.334$      &$0.401$       &$5.325$      &$2.608$     &$-6.499$\\
$V_2^{B_c\to D_{1}^{\prime}}$       &$0.003$      &$-0.076$      &$5.325$      &$0.732$     &$-3.113$\\
\specialrule{0em}{3pt}{3pt}
$A^{B_c\to D_{1}}$                  &$0.118$       &$0.250$       &$5.726$      &$-0.329$     &$-0.344$\\
$V_0^{B_c\to D_{1}}$                &$-0.147$      &$0.988$      &$5.675$      &$-12.231$    &$-152.47$\\
$V_1^{B_c\to D_{1}}$                &$0.176$       &$-0.0004$       &$5.325$      &$3.202$     &$-6.365$\\
$V_2^{B_c\to D_{1}}$                &$0.160$       &$0.463$       &$5.325$      &$-1.074$     &$2.322$\\
\specialrule{0em}{3pt}{3pt}
$A^{B_c\to\chi_{c1}}$               &$0.215$       &$0.303$       &$6.749$        &$-1.771$     &$5.620$\\
$V_0^{B_c\to\chi_{c1}}$             &$0.025$       &$0.066$       &$6.706$        &$-1.462$     &$5.819$\\
$V_1^{B_c\to\chi_{c1}}$             &$0.339$       &$0.129$       &$6.336$        &$11.333$     &$-47.961$\\
$V_2^{B_c\to\chi_{c1}}$             &$0.078$       &$0.097$       &$6.336$        &$-0.044$     &$-1.797$\\
\specialrule{0em}{3pt}{3pt}
$A^{B_c\to h_{c}}$                  &$0.039$       &$0.056$       &$6.749$        &$-0.372$     &$1.332$\\
$V_0^{B_c\to h_{c}}$                &$0.390$       &$0.110$       &$6.706$        &$14.548$     &$181.296$\\
$V_1^{B_c\to h_{c}}$                &$0.298$       &$0.372$       &$6.336$        &$-0.185$     &$-2.383$\\
$V_2^{B_c\to h_{c}}$                &$-0.196$       &$-0.312$       &$6.336$        &$2.783$    &$-12.591$\\
\bottomrule[0.5pt]
\bottomrule[1pt]
\end{tabular*}
\end{table}


With the obtained weak transition form factors, we can future investigate the corresponding semileptonic decays. For the $B_{c}\to M\ell\nu_{\ell}$ processes, the differential decay width can be obtained by
\begin{equation}
\begin{split}
\frac{d^{2}\Gamma}{dq^{2}d\cos\theta_{\ell}}=&\frac{G_{F}^{2}V_{\text{CKM}}^{2}}{512\pi^{3}}
\frac{\sqrt{\lambda(m_{B_{c}}^{2},m_{M}^{2},q^{2})}(1-\hat{m}_{\ell}^{2})}{2m_{B_{c}}^{3}}\\
&\times\Big{(}L_{1}+L_{2}\cos\theta_{\ell}+L_{3}\cos2\theta_{\ell}\Big{)},
\end{split}
\end{equation}
where $G_{F}=1.16637\times10^{-5}\text{GeV}^{-2}$ is the Fermi coupling constant, $V_{\text{CKM}}$ is the Cabibbo-Kobayashi-Maskawa (CKM) matrix element, $\lambda(x,y,z)=x^{2}+y^{2}+z^{2}-2(xy+xz+yz)$ is the K\"{a}llen function, $m_{M}$ is the mass of the daughter meson, and $\hat{m}_{\ell}^{2}=m_{\ell}^{2}/q^{2}$ with $m_{\ell}$ being the lepton mass. The angular coefficients $L_{1,2,3}$ for $M=P(S)$ are
\begin{equation}
\begin{split}
L_{1}=&2(1{\!}-{\!}\hat{m}_{\ell}^{2})\Big{(}2F_{0}^{2}(m_{B_{c}}^2{\!}-{\!}m_{P(S)}^{2})^{2}\hat{m}_{\ell}^{2}\\
&+F_{1}^{2}\lambda(m_{B_{c}}^{2},m_{P(S)}^{2},q^{2})(1+\hat{m}_{\ell}^{2})\Big{)},
\end{split}
\end{equation}

\begin{equation}
L_{2}=-8F_{0}F_{1}\hat{m}_{\ell}^{2}(m_{B_{c}}^2-m_{P(S)}^{2})(1-\hat{m}_{\ell}^{2})\sqrt{\lambda(m_{B_{c}}^{2},m_{P(S)}^{2},q^{2})},\\
\end{equation}

\begin{equation}
L_{3}=-2F_{1}^{2}(1-\hat{m}_{\ell}^{2})^{2}\lambda(m_{B_{c}}^{2},m_{P(S)}^{2},q^{2}),
\end{equation}
and for $M=V(A)$ are
\begin{equation}
\begin{split}
L_{1}=&(1-\hat{m}_{\ell}^{2})\Bigg{(}4A(V)_{0}^{2}\lambda(m_{B_{c}}^{2},m_{V(A)}^{2},q^{2})\hat{m}_{\ell}^{2}+\frac{1+\hat{m}_{\ell}^{2}}{2m_{V(A)}^{2}}\\
&\times\Big{[}A(V)_{1}(m_{B_{c}}\pm m_{V(A)})(m_{B_{c}}^{2}-m_{V(A)}^{2}-q^{2})\\
&-A(V)_{2}\frac{\lambda(m_{B_{c}}^{2},m_{V(A)}^{2},q^{2})}{m_{B_{c}}\pm m_{V(A)}}\Big{]}^{2}\Bigg{)}\\
&+\frac{2(1-\hat{m}_{\ell}^{2})}{(m_{B_{c}}\pm m_{V(A)})^{2}}(3+\hat{m}_{\ell}^{2})q^{2}\big{[}A(V)_{1}^{2}(m_{B_{c}}\pm m_{V(A)})^{4}\\
&+V(A)^{2}\lambda(m_{B_{c}}^{2},m_{V(A)}^{2},q^{2})\big{]},
\label{eq:Bc2VK1}
\end{split}
\end{equation}
\begin{equation}
\begin{split}
L_{2}=&\frac{4A(V)_{0}\hat{m}_{\ell}^{2}(1-\hat{m}_{\ell}^{2})}{m_{V(A)}(m_{B_{c}}\pm m_{V(A)})}\sqrt{\lambda(m_{B_{c}}^{2},m_{V(A)}^{2},q^{2})}\\
&\times\Big{[}A(V)_{2}\lambda(m_{B_{c}}^{2},m_{V(A)}^{2},q^{2})\\
&-A(V)_{1}(m_{B_{c}}\pm m_{V(A)})^{2}(m_{B_{c}}^{2}-m_{V(A)}^{2}-q^{2})\Big{]}\\
&\pm 16A(V)_{1}V(1-\hat{m}_{\ell}^{2})q^{2}\sqrt{\lambda(m_{B_{c}}^{2},m_{V(A)}^{2},q^{2})},
\label{eq:Bc2VK2}
\end{split}
\end{equation}
\begin{equation}
\begin{split}
L_{3}=&-\frac{(1-\hat{m}_{\ell}^{2})^{2}}{2m_{V(A)}^{2}(m_{B_{c}}\pm m_{V(A)})^{2}}\Big{[}A(V)_{1}(m_{B_{c}}\pm m_{V(A)})^{2}\\
&\times(m_{B_{c}}^{2}-m_{V(A)}^{2}-q^{2})-A(V)_{2}\lambda(m_{B_{c}}^{2},m_{V(A)}^{2},q^{2})\Big{]}^{2}\\
&+\frac{2(1-\hat{m}_{\ell}^{2})^{2}q^{2}}{(m_{B_{c}}\pm m_{V(A)})^{2}}\Big{[}A(V)_{1}^{2}(m_{B_{c}}\pm m_{V(A)})^{4}\\
&+V(A)^{2}\lambda(m_{B_{c}}^{2},m_{V(A)}^{2},q^{2})\Big{]}.
\label{eq:Bc2VK3}
\end{split}
\end{equation}

After performing the integral of the angle $\theta_{\ell}$, the differential decay width can be obtained by
\begin{equation}
\begin{split}
\frac{d\Gamma}{dq^2}=&\frac{G_{F}^{2}V_{\text{CKM}}^{2}}{512\pi^{3}}
\frac{\sqrt{\lambda(m_{B_{c}}^{2},m_{M}^{2},q^{2})}(1-\hat{m}_{\ell}^{2})}{2m_{B_{c}}^{3}}
\Big{(}2L_{1}-\frac{2}{3}L_{3}\Big{)}.
\end{split}
\end{equation}
And then, the decay width can be obtained by carrying out the integral of $q^{2}$ in the range of $m_{\ell}^{2}$ to $q^{2}_{\text{max}}$.

In addition, the differential branching decay widths from the longitudinal and transverse polarizations for $B_{c}\to V(A)\ell\nu_{\ell}$ are
\begin{equation}
\begin{split}
\frac{d\Gamma_{L}}{dq^{2}}=&\frac{G_{F}^{2}V_{\text{CKM}}^{2}}{192\pi^{3}m_{B_{c}}^{3}}\frac{\sqrt{\lambda(m_{B_{c}}^{2},m_{V(A)}^{2},q^{2})}}{2}(1-\hat{m}_{\ell}^{2})^{2}\\
&\times\Bigg{\{}3\hat{m}_{\ell}^{2}\lambda(m_{B_{c}}^{2},m_{V(A)}^{2},q^{2})A(V)_{0}^{2}
+\frac{2+\hat{m}_{\ell}^{2}}{4m_{V(A)}^{2}}\\
&\times\Big{[}A(V)_{1}(m_{B_{c}}\pm m_{V(A)})(m_{B_{c}}^{2}-m_{V(A)}^{2}-q^{2})\\
&-A(V)_{2}\frac{\lambda(m_{B_{c}}^{2},m_{V(A)}^{2},q^{2})}{m_{B_{c}}\pm m_{V(A)}}\Big{]}^{2}
\Bigg{\}},
\end{split}
\end{equation}
\begin{equation}
\begin{split}
\frac{d\Gamma_{T}}{dq^{2}}=&\frac{G_{F}^{2}V_{\text{CKM}}^{2}}{192\pi^{3}m_{B_{c}}^{3}}\lambda^{3/2}(m_{B_{c}}^{2},m_{V(A)}^{2},q^{2})
(1-\hat{m}_{\ell}^{2})^{2}(2+\hat{m}_{\ell}^{2})q^{2}\\
&\times\Big{[}\frac{A(V)_{1}^{2}(m_{B_{c}}\pm m_{V(A)})^{2}}{\lambda(m_{B_{c}}^{2},m_{V(A)}^{2},q^{2})}
+\frac{V(A)^{2}}{(m_{B_{c}}\pm m_{V(A)})^{2}}\Big{]}.
\end{split}
\end{equation}

Using the obtained form factors as inputs, we further calculate the corresponding branching ratios and the $\Gamma_{L}/\Gamma_{T}$ ratios. The obtained results are presented in Tables \ref{tab:semileptonP}, \ref{tab:semileptonV}, \ref{tab:semileptonS}, and \ref{tab:semileptonA}. Additionally, we also compare our obtained branching ratios with those obtained from other theoretical works \cite{Wang:2008xt,Wang:2009mi,Shi:2016gqt} in the respective tables, and our obtained results are consistent with those from other theoretical works. The branching ratios of the processes $B_{c}\to B_{s}^{(*)}\ell\nu_{\ell}$ and $B_{c}\to J/\psi(\eta_{c})\ell\nu_{\ell}$ can reach up to the order of magnitude of $10^{-2}$. The measurements of the absolute branching ratios could be reachable at the ongoing LHCb experiment.

Of particular interest is our result for the ratio
\begin{equation*}
\frac{\mathcal{B}(B_{c}\to J/\psi \tau\nu_{\tau})}{\mathcal{B}(B_{c}\to J/\psi \mu\nu_{\mu})}=0.231,
\end{equation*}
which is noticeably smaller than the central value of the experimental measurement $0.71\pm0.17\pm0.18$ reported by the LHCb Collaboration \cite{LHCb:2017vlu}, but is consistent with those from Refs. \cite{Wang:2008xt,Sun:2023iis}. This ratio can be used to test the lepton flavor universality, and the misfit of the theoretical and experimental values maybe indicates the new physics effects beyond the Standard Model. A more precise experiment and more theoretical calculations would greatly aid in testing the lepton flavor universality.

\begin{table*}[htbp]\centering
\caption{The branching ratios of the $B_{c}\to B_{s}(B,D,\eta_{c})\ell\nu_{\ell}$ decays. Additionally, we compare our findings with the branching ratios reported in other theoretical studies.}
\label{tab:semileptonP}
\renewcommand\arraystretch{1.2}
\begin{tabular*}{160mm}{c@{\extracolsep{\fill}}cccc}
\toprule[1pt]
\toprule[0.5pt]
Channels                                                        &$\ell$   &branching ratios        &Ref. \cite{Shi:2016gqt}      &Ref. \cite{Wang:2008xt}\\
\midrule[0.5pt]
\multirow{2}*{\shortstack{$B_{c}\to B_{s}\ell\nu_{\ell}$}}      &$e$       &$1.537\times10^{-2}$       &$1.51\times10^{-2}$          &$1.49^{+0.29}_{-0.30}\times10^{-2}$\\
                                                                &$\mu$     &$1.452\times10^{-2}$       &$1.43\times10^{-2}$          &$1.41^{+0.27}_{-0.28}\times10^{-2}$\\
\specialrule{0em}{3pt}{3pt}
\multirow{2}*{\shortstack{$B_{c}\to B\ell\nu_{\ell}$}}          &$e$       &$0.989\times10^{-3}$       &$1.04\times10^{-3}$          &$1.09^{+0.26}_{-0.26}\times10^{-3}$  \\
                                                                &$\mu$     &$0.944\times10^{-3}$       &$1.00\times10^{-3}$          &$1.04^{+0.24}_{-0.25}\times10^{-3}$\\
\specialrule{0em}{3pt}{3pt}
\multirow{3}*{\shortstack{$B_{c}\to D\ell\nu_{\ell}$}}          &$e$       &$7.024\times10^{-5}$       &$\cdots$                             &$3.0^{+1.0}_{-0.9}\times10^{-5}$       \\
                                                                &$\mu$     &$7.012\times10^{-5}$       &$\cdots$                             &$3.0^{+1.0}_{-0.9}\times10^{-5}$       \\
                                                                &$\tau$    &$3.924\times10^{-5}$       &$\cdots$                             &$2.1^{+0.7}_{-0.7}\times10^{-5}$       \\
\specialrule{0em}{3pt}{3pt}
\multirow{3}*{\shortstack{$B_{c}\to \eta_{c}\ell\nu_{\ell}$}}   &$e$       &$7.899\times10^{-3}$       &$\cdots$                             &$6.7^{+1.1}_{-1.3}\times10^{-3}$     \\
                                                                &$\mu$     &$7.868\times10^{-3}$       &$\cdots$                             &$6.7^{+1.1}_{-1.3}\times10^{-3}$     \\
                                                                &$\tau$    &$2.187\times10^{-3}$       &$\cdots$                             &$1.90^{+0.33}_{-0.34}\times10^{-3}$     \\
\bottomrule[0.5pt]
\bottomrule[1pt]
\end{tabular*}
\end{table*}

\begin{table*}[htbp]\centering
\caption{The branching ratios and the $\Gamma_{L}/\Gamma_{T}$ of the $B_{c}\to B_{s}^{*}(B^{*},D^{*},J/\psi,\psi(2S))\ell\nu_{\ell}$ decays. Additionally, we compare our findings with the branching ratios reported in other theoretical studies.}
\label{tab:semileptonV}
\renewcommand\arraystretch{1.2}
\begin{tabular*}{160mm}{c@{\extracolsep{\fill}}cccc}
\toprule[1pt]
\toprule[0.5pt]
Channels                                                           &$\ell$  &$\text{Branching~ratios}$   &$\Gamma_{L}/\Gamma_{T}$  &Ref. \cite{Wang:2008xt}    \\
\midrule[0.5pt]
\multirow{2}*{\shortstack{$B_{c}\to B_{s}^{*}\ell\nu_{\ell}$}}     &$e$      &$1.416\times10^{-2}$           &$1.123$                  &$1.96^{+0.45}_{-0.44}\times10^{-2}$   \\
                                                                   &$\mu$    &$1.324\times10^{-2}$           &$1.099$                  &$1.83^{+0.43}_{-0.41}\times10^{-2}$   \\
\specialrule{0em}{3pt}{3pt}
\multirow{2}*{\shortstack{$B_{c}\to B^{*}\ell\nu_{\ell}$}}         &$e$      &$0.777\times10^{-3}$           &$1.066$                  &$1.41^{+0.36}_{-0.34}\times10^{-3}$   \\
                                                                   &$\mu$    &$0.736\times10^{-3}$           &$1.049$                  &$1.34^{+0.43}_{-0.32}\times10^{-3}$   \\
\specialrule{0em}{3pt}{3pt}
\multirow{3}*{\shortstack{$B_{c}\to D^{*}\ell\nu_{\ell}$}}         &$e$      &$1.256\times10^{-4}$           &$1.033$                  &$0.45^{+0.16}_{-0.13}\times10^{-4}$    \\
                                                                   &$\mu$    &$1.253\times10^{-4}$           &$1.031$                  &$0.45^{+0.16}_{-0.13}\times10^{-4}$    \\
                                                                   &$\tau$   &$0.601\times10^{-4}$           &$0.773$                  &$0.27^{+0.10}_{-0.08}\times10^{-4}$    \\
\specialrule{0em}{3pt}{3pt}
\multirow{3}*{\shortstack{$B_{c}\to J/\psi\ell\nu_{\ell}$}}        &$e$      &$2.130\times10^{-2}$           &$1.189$                  &$1.49^{+0.27}_{-0.27}\times10^{-2}$    \\
                                                                   &$\mu$    &$2.119\times10^{-2}$           &$1.186$                  &$1.49^{+0.27}_{-0.27}\times10^{-2}$    \\
                                                                   &$\tau$   &$0.489\times10^{-2}$           &$0.838$                  &$0.37^{+0.07}_{-0.07}\times10^{-2}$    \\
\specialrule{0em}{3pt}{3pt}
\multirow{3}*{\shortstack{$B_{c}\to \psi(2S)\ell\nu_{\ell}$}}      &$e$      &$1.311\times10^{-3}$           &$1.856$                  &$\cdots$  \\
                                                                   &$\mu$    &$1.298\times10^{-3}$           &$1.847$                  &$\cdots$  \\
                                                                   &$\tau$   &$0.071\times10^{-3}$           &$0.924$                  &$\cdots$   \\
\bottomrule[0.5pt]
\bottomrule[1pt]
\end{tabular*}
\end{table*}

\begin{table*}[htbp]\centering
\caption{The branching ratios of the $B_{c}\to B_{s0}(B_{0},D_{0},\chi_{c0})\ell\nu_{\ell}$ decays. Additionally, we compare our findings with the branching ratios reported in other theoretical studies.}
\label{tab:semileptonS}
\renewcommand\arraystretch{1.2}
\begin{tabular*}{160mm}{c@{\extracolsep{\fill}}cccc}
\toprule[1pt]
\toprule[0.5pt]
Channels                                                        &$\ell$   &branching ratios      &Ref. \cite{Wang:2009mi}    &Ref. \cite{Shi:2016gqt} \\
\midrule[0.5pt]
\multirow{2}*{\shortstack{$B_{c}\to B_{s0}\ell\nu_{\ell}$}}     &$e$       &$2.203\times10^{-4}$     &$\cdots$                           &$6.58\times10^{-4}$  \\
                                                                &$\mu$     &$1.703\times10^{-4}$     &$\cdots$                           &$5.23\times10^{-4}$ \\
\specialrule{0em}{3pt}{3pt}
\multirow{2}*{\shortstack{$B_{c}\to B_{0}\ell\nu_{\ell}$}}      &$e$       &$0.419\times10^{-4}$     &$\cdots$                           &$4.60\times10^{-5}$  \\
                                                                &$\mu$     &$0.354\times10^{-4}$     &$\cdots$                           &$3.77\times10^{-5}$  \\
\specialrule{0em}{3pt}{3pt}
\multirow{3}*{\shortstack{$B_{c}\to D_{0}\ell\nu_{\ell}$}}      &$e$       &$0.547\times10^{-4}$     &$\cdots$                           &$\cdots$    \\
                                                                &$\mu$     &$0.545\times10^{-4}$     &$\cdots$                           &$\cdots$     \\
                                                                &$\tau$    &$0.202\times10^{-4}$     &$\cdots$                           &$\cdots$     \\
\specialrule{0em}{3pt}{3pt}
\multirow{3}*{\shortstack{$B_{c}\to\chi_{c0}\ell\nu_{\ell}$}}   &$e$       &$1.084\times10^{-3}$     &$2.1^{+0.4}_{-0.5}\times10^{-3}$         &$\cdots$     \\
                                                                &$\mu$     &$1.075\times10^{-3}$     &$2.1^{+0.4}_{-0.5}\times10^{-3}$         &$\cdots$    \\
                                                                &$\tau$    &$0.094\times10^{-3}$     &$0.24^{+0.04}_{-0.05}\times10^{-3}$        &$\cdots$    \\
\bottomrule[0.5pt]
\bottomrule[1pt]
\end{tabular*}
\end{table*}

\begin{table*}[htbp]\centering
\caption{The branching ratios and the $\Gamma_{L}/\Gamma_{T}$ of the $B_{c}\to B_{s1}^{(\prime)}(B_{1}^{(\prime)},D_{1}^{(\prime)},h_{c},\chi_{c1})\ell\nu_{\ell}$ decays. Additionally, we compare our findings with the branching ratios reported in other theoretical studies.}
\label{tab:semileptonA}
\renewcommand\arraystretch{1.2}
\begin{tabular*}{160mm}{c@{\extracolsep{\fill}}ccccc}
\toprule[1pt]
\toprule[0.5pt]
Channels                                                                &$\ell$  &$\text{Branching~ratios}$   &$\Gamma_{L}/\Gamma_{T}$    &Ref. \cite{Wang:2009mi}   &Ref. \cite{Shi:2016gqt}  \\
\midrule[0.5pt]
\multirow{2}*{\shortstack{$B_{c}\to B_{s1}^{\prime}\ell\nu_{\ell}$}}    &$e$      &$3.946\times10^{-5}$           &$2.615$                    &$\cdots$                        &$5.38\times10^{-4}$\\
                                                                        &$\mu$    &$2.793\times10^{-5}$           &$2.314$                    &$\cdots$                        &$3.98\times10^{-4}$\\
\specialrule{0em}{3pt}{3pt}
\multirow{2}*{\shortstack{$B_{c}\to B_{s1}\ell\nu_{\ell}$}}             &$e$      &$5.824\times10^{-5}$           &$1.591$                    &$\cdots$                        &$8.31\times10^{-5}$\\
                                                                        &$\mu$    &$4.230\times10^{-5}$           &$1.379$                    &$\cdots$                        &$6.33\times10^{-5}$\\
\specialrule{0em}{3pt}{3pt}
\multirow{2}*{\shortstack{$B_{c}\to B_{1}^{\prime}\ell\nu_{\ell}$}}     &$e$      &$1.636\times10^{-6}$           &$3.357$                    &$\cdots$                        &$7.70\times10^{-5}$\\
                                                                        &$\mu$    &$1.244\times10^{-6}$           &$3.097$                    &$\cdots$                        &$6.28\times10^{-5}$\\
\specialrule{0em}{3pt}{3pt}
\multirow{2}*{\shortstack{$B_{c}\to B_{1}\ell\nu_{\ell}$}}              &$e$      &$1.743\times10^{-5}$           &$1.708$                    &$\cdots$                        &$1.52\times10^{-5}$\\
                                                                        &$\mu$    &$1.461\times10^{-5}$           &$1.575$                    &$\cdots$                        &$1.28\times10^{-5}$\\
\specialrule{0em}{3pt}{3pt}
\multirow{3}*{\shortstack{$B_{c}\to D_{1}^{\prime}\ell\nu_{\ell}$}}     &$e$      &$3.510\times10^{-5}$           &$1.736$                    &$\cdots$                        &$\cdots$\\
                                                                        &$\mu$    &$3.491\times10^{-5}$           &$1.730$                    &$\cdots$                        &$\cdots$\\
                                                                        &$\tau$   &$1.026\times10^{-5}$           &$0.995$                    &$\cdots$                        &$\cdots$\\
\specialrule{0em}{3pt}{3pt}
\multirow{3}*{\shortstack{$B_{c}\to D_{1}\ell\nu_{\ell}$}}              &$e$      &$0.984\times10^{-5}$           &$1.533$                    &$\cdots$                        &$\cdots$\\
                                                                        &$\mu$    &$0.979\times10^{-5}$           &$1.529$                    &$\cdots$                        &$\cdots$\\
                                                                        &$\tau$   &$0.519\times10^{-5}$           &$2.940$                    &$\cdots$                        &$\cdots$\\
\specialrule{0em}{3pt}{3pt}
\multirow{3}*{\shortstack{$B_{c}\to h_{c}\ell\nu_{\ell}$}}              &$e$      &$1.155\times10^{-3}$           &$11.769$                   &$3.1^{+0.7}_{-0.9}\times10^{-3}$        &$\cdots$\\
                                                                        &$\mu$    &$1.140\times10^{-3}$           &$11.669$                   &$3.1^{+0.7}_{-0.9}\times10^{-3}$        &$\cdots$\\
                                                                        &$\tau$   &$0.051\times10^{-3}$           &$2.877$                    &$0.22^{+0.04}_{-0.05}\times10^{-3}$       &$\cdots$\\
\specialrule{0em}{3pt}{3pt}
\multirow{3}*{\shortstack{$B_{c}\to\chi_{c1}\ell\nu_{\ell}$}}           &$e$      &$2.953\times10^{-4}$           &$0.031$                    &$1.40^{+0.22}_{-0.24}\times10^{-3}$        &$\cdots$\\
                                                                        &$\mu$    &$2.933\times10^{-4}$           &$0.031$                    &$1.40^{+0.22}_{-0.24}\times10^{-3}$        &$\cdots$\\
                                                                        &$\tau$   &$0.256\times10^{-4}$           &$0.050$                    &$0.15^{+0.02}_{-0.03}\times10^{-3}$       &$\cdots$\\
\bottomrule[0.5pt]
\bottomrule[1pt]
\end{tabular*}
\end{table*}

Moreover, utilizing the obtained form factors, we can make predictions for nonleptonic decays. Specifically, we explore the processes $B_{c}\to J/\psi\pi(K)$, $B_{c}\to\psi(2S)\pi$, and $B_{c}\to\chi_{c0}\pi$, since they have corresponding experimental measurements. Assuming the naive factorization assumption, the branching ratios for these decays can be calculated as described in \cite{Zhang:2023ypl}
\begin{widetext}
\begin{equation}
\begin{split}
\mathcal{B}\big{(}B_{c}\to J/\psi(\psi(2S))\pi(K)\big{)}=&\frac{\Big{\vert} G_{F}V_{cb}V_{ud(us)}a_{1}f_{\pi(K)}m_{B_{c}}^{2}A_{0}^{B_{c} J/\psi(\psi(2S))}(m_{\pi(K)}^{2})\Big{\vert}^{2}}{32\pi m_{B_{c}}\Gamma_{B_{c}}}
\Bigg{(}1-\frac{m_{J/\psi(\psi(2S))}^{2}}{m_{B_{c}}^{2}}\Bigg{)},\\
\mathcal{B}\big{(}B_{c}\to \chi_{c0}\pi\big{)}=&\frac{\Big{\vert} G_{F}V_{cb}V_{ud}a_{1}f_{\pi}m_{B_{c}}^{2}F_{0}^{B_{c}\chi_{c0}}(m_{\pi}^{2})\Big{\vert}^{2}}{32\pi m_{B_{c}}\Gamma_{B_{c}}}
\Bigg{(}1-\frac{m_{\chi_{c0}}^{2}}{m_{B_{c}}^{2}}\Bigg{)},
\end{split}
\end{equation}
where $a_{1}=1.07$ and $a_{2}=0.234$ \cite{Zhang:2023ypl}.
\end{widetext}

The concerned branching ratios are determined as
\begin{eqnarray*}
\mathcal{B}\big{(}B_{c}\to J/\psi\pi\big{)}&=&2.785\times10^{-3},\\
\mathcal{B}\big{(}B_{c}\to J/\psi K\big{)}&=&0.213\times10^{-3},\\
\mathcal{B}\big{(}B_{c}\to \psi(2S)\pi\big{)}&=&0.662\times10^{-3},
\end{eqnarray*}
and some ratios of the branching ratios are
\begin{eqnarray*}
R_{J/\psi}^{K/\pi}&=&\frac{\mathcal{B}(B_{c}\to J/\psi K)}{\mathcal{B}(B_{c}\to J/\psi \pi)}=0.077,\\
R_{\pi}^{\psi(2S)/J/\psi}&=&\frac{\mathcal{B}(B_{c}\to \psi(2S)\pi)}{\mathcal{B}(B_{c}\to J/\psi \pi)}=0.238,\\
R_{J/\psi}^{\pi/\mu\nu_{\mu}}&=&\frac{\mathcal{B}(B_{c}\to J/\psi \pi)}{\mathcal{B}(B_{c}\to J/\psi \mu\nu_{\mu})}=0.131.
\end{eqnarray*}
Two of these predictions are in agreement with the experimental values: $R_{J/\psi}^{K/\pi}=0.079\pm0.007\pm0.003$ and $R_{\pi}^{\psi(2S)/J/\psi}=0.268\pm0.032\pm0.007\pm0.006$ \cite{ParticleDataGroup:2022pth}. However, the prediction for $R_{J/\psi}^{\pi/\mu\nu_{\mu}}$ is apparently larger than the experimental value of $(4.69\pm0.28\pm0.46)\times10^{-2}$ \cite{ParticleDataGroup:2022pth}. Additionally, we obtain $\mathcal{B}\big{(}B_{c}\to\chi_{c0}\pi\big{)}=0.633\times10^{-3}$, which is significantly larger than the experimental measurement $(2.4^{+0.9}_{-0.8})\times10^{-5}$ reported by the LHCb Collaboration \cite{LHCb:2016utz}, but is consistent with the Ref. \cite{Zhang:2023ypl}. Further experimental measurements are eagerly awaited to provide additional scrutiny and validation of our theoretical predictions for these weak decays.

{If the discussed hadron has the near-threshold behaviour, the state should be a mixture of $\bar{b}c$ components and two meson components. The meson components should be important because the coupling is given by a strong mechanism, which gives the important decay widths, and they are enhanced by the denominators of the difference between the energy and the threshold energy position. Thus, the decays through the two meson components will be small and the results given in our work can be seen as an upper limit in most cases.}

\section{Summary}\label{sec7}

Although significant progress has been made in observing new hadronic states over the past two decades, the establishment of the $B_c$ meson family remains incomplete, with only $B_c^{+}$ and $B_{c}(2S)^{\pm}$ states listed in the PDG \cite{ParticleDataGroup:2022pth}. With the upgrade being complete of the LHCb experiment in preparation for Run 3 and Run 4 of LHC, we have reason to believe this situation will be changed. Hence, the current work is timely and can provide valuable information for the experimental explorations of $B_c$ mesons. 

According to the previous experience of the studies of hadron spectroscopy \cite{Chao:1992et,Ding:1993uy,Mezoir:2008vx,Li:2009ad,Wang:2018rjg,Wang:2019mhs,Wang:2020prx,Song:2015nia,Duan:2020tsx,Duan:2021alw,Luo:2019qkm,Luo:2021dvj}, the importance of the unquenched effects has been realized step by step. The low mass puzzles involved in the $X(3872)$, $D_{s0}(2317)$, $D_{s1}(2460)$, and $\Lambda_c(2940)$ can be well understood under this scenario. In this work, we provide a complete spectroscopy of the $B_c$ mesons  under the unquenched picture. We first present the mass spectrum of the $B_c$ mesons, where the MGI model was applied to the concrete calculations, which can reflect the unquenched effects. The obtained mass spectrum of the $B_c$ mesons is valuable, but not sufficient for further experimental search for them. Therefore, in this work, we have investigated their various decay behaviors, including the two-body OZI-allowed strong decays, the dipion transitions betweeen $B_c$ mesons, the radiative decays, and some typical weak decays of $B_c(1^3S_0)$. We must emphasize that the discussed decays of $B_c$ mesons are supported by the mass spectrum study, since we simultaneously obtain the information of the numerical spatial wave functions of these focused $B_c$ mesons associated with their masses, which are used as inputs for the calculations of the $B_c$ meson decays. This treatment avoids the parameter dependence in the decay studies. 

With the accumulation of experimental data and enhancement of experimental capabilities, the investigations of hadron spectroscopy will enter a new stage. As an important part of the hadron family, the $B_c$ mesons become focal point, as our knowledge of the $B_c$ meson family is still infufficient. Facing this situation, we have reason to believe that it is full of challenges and opportunities. We expect that our experimental colleagues to seize this opportunity to continue to expand the realm of the observed hadrons.

\section*{Acknowledgment}

This work is supported by the China National Funds for Distinguished Young Scientists under Grant No. 11825503, the National Key Research and Development Program of China under Contract No. 2020YFA0406400, the 111 Project under Grant No. B20063, the National Natural Science Foundation of China under Grant Nos. 12247101 and 12247155, the fundamental Research Funds for the Central Universities, and the project for top-notch innovative talents of Gansu province. F.L.W. is also supported by the China Postdoctoral Science Foundation under Grant No. 2022M721440.


\begin{thebibliography}{}

\bibitem{Godfrey:2008nc}
S.~Godfrey and S.~L.~Olsen,
The exotic $XYZ$ charmonium-like mesons,
\href{https://doi.org/10.1146/annurev.nucl.58.110707.171145}{Ann. Rev. Nucl. Part. Sci. \textbf{58}, 51-73 (2008)}.

\bibitem{Li:2009zu}
B.~Q.~Li and K.~T.~Chao,
Higher charmonia and $X$, $Y$, $Z$ states with screened potential,
\href{https://doi.org/10.1103/PhysRevD.79.094004}{Phys. Rev. D \textbf{79}, 094004 (2009)}.

    \bibitem{Liu:2013waa}
      X.~Liu,
      An overview of $XYZ$ new particles,
      \href{http://dx.doi.org/10.1007/s11434-014-0407-2}{Chin.\ Sci.\ Bull.\  {\bf 59}, 3815 (2014)}.

\bibitem{Yuan:2015kya}
C.~Z.~Yuan [BESIII],
Study of the $XYZ$ states at the BESIII,
\href{https://doi.org/10.1007/s11467-015-0484-y}{Front. Phys. (Beijing) \textbf{10}, 101401 (2015)}.

\bibitem{Chen:2013wva}
W.~Chen, W.~Z.~Deng, J.~He, N.~Li, X.~Liu, Z.~G.~Luo, Z.~F.~Sun and S.~L.~Zhu,
$XYZ$ States,
\href{https://doi.org/10.22323/1.205.0005}{PoS \textbf{Hadron2013}, 005 (2013)}.

\bibitem{HillerBlin:2016odx}
A.~N.~Hiller Blin, C.~Fern\'andez-Ram\'\i{}rez, A.~Jackura, V.~Mathieu, V.~I.~Mokeev, A.~Pilloni and A.~P.~Szczepaniak,
Studying the $P_c(4450)$ resonance in $J/\psi$ photoproduction off protons,
\href{https://doi.org/10.1103/PhysRevD.94.034002}{Phys. Rev. D \textbf{94}, 034002 (2016)}.

\bibitem{Guo:2015umn}
F.~K.~Guo, U.~G.~Mei\ss{}ner, W.~Wang and Z.~Yang,
How to reveal the exotic nature of the $P_c(4450)$,
\href{https://doi.org/10.1103/PhysRevD.92.071502}{Phys. Rev. D \textbf{92}, 071502 (2015)}.

\bibitem{Chen:2016qju}
  H.~X.~Chen, W.~Chen, X.~Liu, and S.~L.~Zhu,
  The hidden-charm pentaquark and tetraquark states,
  \href{http://linkinghub.elsevier.com/retrieve/pii/S037015731630103X}{Phys.\ Rep.\  {\bf 639}, 1 (2016)}.

\bibitem{Guo:2017jvc}
  F.~K.~Guo, C.~Hanhart, U.~G.~Mei$\ss$ner, Q.~Wang, Q.~Zhao, and B.~S.~Zou,
  Hadronic molecules,
  \href{https://journals.aps.org/rmp/abstract/10.1103/RevModPhys.90.015004}{Rev.\ Mod.\ Phys.\  {\bf 90}, 015004 (2018)}.

    \bibitem{Liu:2019zoy}
      Y.~R.~Liu, H.~X.~Chen, W.~Chen, X.~Liu, and S.~L.~Zhu,
      Pentaquark and tetraquark states,
      \href{https://www.sciencedirect.com/science/article/pii/S0146641019300304?via\%3Dihub}{Prog.\ Part.\ Nucl.\ Phys.\  {\bf 107}, 237 (2019)}.

\bibitem{Brambilla:2019esw}
N.~Brambilla, S.~Eidelman, C.~Hanhart, A.~Nefediev, C.~P.~Shen, C.~E.~Thomas, A.~Vairo, and C.~Z.~Yuan,
The $XYZ$ states: Experimental and theoretical status and perspectives,
\href{https://www.sciencedirect.com/science/article/pii/S0370157320301915?via\%3Dihub}{Phys. Rep. \textbf{873}, 1 (2020)}.

\bibitem{Meng:2022ozq}
L.~Meng, B.~Wang, G.~J.~Wang and S.~L.~Zhu,
Chiral perturbation theory for heavy hadrons and chiral effective field theory for heavy hadronic molecules,
\href{https://www.sciencedirect.com/science/article/pii/S0370157323001679?via\%3Dihub}{Phys. Rept. \textbf{1019}, 1-149 (2023)}.

\bibitem{Chen:2022asf}
H.~X.~Chen, W.~Chen, X.~Liu, Y.~R.~Liu and S.~L.~Zhu,
An updated review of the new hadron states,
\href{https://iopscience.iop.org/article/10.1088/1361-6633/aca3b6}{Rept. Prog. Phys. \textbf{86}, no.2, 026201 (2023)}.

\bibitem{LHCb:2019kea}
R.~Aaij \textit{et al.} [LHCb],
Observation of a narrow pentaquark state, $P_c(4312)^+$, and of two-peak structure of the $P_c(4450)^+$,
\href{https://doi.org/10.1103/PhysRevLett.122.222001}{Phys. Rev. Lett. \textbf{122}, 222001 (2019)}.

\bibitem{Chu:2016sjc}
X.~Chu [BESIII],
Hadron spectroscopy at BESIII,
\href{https://doi.org/10.22323/1.274.0028}{PoS \textbf{HQL2016}, 028 (2017)}.

\bibitem{LHCb:2022sfr}
R.~Aaij \textit{et al.} [LHCb],
First observation of a doubly charged tetraquark and its neutral partner,
\href{https://doi.org/10.1103/PhysRevLett.131.041902}{Phys. Rev. Lett. \textbf{131}, 041902 (2023)}.

\bibitem{LHCb:2022lzp}
R.~Aaij \textit{et al.} [LHCb],
Amplitude analysis of $B^{0}\to D^{0}D_{s}^{+}\pi^{-}$ and $B^{+}\to D_{-}D_{s}^{+}\pi^{+}$ decays,
\href{https://doi.org/10.1103/PhysRevD.108.012017}{Phys. Rev. D \textbf{108}, 012017 (2023)}.

\bibitem{Wang:2022zgi}
J.~Wang [LHCb],
Heavy flavor and exotic production at LHCb,
\href{https://doi.org/10.22323/1.414.0487}{PoS \textbf{ICHEP2022}, 487 (2022)}.

\bibitem{Xu:2022kkh}
J.~Xu [LHCb],
LHCb results in charm baryons,
\href{https://doi.org/10.1016/j.nuclphysbps.2022.09.013}{Nucl. Part. Phys. Proc. \textbf{318-323}, 56-60 (2022)}.

\bibitem{Onuki:2022ugx}
Y.~Onuki [Belle II],
Belle II status and prospect,
\href{https://doi.org/10.1016/j.nuclphysbps.2022.09.017}{Nucl. Part. Phys. Proc. \textbf{318-323}, 78-84 (2022)}.

\bibitem{Belle-II:2022fsw}
F.~Abudin\'en \textit{et al.} [Belle II],
Reconstruction of $B \to \rho \ell \nu_\ell$ decays identified using hadronic decays of the recoil $B$ meson in 2019 -- 2021 Belle II data,
\href{https://arxiv.org/abs/2211.15270}{arXiv:2211.15270}.

\bibitem{Belle:2022dyc}
K.~N.~Chu \textit{et al.} [Belle],
Study of $B^+ \to p \overline{n} \pi^0$,
\href{https://arxiv.org/abs/2211.11251}{arXiv:2211.11251}.

\bibitem{Chang:1992bb}
C.~H.~Chang and Y.~Q.~Chen,
The Production of $B_{c}$ or $\bar{B}_{c}$ meson associated with two heavy quark jets in $Z^{0}$ boson decay,
\href{https://doi.org/10.1103/PhysRevD.46.3845}{Phys. Rev. D \textbf{46}, 3845 (1992)}.

\bibitem{Chang:1992jb}
C.~H.~Chang and Y.~Q.~Chen,
The hadronic production of the $B_{c}$ meson at Tevatron, CERN LHC and SSC,
\href{https://doi.org/10.1103/PhysRevD.48.4086}{Phys. Rev. D \textbf{48}, 4086-4091 (1993)}.

\bibitem{CDF:1998ihx}
F.~Abe \textit{et al.} [CDF],
Observation of the $B_c$ meson in $p\bar{p}$ collisions at $\sqrt{s} = 1.8$ TeV,
\href{https://doi.org/10.1103/PhysRevLett.81.2432}{Phys. Rev. Lett. \textbf{81}, 2432-2437 (1998)}.

\bibitem{Lusignoli:1991bn}
M.~Lusignoli, M.~Masetti and S.~Petrarca,
$B_{c}$ production,
\href{https://doi.org/10.1016/0370-2693(91)90757-H}{Phys. Lett. B \textbf{266}, 142-146 (1991)}.

\bibitem{Braaten:1993jn}
E.~Braaten, K.~m.~Cheung and T.~C.~Yuan,
Perturbative QCD fragmentation functions for $B_c$ and $B_{c}^*$ production,
\href{https://doi.org/10.1103/PhysRevD.48.R5049}{Phys. Rev. D \textbf{48}, R5049 (1993)}.

\bibitem{Chang:1996jt}
C.~H.~Chang, Y.~Q.~Chen and R.~J.~Oakes,
Comparative study of the hadronic production of $B_{c}$ mesons,
\href{https://doi.org/10.1103/PhysRevD.54.4344}{Phys. Rev. D \textbf{54}, 4344-4348 (1996)}.

\bibitem{Masetti:1995uk}
M.~Masetti and F.~Sartogo,
Perturbative predictions for $B_{c}$ meson production in hadronic collisions,
\href{https://doi.org/10.1016/0370-2693(95)00908-4}{Phys. Lett. B \textbf{357}, 659-665 (1995)}.

\bibitem{DELPHI:1996vyn}
P.~Abreu \textit{et al.} [DELPHI],
Search for the $B_{c}$ Meson,
\href{https://doi.org/10.1016/S0370-2693(97)00254-2}{Phys. Lett. B \textbf{398}, 207-222 (1997)}.

\bibitem{OPAL:1998gdf}
K.~Ackerstaff \textit{et al.} [OPAL],
Search for the $B_c$ meson in hadronic $Z^0$ decays,
\href{https://doi.org/10.1016/S0370-2693(97)01569-4}{Phys. Lett. B \textbf{420}, 157-168 (1998)}.

\bibitem{ALEPH:1997oob}
R.~Barate \textit{et al.} [ALEPH],
Search for the $B_c$ meson in hadronic $Z$ decays,
\href{https://doi.org/10.1016/S0370-2693(97)00461-9}{Phys. Lett. B \textbf{402}, 213-226 (1997)}.

\bibitem{CDF:1996efe}
F.~Abe \textit{et al.} [CDF],
Measurement of the branching ratio $B(B_u^+ \to J/\psi \pi^+)$ and search for $B_c^+ \to J/\psi \pi^+$,
\href{https://doi.org/10.1103/PhysRevLett.77.5176}{Phys. Rev. Lett. \textbf{77}, 5176-5181 (1996)}.

\bibitem{CDF:2005yjh}
A.~Abulencia \textit{et al.} [CDF],
Evidence for the exclusive decay $B_c^\pm \to J/\psi \pi^\pm$ and measurement of the mass of the $B_c$ meson,
\href{https://doi.org/10.1103/PhysRevLett.96.082002}{Phys. Rev. Lett. \textbf{96}, 082002 (2006)}.

\bibitem{CDF:2007umr}
T.~Aaltonen \textit{et al.} [CDF],
Observation of the decay $B^+_c$ $\to J/\psi \pi^\pm$ and measurement of the $B^+_c$ mass,
\href{https://doi.org/10.1103/PhysRevLett.100.182002}{Phys. Rev. Lett. \textbf{100}, 182002 (2008)}.

\bibitem{D0:2008bqs}
V.~M.~Abazov \textit{et al.} [D0],
Observation of the $B_c$ Meson in the Exclusive Decay $B_c \to J/\psi \pi$,
\href{https://doi.org/10.1103/PhysRevLett.101.012001}{Phys. Rev. Lett. \textbf{101}, 012001 (2008)}.

\bibitem{LHCb:2012ag}
R.~Aaij \textit{et al.} [LHCb],
First observation of the decay $B_c^+ \to J/\psi \pi^+\pi^-\pi^+$,
\href{https://doi.org/10.1103/PhysRevLett.108.251802}{Phys. Rev. Lett. \textbf{108}, 251802 (2012)}.

\bibitem{LHCb:2013vrl}
R.~Aaij \textit{et al.} [LHCb],
Observation of the decay $B_c^+ \to \psi(2S)\pi^+$,
\href{https://doi.org/10.1103/PhysRevD.87.071103}{Phys. Rev. D \textbf{87}, 071103 (2013)}.

\bibitem{LHCb:2013kwl}
R.~Aaij \textit{et al.} [LHCb],
Observation of $B^+_c \to J/\psi D_s^+$ and $B^+_c \to J/\psi D_s^{*+}$ decays,
\href{https://doi.org/10.1103/PhysRevD.87.071103}{Phys. Rev. D \textbf{87}, 112012 (2013)}.

\bibitem{LHCb:2013hwj}
R.~Aaij \textit{et al.} [LHCb],
First observation of the decay $B_{c}^{+}\to J/\psi K^+$,
\href{https://doi.org/10.1007/JHEP09(2013)075}{JHEP \textbf{09}, 075 (2013)}.

\bibitem{LHCb:2013xlg}
R.~Aaij \textit{et al.} [LHCb],
Observation of the decay $B^+_c \to B^0_s\pi^+$,
\href{https://doi.org/10.1103/PhysRevLett.111.181801}{Phys. Rev. Lett. \textbf{111}, 181801 (2013)}.

\bibitem{LHCb:2013rud}
R.~Aaij \textit{et al.} [LHCb],
Observation of the decay $B_c \to J/\psi K^+ K^- \pi^+ $,
\href{https://doi.org/10.1007/JHEP11(2013)094}{JHEP \textbf{11}, 094 (2013)}.

\bibitem{ATLAS:2014lga}
G.~Aad \textit{et al.} [ATLAS],
Observation of an excited $B_c^\pm$ meson state with the ATLAS detector,
\href{https://doi.org/10.1103/PhysRevLett.113.212004}{Phys. Rev. Lett. \textbf{113}, 212004 (2014)}.

\bibitem{CMS:2019uhm}
A.~M.~Sirunyan \textit{et al.} [CMS],
Observation of two excited $B^+_{c}$ states and measurement of the $B^+_\mathrm{c}(2S)$ mass in $pp$ collisions at $\sqrt{s} =$ 13 TeV,
\href{https://doi.org/10.1103/PhysRevLett.122.132001}{Phys. Rev. Lett. \textbf{122}, 132001 (2019)}.

\bibitem{LHCb:2019bem}
R.~Aaij \textit{et al.} [LHCb],
Observation of an excited $B_c^+$ state,
\href{https://doi.org/10.1103/PhysRevLett.122.232001}{Phys. Rev. Lett. \textbf{122}, 232001 (2019)}.

\bibitem{ParticleDataGroup:2022pth}
R.~L.~Workman \textit{et al.} [Particle Data Group],
Review of Particle Physics,
\href{https://doi.org/10.1093/ptep/ptac097}{PTEP \textbf{2022}, 083C01 (2022)}.

\bibitem{Godfrey:1985xj}
S.~Godfrey and N.~Isgur,
Mesons in a relativized quark model with chromodynamics,
\href{https://doi.org/10.1103/PhysRevD.32.189}{Phys. Rev. D \textbf{32}, 189-231 (1985)}.

\bibitem{Eichten:1994gt}
E.~J.~Eichten and C.~Quigg,
Mesons with beauty and charm: Spectroscopy,
\href{https://doi.org/10.1103/PhysRevD.49.5845}{Phys. Rev. D \textbf{49}, 5845-5856 (1994)}.

\bibitem{Gershtein:1994dxw}
S.~S.~Gershtein, V.~V.~Kiselev, A.~K.~Likhoded and A.~V.~Tkabladze,
$B_{c}$ spectroscopy,
\href{https://doi.org/10.1103/PhysRevD.51.3613}{Phys. Rev. D \textbf{51}, 3613-3627 (1995)}.

\bibitem{Fulcher:1998ka}
L.~P.~Fulcher,
Phenomenological predictions of the properties of the $B_c$ system,
\href{https://doi.org/10.1103/PhysRevD.60.074006}{Phys. Rev. D \textbf{60}, 074006 (1999)}.

\bibitem{Ebert:2002pp}
D.~Ebert, R.~N.~Faustov and V.~O.~Galkin,
Properties of heavy quarkonia and $B_c$ mesons in the relativistic quark model,
\href{https://doi.org/10.1103/PhysRevD.67.014027}{Phys. Rev. D \textbf{67}, 014027 (2003)}.

\bibitem{Godfrey:2004ya}
S.~Godfrey,
Spectroscopy of $B_c$ mesons in the relativized quark model,
\href{https://doi.org/10.1103/PhysRevD.70.054017}{Phys. Rev. D \textbf{70}, 054017 (2004)}.

\bibitem{Eichten:2019gig}
E.~J.~Eichten and C.~Quigg,
Mesons with Beauty and Charm: New Horizons in Spectroscopy,
\href{https://doi.org/10.1103/PhysRevD.99.054025}{Phys. Rev. D \textbf{99}, no.5, 054025 (2019)}.

\bibitem{vanBeveren:2003kd}
E.~van Beveren and G.~Rupp,
Observed $D_s(2317)$ and tentative $D(2100\text{--}2300)$ as the charmed cousins of the light scalar nonet,
\href{https://doi.org/10.1103/PhysRevLett.91.012003}{Phys. Rev. Lett. \textbf{91}, 012003 (2003)}.

\bibitem{Dai:2003yg}
Y.~B.~Dai, C.~S.~Huang, C.~Liu and S.~L.~Zhu,
Understanding the $D^+_{sJ}(2317)$ and $D^+_{sJ}(2460)$ with sum rules in HQET,
\href{https://doi.org/10.1103/PhysRevD.68.114011}{Phys. Rev. D \textbf{68}, 114011 (2003)}.

\bibitem{Hwang:2004cd}
D.~S.~Hwang and D.~W.~Kim,
Mass of $D^{*}_{sJ}(2317)$ and coupled channel effect,
\href{https://doi.org/10.1016/j.physletb.2004.09.040}{Phys. Lett. B \textbf{601}, 137-143 (2004)}.

\bibitem{Simonov:2004ar}
Y.~A.~Simonov and J.~A.~Tjon,
The Coupled-channel analysis of the $D$ and $D_{s}$ mesons,
\href{https://doi.org/10.1103/PhysRevD.70.114013}{Phys. Rev. D \textbf{70}, 114013 (2004)}.

\bibitem{Ortega:2016mms}
P.~G.~Ortega, J.~Segovia, D.~R.~Entem and F.~Fernandez,
Molecular components in $P$-wave charmed-strange mesons,
\href{https://doi.org/10.1103/PhysRevD.94.074037}{Phys. Rev. D \textbf{94}, no.7, 074037 (2016)}.

\bibitem{Danilkin:2010cc}
I.~V.~Danilkin and Y.~A.~Simonov,
Dynamical origin and the pole structure of $X(3872)$,
\href{https://doi.org/10.1103/PhysRevLett.105.102002}{Phys. Rev. Lett. \textbf{105}, 102002 (2010)}.

\bibitem{Li:2009ad}
B.~Q.~Li, C.~Meng and K.~T.~Chao,
Coupled-channel and screening effects in charmonium spectrum,
\href{https://doi.org/10.1103/PhysRevD.80.014012}{Phys. Rev. D \textbf{80}, 014012 (2009)}.

\bibitem{Kalashnikova:2005ui}
Y.~S.~Kalashnikova,
Coupled-channel model for charmonium levels and an option for $X(3872)$,
\href{https://doi.org/10.1103/PhysRevD.72.034010}{Phys. Rev. D \textbf{72}, 034010 (2005)}.

\bibitem{Luo:2019qkm}
S.~Q.~Luo, B.~Chen, Z.~W.~Liu and X.~Liu,
Resolving the low mass puzzle of $\Lambda_c(2940)^+$,
\href{https://doi.org/10.1140/epjc/s10052-020-7874-1}{Eur. Phys. J. C \textbf{80}, 301 (2020)}.

\bibitem{Song:2015nia}
Q.~T.~Song, D.~Y.~Chen, X.~Liu and T.~Matsuki,
Charmed-strange mesons revisited: mass spectra and strong decays,
\href{https://doi.org/10.1103/PhysRevD.91.054031}{Phys. Rev. D \textbf{91}, 054031 (2015)}.

\bibitem{Song:2015fha}
Q.~T.~Song, D.~Y.~Chen, X.~Liu and T.~Matsuki,
Higher radial and orbital excitations in the charmed meson family,
\href{https://doi.org/10.1103/PhysRevD.92.074011}{Phys. Rev. D \textbf{92}, 074011 (2015)}.

\bibitem{Wang:2018rjg}
J.~Z.~Wang, Z.~F.~Sun, X.~Liu and T.~Matsuki,
Higher bottomonium zoo,
\href{https://doi.org/10.1140/epjc/s10052-018-6372-1}{Eur. Phys. J. C \textbf{78}, 915 (2018)}.

\bibitem{Wang:2019mhs}
J.~Z.~Wang, D.~Y.~Chen, X.~Liu and T.~Matsuki,
Constructing $J/\psi$ family with updated data of charmoniumlike $Y$ states,
\href{https://doi.org/10.1103/PhysRevD.99.114003}{Phys. Rev. D \textbf{99}, 114003 (2019)}.

\bibitem{Wang:2020prx}
J.~Z.~Wang, R.~Q.~Qian, X.~Liu and T.~Matsuki,
Are the $Y$ states around 4.6 GeV from $e^+e^-$ annihilation higher charmonia?,
\href{https://doi.org/10.1103/PhysRevD.101.034001}{Phys. Rev. D \textbf{101},  034001 (2020)}.

\bibitem{Chao:1992et}
K.~T.~Chao, Y.~B.~Ding and D.~H.~Qin,
Possible phenomenological indication for the string Coulomb term and the color screening effects in the quark-anti-quark potential,
Commun. Theor. Phys. \textbf{18}, 321-326 (1992).

\bibitem{Ding:1993uy}
Y.~B.~Ding, K.~T.~Chao and D.~H.~Qin,
Screened $Q-\bar{Q}$ potential and spectrum of heavy quarkonium,
\href{https://doi.org/10.1088/0256-307X/10/8/004}{Chin. Phys. Lett. \textbf{10}, 460-463 (1993)}.

\bibitem{Duan:2021alw}
M.~X.~Duan and X.~Liu,
Where are 3P and higher P-wave states in the charmonium family?,
\href{https://journals.aps.org/prd/abstract/10.1103/PhysRevD.104.074010}{Phys. Rev. D \textbf{104}, no.7, 074010 (2021)}.

\bibitem{Duan:2020tsx}
M.~X.~Duan, S.~Q.~Luo, X.~Liu and T.~Matsuki,
Possibility of charmoniumlike state $X(3915)$ as $\chi_{c0}(2P)$ state,
\href{https://doi.org/10.1103/PhysRevD.101.054029}{Phys. Rev. D \textbf{101}, 054029 (2020)}.

\bibitem{Micu:1968mk}
L.~Micu,
Decay rates of meson resonances in a quark model,
\href{https://doi.org/10.1016/0550-3213(69)90039-X}{Nucl. Phys. B \textbf{10}, 521-526 (1969)}.

\bibitem{LeYaouanc:1972vsx}
A.~Le Yaouanc, L.~Oliver, O.~Pene and J.~C.~Raynal,
Naive quark pair creation model of strong interaction vertices,
\href{https://doi.org/10.1103/PhysRevD.8.2223}{Phys. Rev. D \textbf{8}, 2223-2234 (1973)}.

\bibitem{Ackleh:1996yt}
E.~S.~Ackleh, T.~Barnes and E.~S.~Swanson,
On the mechanism of open flavor strong decays,
\href{https://doi.org/10.1103/PhysRevD.54.6811}{Phys. Rev. D \textbf{54}, 6811-6829 (1996)}.

\bibitem{Blundell:1996as}
H.~G.~Blundell,
Meson properties in the quark model: A look at some outstanding problems,
\href{https://arxiv.org/abs/hep-ph/9608473}{arXiv:hep-ph/9608473}.

\bibitem{Pang:2017dlw}
C.~Q.~Pang, J.~Z.~Wang, X.~Liu and T.~Matsuki,
A systematic study of mass spectra and strong decay of strange mesons,
\href{https://doi.org/10.1140/epjc/s10052-017-5434-0}{Eur. Phys. J. C \textbf{77}, 861 (2017)}.

\bibitem{Wang:2021abg}
L.~M.~Wang, S.~Q.~Luo and X.~Liu,
Light unflavored vector meson spectroscopy around the mass range of 2.4\ensuremath{\sim}3\,\,GeV and possible experimental evidence,
\href{https://doi.org/10.1103/PhysRevD.105.034011}{Phys. Rev. D \textbf{105},  034011 (2022)}.

\bibitem{Lucha:1991vn}
W.~Lucha, F.~F.~Schoberl and D.~Gromes,
Bound states of quarks,
\href{https://doi.org/10.1016/0370-1573(91)90001-3}{Phys. Rept. \textbf{200}, 127-240 (1991)}.

\bibitem{Eichten:1974af}
E.~Eichten, K.~Gottfried, T.~Kinoshita, J.~B.~Kogut, K.~D.~Lane and T.~M.~Yan,
The spectrum of charmonium,
\href{https://doi.org/10.1103/PhysRevLett.34.369}{Phys. Rev. Lett. \textbf{36}, 1276 (1976)}.

\bibitem{Ding:2021dwh}
R.~Ding, B.~D.~Wan, Z.~Q.~Chen, G.~L.~Wang and C.~F.~Qiao,
Finding $B_c(3S)$ states via their strong decays,
\href{https://doi.org/10.1016/j.physletb.2021.136277}{Phys. Lett. B \textbf{816}, 136277 (2021)}.

\bibitem{Wang:2022cxy}
G.~L.~Wang, T.~Wang, Q.~Li and C.~H.~Chang,
The mass spectrum and wave functions of the $B_{c}$ system,
\href{https://doi.org/10.1007/JHEP05(2022)006}{JHEP \textbf{05}, 006 (2022)}.

\bibitem{LeYaouanc:1973ldf}
A.~Le Yaouanc, L.~Oliver, O.~Pene and J.~C.~Raynal,
Naive quark pair creation model and baryon decays,
\href{https://doi.org/10.1103/PhysRevD.9.1415}{Phys. Rev. D \textbf{9}, 1415-1419 (1974)}.

\bibitem{LeYaouanc:1974cvx}
A.~Le Yaouanc, L.~Oliver, O.~Pene and J.~C.~Raynal,
Resonant Partial wave amplitudes in $\pi N\to \pi\pi N$ according to the naive quark pair creation model,
\href{https://doi.org/10.1103/PhysRevD.11.1272}{Phys. Rev. D \textbf{11}, 1272 (1975)}.

\bibitem{LeYaouanc:1977fsz}
A.~Le Yaouanc, L.~Oliver, O.~Pene and J.~C.~Raynal,
Strong decays of $\psi^{\prime\prime}(4.028)$ as a radial excitation of charmonium,
\href{https://doi.org/10.1016/0370-2693(77)90250-7}{Phys. Lett. B \textbf{71}, 397-399 (1977)}.

\bibitem{Jacob:1959at}
M.~Jacob and G.~C.~Wick,
On the general theory of collisions for particles with spin,
\href{https://doi.org/10.1016/0003-4916(59)90051-X}{Annals Phys. \textbf{7}, 404-428 (1959)}.

\bibitem{Hayne:1981zy}
C.~Hayne and N.~Isgur,
Beyond the wave function at the origin: Some momentum dependent effects in the nonrelativistic quark model,
\href{https://doi.org/10.1103/PhysRevD.25.1944}{Phys. Rev. D \textbf{25}, 1944 (1982)}.

\bibitem{Li:2019tbn}
Q.~Li, M.~S.~Liu, L.~S.~Lu, Q.~F.~L\"u, L.~C.~Gui and X.~H.~Zhong,
Excited bottom-charmed mesons in a nonrelativistic quark model,
\href{https://doi.org/10.1103/PhysRevD.99.096020}{Phys. Rev. D \textbf{99},  096020 (2019)}.

\bibitem{Yan:1980uh}
T.~M.~Yan,
Hadronic transitions between heavy quark states in quantum chromodynamics,
\href{https://doi.org/10.1103/PhysRevD.22.1652}{Phys. Rev. D \textbf{22}, 1652 (1980)}.

\bibitem{Zhou:1990ik}
H.~Y.~Zhou and Y.~P.~Kuang,
Coupled channel effects in hadronic transitions in heavy quarkonium systems,
\href{https://doi.org/10.1103/PhysRevD.44.756}{Phys. Rev. D \textbf{44}, 756-769 (1991)}.

\bibitem{Kuang:1981se}
Y.~P.~Kuang and T.~M.~Yan,
Predictions for hadronic transitions in the $b\bar{b}$ system,
\href{https://doi.org/10.1103/PhysRevD.24.2874}{Phys. Rev. D \textbf{24}, 2874 (1981)}.

\bibitem{Kuang:2006me}
Y.~P.~Kuang,
QCD multipole expansion and hadronic transitions in heavy quarkonium systems,
\href{https://doi.org/10.1007/s11467-005-0012-6}{Front. Phys. China \textbf{1}, 19-37 (2006)}.

\bibitem{Godfrey:2015dia}
S.~Godfrey and K.~Moats,
Bottomonium mesons and strategies for their observation,
\href{https://doi.org/10.1103/PhysRevD.92.054034}{Phys. Rev. D \textbf{92}, 054034 (2015)}.

\bibitem{Segovia:2016xqb}
J.~Segovia, P.~G.~Ortega, D.~R.~Entem and F.~Fern\'andez,
Bottomonium spectrum revisited,
\href{https://doi.org/10.1103/PhysRevD.93.074027}{Phys. Rev. D \textbf{93}, 074027 (2016)}.

\bibitem{Wang:2015xsa}
B.~Wang, H.~Xu, X.~Liu, D.~Y.~Chen, S.~Coito and E.~Eichten,
Using $X(3823)\to J/\psi\pi^+\pi^-$ to identify coupled-channel effects,
\href{https://doi.org/10.1007/s11467-016-0564-7}{Front. Phys. (Beijing) \textbf{11}, 111402 (2016)}.

\bibitem{Giles:1977mp}
R.~Giles and S.~H.~H.~Tye,
The application of the quark-confining string to the $\psi$ spectroscopy,
\href{https://doi.org/10.1103/PhysRevD.16.1079}{Phys. Rev. D \textbf{16}, 1079 (1977)}.

\bibitem{Tye:1975fz}
S.~H.~H.~Tye,
A quark-binding string,
\href{https://doi.org/10.1103/PhysRevD.13.3416}{Phys. Rev. D \textbf{13}, 3416 (1976)}.

\bibitem{Buchmuller:1979gy}
W.~Buchmuller and S.~H.~H.~Tye,
Vibrational states in the $\Upsilon$ spectroscopy,
\href{https://doi.org/10.1103/PhysRevLett.44.850}{Phys. Rev. Lett. \textbf{44}, 850-853 (1980)}.

\bibitem{Capstick:1985xss}
S.~Capstick and N.~Isgur,
Baryons in a relativized quark model with chromodynamics,
\href{https://doi.org/10.1063/1.35361}{AIP Conf. Proc. \textbf{132}, 267-271 (1985)}.

\bibitem{Martin-Gonzalez:2022qwd}
B.~Mart\'\i{}n-Gonz\'alez, P.~G.~Ortega, D.~R.~Entem, F.~Fern\'andez and J.~Segovia,
Toward the discovery of novel $B_c$ states: Radiative and hadronic transitions,
\href{https://doi.org/10.1103/PhysRevD.106.054009}{Phys. Rev. D \textbf{106},  054009 (2022)}.

\bibitem{CLEO:2008kwj}
H.~Mendez \textit{et al.} [CLEO],
Branching fractions for transitions of $\psi(2S)$ to $J/\psi$,
\href{https://doi.org/10.1103/PhysRevD.78.011102}{Phys. Rev. D \textbf{78}, 011102 (2008)}.

\bibitem{CLEO:2005zky}
N.~E.~Adam \textit{et al.} [CLEO],
Observation of $\psi(3770)\to\pi\pi J/\psi$ and measurement of $\Gamma_{ee}[\psi(2S)]$,
\href{https://doi.org/10.1103/PhysRevLett.96.082004}{Phys. Rev. Lett. \textbf{96}, 082004 (2006)}.

\bibitem{CLEO:1993fsd}
F.~Butler \textit{et al.} [CLEO],
Analysis of hadronic transitions in $\Upsilon(3S)$ decays,
\href{https://doi.org/10.1103/PhysRevD.49.40}{Phys. Rev. D \textbf{49}, 40-57 (1994)}.

\bibitem{CLEO:1998fxo}
J.~P.~Alexander \textit{et al.} [CLEO],
The hadronic transitions $\Upsilon(2S)\to \Upsilon(1S)$,
\href{https://doi.org/10.1103/PhysRevD.58.052004}{Phys. Rev. D \textbf{58}, 052004 (1998)}.

\bibitem{Belle:2017vat}
E.~Guido \textit{et al.} [Belle],
Study of $\eta$ and dipion transitions in $\Upsilon(4S)$ decays to lower bottomonia,
\href{https://doi.org/10.1103/PhysRevD.96.052005}{Phys. Rev. D \textbf{96}, 052005 (2017)}.

\bibitem{Chen:2011jp}
D.~Y.~Chen, X.~Liu and X.~Q.~Li,
Anomalous dipion invariant mass distribution of the $\Upsilon(4S)$ decays into $\Upsilon(1S) \pi^{+} \pi^{-}$ and $\Upsilon(2S) \pi^{+} \pi^{-}$,
\href{https://doi.org/10.1140/epjc/s10052-011-1808-x}{Eur. Phys. J. C \textbf{71}, 1808 (2011)}.

\bibitem{Chen:2011qx}
D.~Y.~Chen, J.~He, X.~Q.~Li and X.~Liu,
Dipion invariant mass distribution of the anomalous $\Upsilon(1S) \pi^{+} \pi^{-}$ and $\Upsilon(2S) \pi^{+} \pi^{-}$ production near the peak of $\Upsilon(10860)$,
\href{https://doi.org/10.1103/PhysRevD.84.074006}{Phys. Rev. D \textbf{84}, 074006 (2011)}.

\bibitem{Bai:2022cfz}
Z.~Y.~Bai, Y.~S.~Li, Q.~Huang, X.~Liu and T.~Matsuki,
$\Upsilon(10753)\to\Upsilon(nS)\pi^{+}\pi^{-}$ decays induced by hadronic loop mechanism,
\href{https://doi.org/10.1103/PhysRevD.105.074007}{Phys. Rev. D \textbf{105}, 074007 (2022)}.

\bibitem{Moxhay:1988ri}
P.~Moxhay,
Coupled channel effects in the decay $\Upsilon(3S)\to\Upsilon(1S)\pi^{+}\pi^{-}$,
\href{https://doi.org/10.1103/PhysRevD.39.3497}{Phys. Rev. D \textbf{39}, 3497 (1989)}.

\bibitem{Meng:2007tk}
C.~Meng and K.~T.~Chao,
Scalar resonance contributions to the dipion transition rates of $\Upsilon(4S,5S)$ in the re-scattering model,
\href{https://doi.org/10.1103/PhysRevD.77.074003}{Phys. Rev. D \textbf{77}, 074003 (2008)}.

\bibitem{Meng:2008dd}
C.~Meng and K.~T.~Chao,
Peak shifts due to $B^{(*)}$-$\bar{B}^{(*)}$ rescattering in $\Upsilon(5S)$ dipion transitions,
\href{https://doi.org/10.1103/PhysRevD.78.034022}{Phys. Rev. D \textbf{78}, 034022 (2008)}.

\bibitem{Chen:2011zv}
D.~Y.~Chen, X.~Liu and S.~L.~Zhu,
Charged bottomonium-like states $Z_b(10610)$ and $Z_b(10650)$ and the $\Upsilon(5S)\to \Upsilon(2S)\pi^+\pi^-$ decay,
\href{https://doi.org/10.1103/PhysRevD.84.074016}{Phys. Rev. D \textbf{84}, 074016 (2011)}.

\bibitem{Danilkin:2011sh}
I.~V.~Danilkin, V.~D.~Orlovsky and Y.~A.~Simonov,
Hadron interaction with heavy quarkonia,
\href{https://doi.org/10.1103/PhysRevD.85.034012}{Phys. Rev. D \textbf{85}, 034012 (2012)}.

\bibitem{Chen:2011pu}
D.~Y.~Chen, X.~Liu and T.~Matsuki,
Charged bottomonium-like structures in the hidden-bottom dipion decays of $\Upsilon(11020)$,
\href{https://doi.org/10.1103/PhysRevD.84.074032}{Phys. Rev. D \textbf{84}, 074032 (2011)}.

\bibitem{Chen:2012nva}
D.~Y.~Chen, X.~Liu and T.~Matsuki,
$\eta$ transitions between charmonia with meson loop contributions,
\href{https://doi.org/10.1103/PhysRevD.87.054006}{Phys. Rev. D \textbf{87}, 054006 (2013)}.

\bibitem{Chen:2014ccr}
D.~Y.~Chen, X.~Liu and T.~Matsuki,
Explaining the anomalous $\Upsilon(5S)\to \chi_{bJ}\omega$ decays through the hadronic loop effect,
\href{https://doi.org/10.1103/PhysRevD.90.034019}{Phys. Rev. D \textbf{90}, 034019 (2014)}.

\bibitem{Chen:2014sra}
D.~Y.~Chen, X.~Liu and T.~Matsuki,
Observation of $e^+e^-\to \chi_{c0}\omega$ and missing higher charmonium $\psi(4S)$,
\href{https://doi.org/10.1103/PhysRevD.91.094023}{Phys. Rev. D \textbf{91}, 094023 (2015)}.

\bibitem{Wang:2016qmz}
B.~Wang, X.~Liu and D.~Y.~Chen,
Prediction of anomalous $\Upsilon(5S)\to\Upsilon(1^3D_J)\eta$ transitions,
\href{https://doi.org/10.1103/PhysRevD.94.094039}{Phys. Rev. D \textbf{94}, 094039 (2016)}.

\bibitem{Huang:2017kkg}
Q.~Huang, B.~Wang, X.~Liu, D.~Y.~Chen and T.~Matsuki,
Exploring the $\Upsilon (6S)\rightarrow \chi_{bJ}\phi $ and $\Upsilon (6S)\rightarrow \chi_{bJ}\omega $ hidden-bottom hadronic transitions,
\href{https://doi.org/10.1140/epjc/s10052-017-4726-8}{Eur. Phys. J. C \textbf{77}, 165 (2017)}.

\bibitem{Huang:2018cco}
Q.~Huang, H.~Xu, X.~Liu and T.~Matsuki,
Potential observation of the $\Upsilon(6S) \to \Upsilon(1^3D_J) \eta$ transitions at Belle II,
\href{https://doi.org/10.1103/PhysRevD.97.094018}{Phys. Rev. D \textbf{97}, 094018 (2018)}.

\bibitem{Huang:2018pmk}
Q.~Huang, X.~Liu and T.~Matsuki,
Proposal of searching for the $\Upsilon(6S)$ hadronic decays into $\Upsilon(nS)$ plus $\eta^{(\prime)}$,
\href{https://doi.org/10.1103/PhysRevD.98.054008}{Phys. Rev. D \textbf{98}, 054008 (2018)}.

\bibitem{Li:2021jjt}
Y.~S.~Li, Z.~Y.~Bai, Q.~Huang and X.~Liu,
Hidden-bottom hadronic decays of $\Upsilon(10753)$ with a $\eta^{(\prime)}$ or $\omega$ emission,
\href{https://doi.org/10.1103/PhysRevD.104.034036}{Phys. Rev. D \textbf{104}, 034036 (2021)}.

\bibitem{Li:2022leg}
Y.~S.~Li, Z.~Y.~Bai and X.~Liu,
Investigating the $\Upsilon(10753)\to\Upsilon(1^{3}D_{J})\eta$ transitions,
\href{https://doi.org/10.1103/PhysRevD.105.114041}{Phys. Rev. D \textbf{105}, 114041 (2022)}.

\bibitem{Brodsky:1968ea}
S.~J.~Brodsky and J.~R.~Primack,
The electromagnetic interactions of composite systems,
\href{https://doi.org/10.1016/0003-4916(69)90264-4}{Annals Phys. \textbf{52}, 315-365 (1969)}.

\bibitem{Kwong:1988ae}
W.~Kwong and J.~L.~Rosner,
$D$ wave quarkonium levels of the $\Upsilon$ family,
\href{https://doi.org/10.1103/PhysRevD.38.279}{Phys. Rev. D \textbf{38}, 279 (1988)}.

\bibitem{Novikov:1977dq}
V.~A.~Novikov, L.~B.~Okun, M.~A.~Shifman, A.~I.~Vainshtein, M.~B.~Voloshin and V.~I.~Zakharov,
Charmonium and gluons: Basic experimental facts and theoretical introduction,
\href{https://doi.org/10.1016/0370-1573(78)90120-5}{Phys. Rept. \textbf{41}, 1-133 (1978)}.

\bibitem{Schlumpf:1993rm}
F.~Schlumpf,
Magnetic moments of the baryon decuplet in a relativistic quark model,
\href{https://journals.aps.org/prd/abstract/10.1103/PhysRevD.48.4478}{Phys. Rev. D \textbf{48}, 4478-4480 (1993)}.

\bibitem{Kumar:2005ei}
S.~Kumar, R.~Dhir and R.~C.~Verma,
Magnetic moments of charm baryons using effective mass and screened charge of quarks,
\href{https://iopscience.iop.org/article/10.1088/0954-3899/31/2/006}{J. Phys. G \textbf{31}, 141-147 (2005)}.

\bibitem{Ramalho:2009gk}
G.~Ramalho, K.~Tsushima and F.~Gross,
A Relativistic quark model for the Omega-electromagnetic form factors,
\href{https://journals.aps.org/prd/abstract/10.1103/PhysRevD.80.033004}{Phys. Rev. D \textbf{80}, 033004 (2009)}.

\bibitem{Li:2021ryu}
M.~W.~Li, Z.~W.~Liu, Z.~F.~Sun and R.~Chen,
Magnetic moments and transition magnetic moments of $P_c$ and $P_{cs}$ states,
\href{https://journals.aps.org/prd/abstract/10.1103/PhysRevD.104.054016}{Phys. Rev. D \textbf{104}, 054016 (2021)}.

\bibitem{Wang:2022tib}
F.~L.~Wang, H.~Y.~Zhou, Z.~W.~Liu and X.~Liu,
What can we learn from the electromagnetic properties of hidden-charm molecular pentaquarks with single strangeness?,
\href{https://journals.aps.org/prd/abstract/10.1103/PhysRevD.106.054020}{Phys. Rev. D \textbf{106}, 054020 (2022)}.

\bibitem{Wang:2022ugk}
F.~L.~Wang, H.~Y.~Zhou, Z.~W.~Liu and X.~Liu,
Exploring the electromagnetic properties of the $\Xi_c^{(\prime,\,*)} \bar D_s^*$ and $\Omega_c^{(*)} \bar D_s^*$ molecular states,
\href{https://journals.aps.org/prd/abstract/10.1103/PhysRevD.108.034006}{Phys. Rev. D \textbf{108}, 034006 (2023)}.

\bibitem{Deng:2021gnb}
C.~Deng and S.~L.~Zhu,
$T_{cc}^+$ and its partners,
\href{https://journals.aps.org/prd/abstract/10.1103/PhysRevD.105.054015}{Phys. Rev. D \textbf{105}, 054015 (2022)}.

\bibitem{Gao:2021hmv}
F.~Gao and H.~S.~Li,
Magnetic moments of the hidden-charm strange pentaquark states,
\href{https://iopscience.iop.org/article/10.1088/1674-1137/ac8651}{Chin. Phys. C \textbf{46}, 123111 (2022)}.

\bibitem{Zhu:2004xa}
S.~L.~Zhu,
Pentaquarks,
\href{https://www.worldscientific.com/doi/abs/10.1142/S0217751X04019676}{Int. J. Mod. Phys. A \textbf{19}, 3439-3469 (2004)}.

\bibitem{Haghpayma:2006hu}
A.~R.~Haghpayma,
Magnetic moment of the pentaquark $\Theta^+$ state,
\href{https://arxiv.org/abs/hep-ph/0609253}{arXiv:hep-ph/0609253}.

\bibitem{Schlumpf:1992vq}
F.~Schlumpf,
Relativistic constituent quark model of electroweak properties of baryons,
\href{https://journals.aps.org/prd/abstract/10.1103/PhysRevD.47.4114}{Phys. Rev. D \textbf{47}, 4114 (1993)};
\href{https://journals.aps.org/prd/abstract/10.1103/PhysRevD.49.6246.2}{erratum: Phys. Rev. D \textbf{49}, 6246 (1994)}.

\bibitem{Cheng:1997kr}
T.~P.~Cheng and L.~F.~Li,
Why naive quark model can yield a good account of the baryon magnetic moments,
\href{https://journals.aps.org/prl/abstract/10.1103/PhysRevLett.80.2789}{Phys. Rev. Lett. \textbf{80}, 2789-2792 (1998)}.

\bibitem{Ha:1998gf}
P.~Ha and L.~Durand,
Baryon magnetic moments in a QCD based quark model with loop corrections,
\href{https://journals.aps.org/prd/abstract/10.1103/PhysRevD.58.093008}{Phys. Rev. D \textbf{58}, 093008 (1998)}.

\bibitem{Dhir:2009ax}
R.~Dhir and R.~C.~Verma,
Magnetic moments of ($J^P = 3/2^+$) heavy baryons using effective mass scheme,
\href{https://link.springer.com/article/10.1140/epja/i2009-10872-8}{Eur. Phys. J. A \textbf{42}, 243-249 (2009)}.

\bibitem{Majethiya:2009vx}
A.~Majethiya, B.~Patel and P.~C.~Vinodkumar,
Radiative decays of single heavy flavour baryons,
\href{https://link.springer.com/article/10.1140/epja/i2009-10880-8}{Eur. Phys. J. A \textbf{42}, 213-218 (2009)}.

\bibitem{Sharma:2010vv}
N.~Sharma, H.~Dahiya, P.~K.~Chatley and M.~Gupta,
Spin $\frac{1}{2}^+$, spin $\frac{3}{2}^+$ and transition magnetic moments of low lying and charmed baryons,
\href{https://journals.aps.org/prd/abstract/10.1103/PhysRevD.81.073001}{Phys. Rev. D \textbf{81}, 073001 (2010)}.

\bibitem{Sharma:2012jqz}
N.~Sharma, A.~Martinez Torres, K.~P.~Khemchandani and H.~Dahiya,
Magnetic moments of the low-lying ${1/2}^-$ octet baryon resonances,
\href{https://link.springer.com/article/10.1140/epja/i2013-13011-2}{Eur. Phys. J. A \textbf{49}, 11 (2013)}.

\bibitem{Dhir:2013nka}
R.~Dhir, C.~S.~Kim and R.~C.~Verma,
Magnetic moments of bottom baryons: Effective mass and screened charge,
\href{https://journals.aps.org/prd/abstract/10.1103/PhysRevD.88.094002}{Phys. Rev. D \textbf{88}, 094002 (2013)}.

\bibitem{Ghalenovi:2014swa}
Z.~Ghalenovi, A.~A.~Rajabi, S.~X.~Qin and D.~H.~Rischke,
Ground-state masses and magnetic moments of heavy baryons,
\href{https://www.worldscientific.com/doi/abs/10.1142/S0217732314501065}{Mod. Phys. Lett. A \textbf{29}, 1450106 (2014)}.

\bibitem{Girdhar:2015gsa}
A.~Girdhar, H.~Dahiya and M.~Randhawa,
Magnetic moments of $J^P=\frac{3}{2}^+$ decuplet baryons using effective quark masses in chiral constituent quark model,
\href{https://journals.aps.org/prd/abstract/10.1103/PhysRevD.92.033012}{Phys. Rev. D \textbf{92}, 033012 (2015)}.

\bibitem{Majethiya:2011ry}
A.~Majethiya, K.~Thakkar and P.~C.~Vinodkumar,
Spectroscopy and decay properties of  $\Sigma_{b}, \Lambda_{b}$ baryons in quark-diquark model,
\href{https://www.sciencedirect.com/science/article/pii/S057790731630243X?via\%3Dihub}{Chin. J. Phys. \textbf{54}, 495-502 (2016)}.

\bibitem{Thakkar:2016sog}
K.~Thakkar, A.~Majethiya and P.~C.~Vinodkumar,
Magnetic moments of baryons containing all heavy quarks in the quark-diquark model,
\href{https://link.springer.com/article/10.1140/epjp/i2016-16339-4}{Eur. Phys. J. Plus \textbf{131}, 339 (2016)}.

\bibitem{Shah:2016nxi}
Z.~Shah, K.~Thakkar, A.~K.~Rai and P.~C.~Vinodkumar,
Mass spectra and Regge trajectories of $\Lambda_{c}^{+}$, $\Sigma_{c}^{0}$, $\Xi_{c}^{0}$ and $\Omega_{c}^{0}$ baryons,
\href{https://iopscience.iop.org/article/10.1088/1674-1137/40/12/123102}{Chin. Phys. C \textbf{40}, 123102 (2016)}.

\bibitem{Shah:2016vmd}
Z.~Shah, K.~Thakkar and A.~K.~Rai,
Excited state mass spectra of doubly heavy baryons $\Omega_{cc}$, $\Omega_{bb}$ and $\Omega_{bc}$,
\href{https://link.springer.com/article/10.1140/epjc/s10052-016-4379-z}{Eur. Phys. J. C \textbf{76}, 530 (2016)}.

\bibitem{Kaur:2016kan}
A.~Kaur, P.~Gupta and A.~Upadhyay,
Properties of $J^{P}=1/2^{+}$ baryon octets at low energy,
\href{https://doi.org/10.1093/ptep/ptx068}{PTEP \textbf{2017}, 063B02 (2017)}.

\bibitem{Shah:2018bnr}
Z.~Shah and A.~Kumar Rai,
Spectroscopy of the $\Omega_{ccb}$ baryon in the hypercentral constituent quark model,
\href{https://iopscience.iop.org/article/10.1088/1674-1137/42/5/053101}{Chin. Phys. C \textbf{42}, 053101 (2018)}.

\bibitem{Gandhi:2018lez}
K.~Gandhi, Z.~Shah and A.~K.~Rai,
Decay properties of singly charmed baryons,
\href{https://link.springer.com/article/10.1140/epjp/i2018-12318-1}{Eur. Phys. J. Plus \textbf{133}, 512 (2018)}.

\bibitem{Dahiya:2018ahb}
H.~Dahiya,
Transition magnetic moments of $J^P=\frac{3}{2}^+$ decuplet to $J^P=\frac{1}{2}^+$ octet baryons in the chiral constituent quark model,
\href{https://iopscience.iop.org/article/10.1088/1674-1137/42/9/093102}{Chin. Phys. C \textbf{42}, 093102 (2018)}.

\bibitem{Simonis:2018rld}
V.~\v{S}imonis,
Improved predictions for magnetic moments and M1 decay widths of heavy hadrons,
\href{https://arxiv.org/abs/1803.01809}{arXiv:1803.01809}.

\bibitem{Ghalenovi:2018fxh}
Z.~Ghalenovi and M.~Moazzen Sorkhi,
Mass spectra and decay properties of $\Sigma_{{b}}^{}$ and $\Lambda_{{b}}^{}$ baryons in a quark model,
\href{https://link.springer.com/article/10.1140/epjp/i2018-12111-2}{Eur. Phys. J. Plus \textbf{133}, 301 (2018)}.

\bibitem{Gandhi:2019bju}
K.~Gandhi and A.~K.~Rai,
Spectrum of strange singly charmed baryons in the constituent quark model,
\href{https://www.sciencedirect.com/science/article/abs/pii/0370269394914664?via\%3Dihub}{Eur. Phys. J. Plus \textbf{135}, 213 (2020)}.

\bibitem{Rahmani:2020pol}
S.~Rahmani, H.~Hassanabadi and H.~Sobhani,
Mass and decay properties of double heavy baryons with a phenomenological potential model,
\href{https://link.springer.com/article/10.1140/epjc/s10052-020-7867-0}{Eur. Phys. J. C \textbf{80}, 312 (2020)}.

\bibitem{Hazra:2021lpa}
A.~Hazra, S.~Rakshit and R.~Dhir,
Radiative M1 transitions of heavy baryons: Effective quark mass scheme,
\href{https://journals.aps.org/prd/abstract/10.1103/PhysRevD.104.053002}{Phys. Rev. D \textbf{104}, 053002 (2021)}.

\bibitem{Menapara:2021dzi}
C.~Menapara and A.~K.~Rai,
Spectroscopic investigation of light strange $S=-1$ $\Lambda$, $\Sigma$ and $S=-2$ $\Xi$ baryons,
\href{https://iopscience.iop.org/article/10.1088/1674-1137/abf4f4}{Chin. Phys. C \textbf{45}, 063108 (2021)}.

\bibitem{Menapara:2021vug}
C.~Menapara and A.~K.~Rai,
Spectroscopic study of $\mathrm{Strangeness}=-3$ $\Omega^{-}$ baryon,
\href{https://iopscience.iop.org/article/10.1088/1674-1137/ac78d1}{Chin. Phys. C \textbf{46}, 103102 (2022)}.

\bibitem{Mutuk:2021epz}
H.~Mutuk,
The status of $\Xi_\mathrm{{cc}}^{++}$ baryon: investigating quark-diquark model,
\href{https://link.springer.com/article/10.1140/epjp/s13360-021-02256-4}{Eur. Phys. J. Plus \textbf{137}, 10 (2022)}.

\bibitem{Menapara:2022ksj}
C.~Menapara and A.~K.~Rai,
Spectroscopy of light baryons: $\Delta$ resonances,
\href{https://www.worldscientific.com/doi/10.1142/S0217751X22501779}{Int. J. Mod. Phys. A \textbf{37}, 2250177 (2022)}.

\bibitem{Mohan:2022sxm}
B.~Mohan, T.~M.~S., A.~Hazra and R.~Dhir,
Screening of the quark charge and mixing effects on transition moments and M1 decay widths of baryons,
\href{https://journals.aps.org/prd/abstract/10.1103/PhysRevD.106.113007}{Phys. Rev. D \textbf{106}, 113007 (2022)}.

\bibitem{An:2022qpt}
H.~T.~An, S.~Q.~Luo, Z.~W.~Liu and X.~Liu,
Spectroscopy behavior of fully heavy tetraquarks,
\href{https://arxiv.org/abs/2208.03899}{arXiv:2208.03899}.

\bibitem{Wu:2022gie}
T.~W.~Wu and Y.~L.~Ma,
Doubly heavy tetraquark multiplets as heavy antiquark-diquark symmetry partners of heavy baryons,
\href{https://arxiv.org/abs/2211.15094}{arXiv:2211.15094}.

\bibitem{Kakadiya:2022pin}
A.~Kakadiya, Z.~Shah and A.~K.~Rai,
Spectroscopy of $\Omega_{ccc}$ and $\Omega_{bbb}$ baryons,
\href{https://www.worldscientific.com/doi/10.1142/S0217751X22502256}{Int. J. Mod. Phys. A \textbf{37}, 2250225 (2022)}.

\bibitem{Wang:2023bek}
F.~L.~Wang, S.~Q.~Luo and X.~Liu,
Radiative decays and magnetic moments of the predicted $B_c$-like molecules,
\href{https://journals.aps.org/prd/abstract/10.1103/PhysRevD.107.114017}{Phys. Rev. D \textbf{107}, no.11, 114017 (2023)}.
\bibitem{Zhou:2022gra}
H.~Y.~Zhou, F.~L.~Wang, Z.~W.~Liu and X.~Liu,
Probing the electromagnetic properties of the $\Sigma_c^{(*)}D^{(*)}$-type doubly charmed molecular pentaquarks,
\href{https://journals.aps.org/prd/abstract/10.1103/PhysRevD.106.034034}{Phys. Rev. D \textbf{106}, 034034 (2022)}.

\bibitem{Wang:2016dzu}
G.~J.~Wang, R.~Chen, L.~Ma, X.~Liu and S.~L.~Zhu,
Magnetic moments of the hidden-charm pentaquark states,
\href{https://journals.aps.org/prd/abstract/10.1103/PhysRevD.94.094018}{Phys. Rev. D \textbf{94}, 094018 (2016)}.

\bibitem{Liu:2003ab}
Y.~R.~Liu, P.~Z.~Huang, W.~Z.~Deng, X.~L.~Chen and S.~L.~Zhu,
Pentaquark magnetic moments in different models,
\href{https://journals.aps.org/prc/abstract/10.1103/PhysRevC.69.035205}{Phys. Rev. C \textbf{69}, 035205 (2004)}.

\bibitem{Huang:2004tn}
P.~Z.~Huang, Y.~R.~Liu, W.~Z.~Deng, X.~L.~Chen and S.~L.~Zhu,
Heavy pentaquarks,
\href{https://journals.aps.org/prd/abstract/10.1103/PhysRevD.70.034003}{Phys. Rev. D \textbf{70}, 034003 (2004)}.

\bibitem{Lichtenberg:1976fi}
D.~B.~Lichtenberg,
Magnetic moments of charmed baryons in the quark model,
\href{https://journals.aps.org/prd/abstract/10.1103/PhysRevD.15.345}{Phys. Rev. D \textbf{15}, 345 (1977)}.

\bibitem{Li:2017cfz}
H.~S.~Li, L.~Meng, Z.~W.~Liu and S.~L.~Zhu,
Magnetic moments of the doubly charmed and bottom baryons,
\href{https://journals.aps.org/prd/pdf/10.1103/PhysRevD.96.076011}{Phys. Rev. D \textbf{96}, 076011 (2017)}.

\bibitem{Meng:2017dni}
L.~Meng, H.~S.~Li, Z.~W.~Liu and S.~L.~Zhu,
Magnetic moments of the spin-$\frac{3}{2}$ doubly heavy baryons,
\href{https://link.springer.com/article/10.1140/epjc/s10052-017-5447-8}{Eur. Phys. J. C \textbf{77}, 869 (2017)}.

\bibitem{Li:2017pxa}
H.~S.~Li, L.~Meng, Z.~W.~Liu and S.~L.~Zhu,
Radiative decays of the doubly charmed baryons in chiral perturbation theory,
\href{https://www.sciencedirect.com/science/article/pii/S0370269317310080?via\%3Dihub}{Phys. Lett. B \textbf{777}, 169-176 (2018)}.

\bibitem{Wang:2019mhm}
B.~Wang, B.~Yang, L.~Meng and S.~L.~Zhu,
Radiative transitions and magnetic moments of the charmed and bottom vector mesons in chiral perturbation theory,
\href{https://journals.aps.org/prd/abstract/10.1103/PhysRevD.100.016019}{Phys. Rev. D \textbf{100}, 016019 (2019)}.

\bibitem{Simonis:2016pnh}
V.~\v{S}imonis,
Magnetic properties of ground-state mesons,
\href{https://doi.org/10.1140/epja/i2016-16090-5}{Eur. Phys. J. A \textbf{52}, 90 (2016)}.

\bibitem{Sun:2012sy}
Z.~F.~Sun, X.~Liu, M.~Nielsen and S.~L.~Zhu,
Hadronic molecules with both open charm and bottom,
\href{https://journals.aps.org/prd/abstract/10.1103/PhysRevD.85.094008}{Phys. Rev. D \textbf{85}, 094008 (2012)}.

\bibitem{Lytle:2016ixw}
A.~Lytle, B.~Colquhoun, C.~Davies, J.~Koponen and C.~McNeile,
Semileptonic $B_c$ decays from full lattice QCD,
\href{https://doi.org/10.22323/1.273.0069}{PoS \textbf{BEAUTY2016}, 069 (2016)}.

\bibitem{Harrison:2020gvo}
J.~Harrison \textit{et al.} [HPQCD],
$B_c \to J/\psi$ form factors for the full $q^2$ range from lattice QCD,
\href{https://doi.org/10.1103/PhysRevD.102.094518}{Phys. Rev. D \textbf{102}, 094518 (2020)}.

\bibitem{Hu:2019qcn}
X.~Q.~Hu, S.~P.~Jin and Z.~J.~Xiao,
Semileptonic decays $B_c \to (\eta_c,J/\psi) l \bar{\nu}_l $ in the ``PQCD + Lattice'' approach,
\href{https://doi.org/10.1088/1674-1137/44/2/023104}{Chin. Phys. C \textbf{44}, 023104 (2020)}.

\bibitem{Liu:2023kxr}
X.~Liu,
The $B_c$-meson decays into $J/\psi$ plus a light meson in the iPQCD formalism,
\href{https://arxiv.org/abs/2305.00713}{arXiv:2305.00713}.

\bibitem{Leljak:2019eyw}
D.~Leljak, B.~Melic and M.~Patra,
On lepton flavour universality in semileptonic $B_{c}\to \eta_{c}$, $J/\psi$ decays,
\href{https://doi.org/10.1007/JHEP05(2019)094}{JHEP \textbf{05}, 094 (2019)}.

\bibitem{Kiselev:2002vz}
V.~V.~Kiselev,
Exclusive decays and lifetime of $B_c$ meson in QCD sum rules,
\href{https://arxiv.org/abs/hep-ph/0211021}{arXiv:hep-ph/0211021}.

\bibitem{Huang:2007kb}
T.~Huang and F.~Zuo,
Semileptonic $B_c$ decays and charmonium distribution amplitude,
\href{https://doi.org/10.1140/epjc/s10052-007-0333-4}{Eur. Phys. J. C \textbf{51}, 833-839 (2007)}.

\bibitem{Scora:1995ty}
D.~Scora and N.~Isgur,
Semileptonic meson decays in the quark model: An update,
\href{https://doi.org/10.1103/PhysRevD.52.2783}{Phys. Rev. D \textbf{52}, 2783-2812 (1995)}.

\bibitem{Colangelo:1999zn}
P.~Colangelo and F.~De Fazio,
Using heavy quark spin symmetry in semileptonic $B_c$ decays,
\href{https://doi.org/10.1103/PhysRevD.61.034012}{Phys. Rev. D \textbf{61}, 034012 (2000)}.

\bibitem{Ivanov:2000aj}
M.~A.~Ivanov, J.~G.~Korner and P.~Santorelli,
The Semileptonic decays of the $B_c$ meson,
\href{https://doi.org/10.1103/PhysRevD.63.074010}{Phys. Rev. D \textbf{63}, 074010 (2001)}.

\bibitem{Ebert:2003cn}
D.~Ebert, R.~N.~Faustov and V.~O.~Galkin,
Weak decays of the $B_c$ meson to charmonium and $D$ mesons in the relativistic quark model,
\href{https://doi.org/10.1103/PhysRevD.68.094020}{Phys. Rev. D \textbf{68}, 094020 (2003)}.

\bibitem{Ivanov:2006ni}
M.~A.~Ivanov, J.~G.~Korner and P.~Santorelli,
Exclusive semileptonic and nonleptonic decays of the $B_c$ meson,
\href{https://doi.org/10.1103/PhysRevD.73.054024}{Phys. Rev. D \textbf{73}, 054024 (2006)}.

\bibitem{Wang:2007sxa}
W.~Wang, Y.~L.~Shen and C.~D.~Lu,
The study of $B_{c}^{-}\to X(3872)\pi^{-}(K^{-})$ decays in the covariant light-front approach,
\href{https://doi.org/10.1140/epjc/s10052-007-0334-3}{Eur. Phys. J. C \textbf{51}, 841-847 (2007)}.

\bibitem{Wang:2008xt}
W.~Wang, Y.~L.~Shen and C.~D.~Lu,
Covariant light-front approach for $B_{c}$ transition form factors,
\href{https://doi.org/10.1103/PhysRevD.79.054012}{Phys. Rev. D \textbf{79}, 054012 (2009)}.

\bibitem{Wang:2009mi}
X.~X.~Wang, W.~Wang and C.~D.~Lu,
$B_{c}$ to $p$-wave charmonia transitions in covariant light-front approach,
\href{https://doi.org/10.1103/PhysRevD.79.114018}{Phys. Rev. D \textbf{79}, 114018 (2009)}.

\bibitem{Ke:2013yka}
H.~W.~Ke, T.~Liu and X.~Q.~Li,
Transitions of $B_c\to \psi(1S,2S)$ and the modified harmonic oscillator wave function in LFQM,
\href{https://doi.org/10.1103/PhysRevD.89.017501}{Phys. Rev. D \textbf{89}, 017501 (2014)}.

\bibitem{Shi:2016gqt}
Y.~J.~Shi, W.~Wang and Z.~X.~Zhao,
$B_c\to B_{sJ}$ form factors and $B_c$ decays into $B_{sJ}$ in covariant light-front approach,
\href{https://doi.org/10.1140/epjc/s10052-016-4405-1}{Eur. Phys. J. C \textbf{76}, 555 (2016)}.

\bibitem{Chen:2021ywv}
L.~Chen, Y.~W.~Ren, L.~T.~Wang and Q.~Chang,
Form factors of $P\to T$ transition within the light-front quark models,
\href{https://doi.org/10.1140/epjc/s10052-022-10391-0}{Eur. Phys. J. C \textbf{82}, 451 (2022)}.

\bibitem{Zhang:2023ypl}
Z.~Q.~Zhang, Z.~J.~Sun, Y.~C.~Zhao, Y.~Y.~Yang and Z.~Y.~Zhang,
Covariant light-front approach for $B_c$ decays into charmonium: Implications on form factors and branching ratios,
\href{https://arxiv.org/abs/2301.11107}{arXiv:2301.11107}.

\bibitem{Sun:2023iis}
Z.~J.~Sun, S.~Y.~Wang, Z.~Q.~Zhang, Y.~Y.~Yang and Z.~Y.~Zhang,
Semileptonic $B_{c}$ meson decays to S-wave charmonia and $X(3872)$ within the covariant light-front approach,
\href{https://arxiv.org/abs/2308.03114}{arXiv:2308.03114}.

\bibitem{Yao:2021pyf}
Z.~Q.~Yao, D.~Binosi, Z.~F.~Cui and C.~D.~Roberts,
Semileptonic $B_c \to \eta_c J/\psi$ transitions,
\href{https://doi.org/10.1016/j.physletb.2021.136344}{Phys. Lett. B \textbf{818}, 136344 (2021)}.

\bibitem{Wirbel:1985ji}
M.~Wirbel, B.~Stech and M.~Bauer,
Exclusive Semileptonic Decays of Heavy Mesons,
\href{https://doi.org/10.1007/BF01560299}{Z. Phys. C \textbf{29}, 637 (1985)}.

\bibitem{Jaus:1999zv}
W.~Jaus,
Covariant analysis of the light front quark model,
\href{https://doi.org/10.1103/PhysRevD.60.054026}{Phys. Rev. D \textbf{60}, 054026 (1999)}.

\bibitem{Cheng:2003sm}
H.~Y.~Cheng, C.~K.~Chua and C.~W.~Hwang,
Covariant light front approach for $s$-wave and $p$-wave mesons: Its application to decay constants and form-factors,
\href{https://doi.org/10.1103/PhysRevD.69.074025}{Phys. Rev. D \textbf{69}, 074025 (2004)}.

\bibitem{Verma:2011yw}
R.~C.~Verma,
Decay constants and form factors of $s$-wave and $p$-wave mesons in the covariant light-front quark model,
\href{https://doi.org/10.1088/0954-3899/39/2/025005}{J. Phys. G \textbf{39}, 025005 (2012)}.

\bibitem{Bourrely:2008za}
C.~Bourrely, I.~Caprini and L.~Lellouch,
Model-independent description of $B\to \pi\ell\nu$ decays and a determination of $\vert V_{ub} \vert$,
\href{https://doi.org/10.1103/PhysRevD.82.099902}{Phys. Rev. D \textbf{82}, 099902 (2010)}.

\bibitem{Chen:2017vgi}
K.~Chen, H.~W.~Ke, X.~Liu and T.~Matsuki,
Estimating the production rates of $D$-wave charmed mesons via the semileptonic decays of bottom mesons,
\href{https://doi.org/10.1088/1674-1137/43/2/023106}{Chin. Phys. C \textbf{43}, 023106 (2019)}.

\bibitem{Khodjamirian:2010vf}
A.~Khodjamirian, T.~Mannel, A.~A.~Pivovarov and Y.~M.~Wang,
Charm-loop effect in $B \to K^{(*)} \ell^{+} \ell^{-}$ and $B\to K^*\gamma$,
\href{https://doi.org/10.1007/JHEP09(2010)089}{JHEP \textbf{09}, 089 (2010)}.

\bibitem{Cheng:2017smj}
S.~Cheng, A.~Khodjamirian and J.~Virto,
$B\to\pi\pi$ form factors from light-cone sum rules with $B$-meson distribution amplitudes,
\href{https://doi.org/10.1007/JHEP05(2017)157}{JHEP \textbf{05}, 157 (2017)}.

\bibitem{LHCb:2017vlu}
R.~Aaij \textit{et al.} [LHCb],
Measurement of the ratio of branching ratios $\mathcal{B}(B_c^+\,\to\,J/\psi\tau^+\nu_\tau)$/$\mathcal{B}(B_c^+\,\to\,J/\psi\mu^+\nu_\mu)$,
\href{https://doi.org/10.1103/PhysRevLett.120.121801}{Phys. Rev. Lett. \textbf{120}, 121801 (2018)}.

\bibitem{LHCb:2016utz}
R.~Aaij \textit{et al.} [LHCb],
Study of $B_c^+$ decays to the $K^+K^-\pi^+$ final state and evidence for the decay $B_c^+\to\chi_{c0}\pi^+$,
\href{https://doi.org/10.1103/PhysRevD.94.091102}{Phys. Rev. D \textbf{94}, 091102 (2016)}.

\bibitem{Mezoir:2008vx}
E.~H.~Mezoir and P.~Gonzalez,
Is the spectrum of highly excited mesons purely coulombian?,
\href{https://doi.org/10.1103/PhysRevLett.101.232001}{Phys. Rev. Lett. \textbf{101}, 232001 (2008)}.

\bibitem{Luo:2021dvj}
S.~Q.~Luo, B.~Chen, X.~Liu and T.~Matsuki,
Predicting a new resonance as charmed-strange baryonic analog of $D^*_{s0}(2317)$,
\href{https://doi.org/10.1103/PhysRevD.103.074027}{Phys. Rev. D \textbf{103}, 074027 (2021)}.

\end{thebibliography}
\end{document}